\documentclass[3p,number,sort&compress]{elsarticle}

\usepackage{amsfonts,amsmath,bm}
\usepackage{booktabs}
\usepackage{lineno,hyperref}
\modulolinenumbers[5]
\hypersetup{colorlinks=true}
\usepackage{siunitx}
\usepackage{multirow}
\usepackage{subcaption}
\usepackage{floatrow}
\newfloatcommand{capbtabbox}{table}[][\FBwidth]
\journal{Computer Methods in Applied Mechanics and Engineering}

\newcommand{\LF}{\mathrm{LF}}
\newcommand{\HF}{\mathrm{HF}}
\newcommand{\MC}{\mathrm{MC}}
\newcommand{\ML}{\mathrm{ML}}

\newlength{\tempwidth}

\newcommand{\columnname}[1]
{\makebox[\tempwidth][c]{#1}}

\begin{document}

\begin{frontmatter}

\title{Multilevel and multifidelity uncertainty quantification for cardiovascular hemodynamics\tnoteref{t1,t2}}
\tnotetext[t1]{\url{https://doi.org/10.1016/j.cma.2020.113030}}
\tnotetext[t2]{\textcopyright \hspace{2pt}2020. This manuscript version is made available under the \href{http://creativecommons.org/licenses/by-nc-nd/4.0/}{CC-BY-NC-ND 4.0 license.}}

\author[icme]{Casey~M.~Fleeter}

\author[snl]{Gianluca~Geraci}

\author[und]{Daniele~E.~Schiavazzi}

\author[ucsd]{Andrew~M.~Kahn}

\author[icme,peds]{Alison~L.~Marsden\corref{cor1}}
\ead{amarsden@stanford.edu}

\cortext[cor1]{Corresponding author}
\address[icme]{Institute for Computational and Mathematical Engineering, Stanford University, Stanford, CA, USA}
\address[snl]{Center for Computing Research, Sandia National Laboratories, Albuquerque, NM, USA}
\address[und]{Department of Applied and Computational Mathematics and Statistics, University of Notre Dame, Notre Dame, IN, USA}
\address[ucsd]{Department of Medicine, University of California San Diego, La Jolla, CA, USA}
\address[peds]{Departments of Pediatrics and Bioengineering, Stanford University, Stanford, CA, USA}

\begin{abstract}
Standard approaches for uncertainty quantification in cardiovascular modeling pose challenges due to the large number of uncertain inputs and the significant computational cost of realistic three-dimensional simulations. 
We propose an efficient uncertainty quantification framework utilizing a multilevel multifidelity Monte Carlo (MLMF) estimator to improve the accuracy of hemodynamic quantities of interest while maintaining reasonable computational cost. This is achieved by leveraging three cardiovascular model fidelities, each with varying spatial resolution to rigorously quantify the variability in hemodynamic outputs. We employ two low-fidelity models (zero- and one-dimensional) to construct several different estimators. Our goal is to investigate and compare the efficiency of estimators built from combinations of these two low-fidelity model alternatives and our high-fidelity three-dimensional models.
We demonstrate this framework on healthy and diseased models of aortic and coronary anatomy, including uncertainties in material property and boundary condition parameters.
Our goal is to demonstrate that for this application it is possible to accelerate the convergence of the estimators by utilizing a MLMF paradigm. Therefore, we compare our approach to single fidelity Monte Carlo estimators and to a multilevel Monte Carlo approach based only on three-dimensional simulations, but leveraging multiple spatial resolutions. We demonstrate significant, on the order of 10 to 100 times, 
reduction in total computational cost with the MLMF estimators. We also examine the differing properties of the MLMF estimators in healthy versus diseased models, as well as global versus local quantities of interest. As expected, global quantities such as outlet pressure and flow show larger reductions than local quantities, such as those relating to wall shear stress, as the latter rely more heavily on the highest fidelity model evaluations. Similarly, healthy models show larger reductions than diseased models. In all cases, our workflow coupling Dakota's MLMF estimators with the SimVascular cardiovascular modeling framework makes uncertainty quantification feasible for constrained computational budgets.
\end{abstract}


\end{frontmatter}

\section{Introduction} \label{sec:Intro}

Cardiovascular disease is the leading cause of mortality worldwide, as well as the leading cause of death for both men and women in the United States~\cite{GSB_2017,HERON_2018}. Computational models of the cardiovascular system are increasingly adopted due to the hemodynamic insights they provide, which can be beneficial in diagnosis, assessment of disease progression risk, and treatment planning for cardiovascular disease. These models can be generated in a non-invasive fashion using routinely collected clinical data~\cite{TAYLOR_1998}, and their applications cover a wide range of diseases and anatomies. Coronary artery disease is the most prevalent cause of death in the United States~\cite{Benajmin_2018}, and has thus been the subject of many modeling studies. These include computing fractional flow reserve for patients with coronary stenoses~\cite{TAYLOR_2013} and assessing differences in the hemodynamic and mechanical response of arterial and venous grafts towards determining cause of vein graft failure~\cite{Ramachandra_2016}. 
Computational models have also been applied to congenital heart disease for prognosis and treatment planning, from testing designs for surgical interventions of single ventricle congenital heart disease~\cite{KUNG2013,Schiavazzi_Hemodynamic_2015,Verma_ABG_2018} to thrombotic risk assessment for Kawasaki disease~\cite{Sengupta2012,GRANDE_GUTIERREZ_2017,Grande2019}, and assessment of disease progression and mechanical stimuli in pulmonary hypertension~\cite{Yang_2018,Yang2019}. Hemodynamic studies are also used for additional disease cases, such as analyzing blood flow patterns in abdominal aortic aneurysms~\cite{Suh2011,Arzani2012} and predicting the development and rupture of cerebral aneurysms~\cite{Piccinelli2009,Cebral264,Cebral2550}.

Realistic three-dimensional hemodynamic simulations require segmentation of vascular anatomies from medical image data, followed by numerical solution of the equations governing blood flow in elastically deformable vessels. This is typically achieved through a complex and expensive workflow. 
However, various simplifying assumptions can be made to generate models of intermediate complexity. 
\emph{One-dimensional} hemodynamic models are formulated by integrating the Navier-Stokes equations across the vessel cross sections, with additional assumptions on the properties of the fluid and the vascular walls~\cite{HUGHES1973,Mirramezani2019,QUARTERONI2016193}. 
A linearization of the incompressible Navier-Stokes equations around rest conditions leads to an even simpler \emph{zero-dimensional} (or lumped parameter) formulation where vascular networks are analyzed using analogous electrical circuits~\cite{MOGHADAM2013,Quarteroni2001,Migliavacca2001,Moghadam2001}.
These simplified representations are computationally inexpensive, especially when compared to the cost of a full three-dimensional model. 

Unfortunately, all models of the cardiovascular system, as well as biological and biomedical systems more generally, are intrinsically affected by model and data uncertainty~\cite{schiavazzi2016uncertainty}. These uncertainties arise from modeling choices, measurement errors, and intra- and inter-patient variability.
Therefore, relying on a solely deterministic framework provides limited information, as results cannot include statistical distributions or confidence intervals, and hinders their application in the clinical setting.
Here, we aim to account for these uncertainties and provide confidence intervals through uncertainty quantification (UQ). By adopting a stochastic framework, model parameters are sampled from appropriate probability distributions which are either assumed, relying on existing literature and clinical data, or assimilated from available patient-specific data.

Recently, UQ has gained traction in the field of cardiovascular modeling, primarily utilizing UQ techniques centered around a single model complexity, primarily three-dimensional computational fluid dynamics (CFD) simulations or lower complexity one-dimensional models. Recent studies have investigated the impact of geometry and boundary condition parameters on coronary fractional flow reserve~\cite{SANKARAN_2016}, demonstrated a multi-resolution approach to quantify boundary condition uncertainties~\cite{SCHIAVAZZI_2017196} and in conjugation with random fields~\cite{TRAN_2019402} for coronary artery bypass grafts, and performed stochastic collocation in a one-dimensional arterial network~\cite{Chen_2013}, along with others ~\cite{Sankaran_2010,ECK_2017188,TRAN_2017128,Schiavazzi_2017,Biehler_2018,MARQUIS_20189,Brault2017,BOCCADIFUOCO2018}.

Several challenges arise when quantifying uncertainty in cardiovascular simulations.
First, there are typically multiple sources of uncertainty to account for and propagate through the model. 
Such uncertainties include, but are not limited to, boundary condition parameters, spatial variability of material properties in vascular tissue, and operator dependent vessel lumen segmentation leading to an uncertain vascular model anatomy.
With myriad sources of uncertainty, it follows that any UQ scheme must account for a large number of possibly heterogeneous random inputs; this leads to the so-called \emph{curse of dimensionality}, which poses significant challenges for generalized polynomial chaos expansions~\cite{xiu2002wiener} and stochastic collocation~\cite{babuvska2007stochastic} due to the fast increase of the computational complexity in high-dimensions. 
Second, for any realization of uncertain input parameters, a three-dimensional model solution is computationally expensive. Three-dimensional simulations involve discrete representations with millions of degrees of freedom that are solved in parallel over multiple computational nodes. Maintaining a reasonable computational cost therefore becomes challenging when relying solely on three-dimensional simulations.

Monte Carlo estimators~\cite{metropolis1949monte} have, in this context, many desirable properties. 
They are unbiased, offer flexibility with respect to heterogeneous input sources, and the associated variance depends only on the true variance and number of model evaluations, not on the problem dimensionality or regularity. However, one can typically afford only a small number of these evaluations when working with expensive deterministic solvers.
Multilevel and multifidelity estimators lead to a reduced variance compared to their ``vanilla'' Monte Carlo analogues for the same computational cost. The interested reader is referred to~\cite{Giles_2015,Peher_2016} for an in depth discussion of these methods. Another alternative to multilevel estimators, which rely on spatial resolution, and multifidelity estimators, which are based on a control variate approach, are so-called multilevel multifidelity (MLMF) methods~\cite{Nobile2015,Geraci_A_2015,Geraci_A_2017,Fairbanks2016} which combine the efficiency of multilevel Monte Carlo across discretization levels with embedded control variates based on a low-fidelity model at each level. In this way, MLMF estimators leverage a cascade of varying-fidelity  models ranging from computationally inexpensive, patient-agnostic lumped parameter models to computationally intensive, patient-specific models of blood flow. We provide an overview of these approaches in~\autoref{sec:UQ Methods}. 

The main contribution of this paper is to introduce the integration of multilevel multifidelity estimators for uncertainty quantification with cardiovascular models. This framework integrates and manages multiple hemodynamic solvers from the SimVascular open source package~\cite{Updegrove_SimVascular_2016} with Sandia National Laboratories' Dakota toolkit~\cite{Dakota_Adams_Version} for uncertainty analysis in a manner designed to automatically compute MLMF estimators for a variety of quantities of interest (QoIs). Due to the much lower computational cost of one- and zero-dimensional models, the MLMF estimators allow stochastic analysis to be performed at a drastically reduced computational cost. The specific contributions of this paper are
\begin{enumerate}
	\item Definition of model fidelities (3D, 1D and 0D) with corresponding discretization levels obtained after a thorough convergence study.
	\item Creation of a semi-automated framework enabling Dakota to directly manage and invoke the SimVascular modeling simulations.
	\item Numerical experiments utilizing the above framework to demonstrate the performance of the estimators for cardiovascular hemodynamics on pulsatile simulations with realistic boundary conditions, healthy and diseased models, and a large number of quantities of interest not previously addressed in the literature.
\end{enumerate}

This paper is organized as follows. \hyperref[sec:Cardio Modeling Background]{Sections~\ref*{sec:Cardio Modeling Background}} and~\ref{sec:UQ Methods} provide background on the cardiovascular modeling and uncertainty quantification techniques used in the study, respectively. \hyperref[sec:Methods]{Section~\ref*{sec:Methods}} describes the workflow, including further details on model construction as well as the selection of uncertain parameters. \hyperref[sec:Results]{Section~\ref*{sec:Results}} provides UQ results for four patient anatomies, providing comparisons of the performance of uncertainty quantification estimators for healthy versus diseased models and local versus global QoIs, as well as comparisons of different uncertainty quantification estimators. \hyperref[sec:Discussion]{Sections~\ref*{sec:Discussion}} and~\ref{sec:Conclusion} provide discussions and draw conclusions.

\section{Cardiovascular Modeling} \label{sec:Cardio Modeling Background}

\begin{figure}[!ht]
\centering
\includegraphics[width=0.6\textwidth]{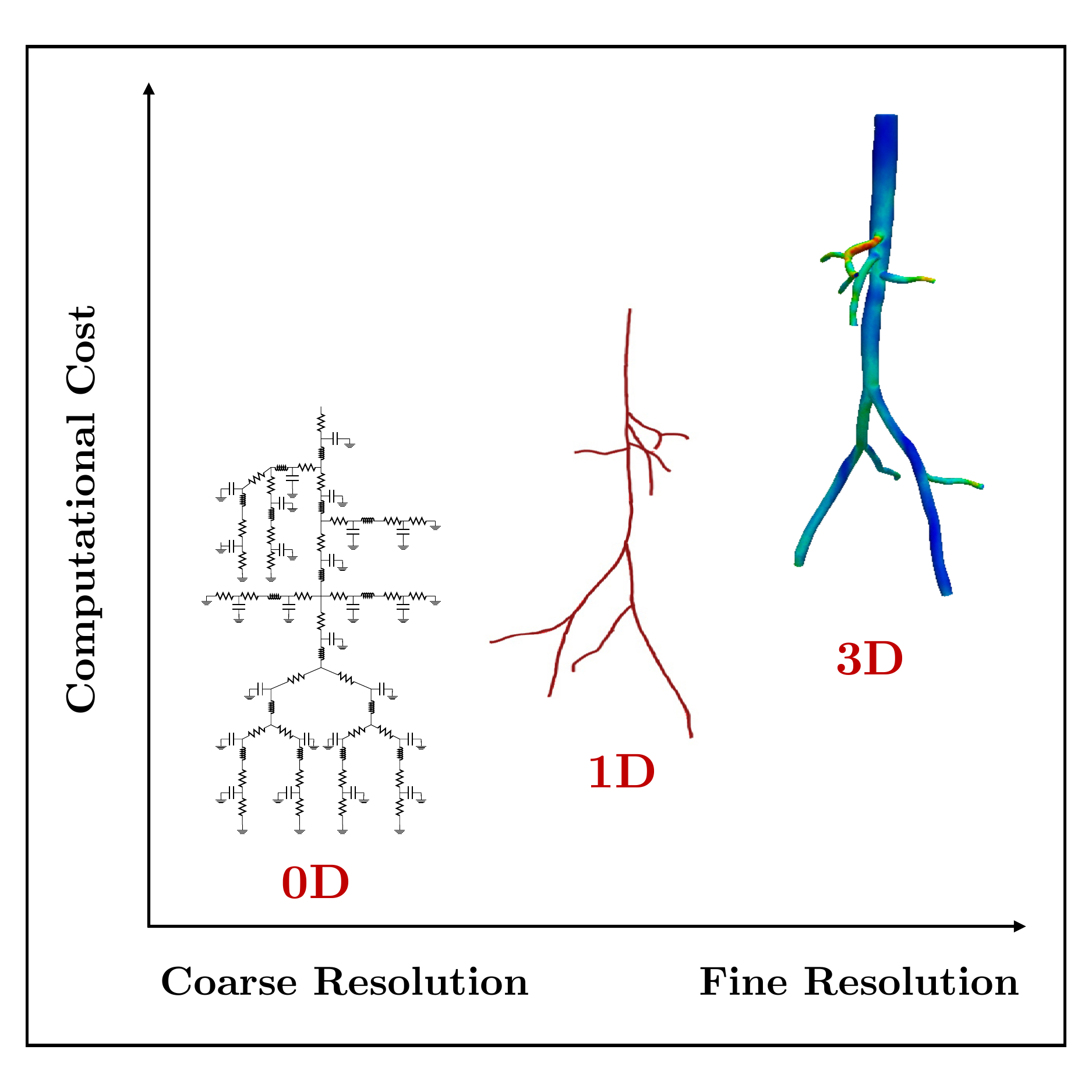}
\caption{Improved fine-scale resolution for quantities of interest comes at the expense of higher computational cost when solving the model hemodynamics with different fidelity methods.}
\label{fig:FidelityTrade}
\end{figure}

We first illustrate the three model formulations in decreasing order of complexity (3D, 1D and 0D) and discuss their trade-offs (\autoref{fig:FidelityTrade}). While zero-dimensional formulations are computationally inexpensive, these models provide only a coarse spatial representation of the quantities of interest. One-dimensional models, while somewhat more expensive, provide finer grained resolution for QoIs along the vessel length, but not across vessel cross-sections. Finally, while the most computationally expensive by far, the three-dimensional CFD models are well-resolved even for local flow features, allowing detailed examination of quantities of interest such as wall shear stress (WSS) in specific regions of interest.

\subsection{Three-dimensional models} \label{sec:Cardio Modeling 3D}

Three-dimensional anatomical models are typically constructed from medical image data of specific patients. In our workflow, these images are imported into the SimVascular open source platform~\cite{Updegrove_SimVascular_2016}, which provides a pipeline for model creation, generation of tetrahedral meshes, application of physiologic boundary conditions, and finite element solution. More specifically, the modeling pipeline is as follows. First, the user obtains patient-specific medical image data. Next, centerline paths are generated for the vessels of interest. The lumen is then segmented along, but perpendicular to, the centerline path in a semi-automated fashion. Next, the model is lofted using boolean operations to merge the vessels together. Meshing then occurs, with many specific options for local mesh{} refinement and model smoothing. At this point, the model is ready for the application of boundary conditions, which range from prescribed flow or pressure values to a geometrically multi-scale approach~\cite{QUARTERONI2016193,Formaggia1999} where zero-dimensional lumped parameter networks are used to model the upstream and downstream circulation~\cite{MOGHADAM2013,VIGNONCLEMENTEL2006}. Finally, the model is solved with a stabilized incompressible Navier-Stokes finite element solver ~\cite{EsmailyMoghadam2011,ESMAILYMOGHADAM201540} with linear tetrahedral elements and second order implicit generalized-$\alpha$ time integration~\cite{JANSEN2000305}. A custom linear solver with specialized preconditioning increases solver efficiency~\cite{EsmailyMoghadam2013,Seo2019}.
SimVascular post-processing tools are then used to extract hemodynamic indicators of interest, such as pressures and wall shear stress, known to be correlated in the literature with endothelial damage and thrombus formation~\cite{chatzizisis2007role}. 
Hemodynamic simulations can be run with rigid or deformable walls, i.e., with or without accounting for the mutual interaction between fluid and structure. In this study the coupled momentum method (CMM)~\cite{Figueroa_A_2006}, which assumes a linear elastic constitutive model characterized by membrane and shear stiffness, is selected to model fluid-structure interaction (FSI). By using a monolithic coupling approach, CMM FSI results in a monolithic algorithm in which the solid momentum contributions are embedded, using the same degrees of freedom, in the fluid equations. The CMM formulation is available in the open source solver package svSolver, distributed as part of SimVascular.

Following the derivation in~\cite{Figueroa_A_2006}, blood flow is approximated as flow of incompressible Newtonian fluid in a domain $D \subset \mathbb{R}^3$. The boundary $\Gamma$ of domain $D$ is considered to be divided into three partitions $\Gamma = \partial D = \partial D_g \cup \partial D_h \cup \partial D_s$, where Dirichlet, Neumann, and wall boundary conditions are applied, respectively. The mesh on $D$ consists of $n_{el}$ linear tetrahedral elements $D_e$, $e = 1, \dots, n_{el}$. To prevent rigid body motion, the perimeters of all inlet and outlet vessels (denoted as $\Gamma_s$) are mechanically fixed by applying zero relative displacement and zero velocity boundary conditions. Therefore, the combined weak form for the coupling of the stabilized incompressible Navier-Stokes equations and the CMM FSI method can be written as
\begin{equation}\label{equ:3DNS}
\begin{split}
\int_D \Bigl[\vec{w} \cdot (\rho \frac{\partial\vec{v}}{\partial t} + \rho\vec{v} \cdot \nabla\vec{v} - \vec{f}) + \nabla\vec{w} : (-p\underline{I} + \underline{\tau}) - \nabla q \cdot\vec{v} \Bigr]d\vec{x} - \int_{\partial D_h} \vec{w}\cdot\vec{h}ds + \int_{\partial D_h}qv_nds + \\
\int_{\partial D_g}qv_nds + \sum_{e=1}^{n_{el}}\int_{D_e}\nabla q \cdot \frac{\tau_M}{\rho}\vec{\mathcal{L}}(\vec{v},p)d\vec{x} + \sum_{e=1}^{n_{el}}\int_{D_e} \Bigl[(\vec{v}\cdot\nabla)\vec{w} \cdot \tau_M\vec{\mathcal{L}}(\vec{v},p)+\nabla\cdot\vec{w}\tau_C\nabla\cdot\vec{v}\Bigr]d\vec{x} \\
+ \sum_{e=1}^{n_{el}}\int_{D_e}\Bigl[ \vec{w}\cdot(\rho\hat{\vec{v}}\cdot\nabla\vec{v}) + (\vec{\mathcal{L}}(\vec{v},p)\cdot\nabla\vec{w})(\overline{\tau}\vec{\mathcal{L}}(\vec{v},p)\cdot\vec{v}) \Bigr]d\vec{x} \\
+ \zeta\int_{\partial D_s} \Bigl[\vec{w}\cdot\rho_s\frac{\partial\vec{v}}{\partial t} + \nabla\vec{w} : \underline{\sigma^s}(\vec{u}) \Bigr]ds - \zeta\int_{\Gamma_s}\vec{w}\cdot\vec{h_s}dl + \int_{\partial D_s}qv_nds = 0,
\end{split}
\end{equation} 
where $\vec{v}$ is velocity, $p$ is pressure, $\rho$ is blood density, $\underline{\tau} = \mu(\nabla\vec{v} + (\nabla\vec{v})^T)$ is the viscous stress tensor where $\mu$ is the dynamic viscosity, and $\zeta$ and $\underline{\sigma^s}$ are the prescribed vessel wall thickness and stress tensor, respectively. Vectors and matrices are denoted by $\vec{\cdot}$ and $\underline{\cdot}$, respectively. These equations lend themselves to the choice of uncertain parameters we will discuss in \autoref{sec:Params}.

\subsection{One-dimensional models} \label{sec:Cardio Modeling 1D}

One dimensional formulations for cardiovascular hemodynamics result from a long history of research and experiments, whose origins can be traced back to Euler's work on blood flow in elastic arteries~\cite{Euler}. For our purposes, an in-house stabilized finite element solver is used for the one-dimensional hemodynamics.
Its specific formulation is adapted from Hughes and Lubliner~\cite{HUGHES1973}, with implementation details discussed in~\cite{Vignon_A_2004,VignonPhD_2006,Wan_2002}. 
Blood is simulated as a Newtonian fluid, assuming velocity in the axial direction ($z$) of each cylindrical branch, constant pressure over the vessel cross section, and a non-slip boundary condition applied at the vessel lumen. 
The governing equations are obtained by integrating the incompressible Navier-Stokes equations over the cross section of a deformable cylindrical domain.
A constitutive model relates pressure to change in cross-sectional area.
The resulting conservation equations for mass and momentum are, respectively~\cite{HUGHES1973},
\begin{align}
\begin{split}
\frac{\partial A}{\partial t} + \frac{\partial Q}{\partial z} &= -\psi \\[10pt]
\frac{\partial Q}{\partial t} + \frac{\partial }{\partial z}\left[(1+\delta)\frac{Q^2}{A}\right] + \frac{A}{\rho}\frac{\partial p}{\partial z} &= Af + N \frac{Q}{A} + v\frac{\partial^2 Q}{\partial z^2}.
\end{split}
\end{align}

The solution variables are the cross-sectional area $A$ and flow rate $Q$, and model parameters are blood density $\rho$, external force $f$, kinematic viscosity $\nu$, and an outflow function $\psi$ for permeable vessels. To prescribe a velocity profile over the cross-section, parameters $\delta$ and $N$ are defined as
\begin{align}
\begin{split}
\delta = \frac{1}{A}\int_A(\phi^2-1)\,ds \quad\text{and}\quad N &= \nu \int_{\partial A}\frac{\partial \phi}{\partial m}\,dl.
\end{split}
\end{align}

Pressure continuity and conservation of mass are enforced at the bifurcations via Lagrange multipliers. A single branch point is presented without loss of generality, as this case easily generalizes to multiple branch points. For the case of one branch point $l$ with $m$ inlets and $n$ outlets, the following $m+n$ constraints are enforced:
\begin{align}
\begin{split}
\sum_{C=1}^m Q_C^{\text{in}} - \sum_{C=1}^n Q_C^{\text{out}} = 0 \\[2pt]
p_C^{\text{in}} - p_l^{\text{in}} = 0 \quad C = 2\dots m \\[2pt]
p_C^{\text{out}} - p_l^{\text{in}} = 0 \quad C = 1\dots n
\end{split}
\end{align}
The first equation enforces conservation of mass. The second equation enforces pressure continuity among the $m$ inlet vessels, with one vessel chosen without loss of generality as a reference vessel. The pressure continuity is enforced between this reference branch and each of the $m-1$ inlet daughter branches. The third equation enforces pressure continuity between the native vessel to the $n$ outlet vessels, with one constraint for each outlet vessel. With Lagrange multipliers, these constraints are formulated as a potential function $Z$ as
\begin{align}
Z &= \lambda_Q\Bigg[\sum_{C=1}^m Q_C^{\text{in}} - \sum_{C=1}^n Q_C^{\text{out}}\Bigg] +\sum_{C=2}^m \big[\lambda_{p_{(C-1)}} (p_C^{\text{in}} - p_l^{\text{in}})\big] + \sum_{C=1}^n \big[\lambda_{p_{(C-1+m)}} (p_C^{\text{out}} - p_l^{\text{in}})\big].
\end{align}
The details of the numerical implementation of the Lagrange multipliers and potential function are available in ~\cite{Wan_2002}. Branch losses are not included here, though empirical relations for branch losses have been considered in prior work and could be easily included in future studies. Such minor loss values at branch junctions are implemented and explained in~\cite{Steele2001}.

Finally, a constitutive relationship is needed to complete the system of equations, relating the pressure $p$ to the cross-sectional area: 
\begin{equation}
p(z,t) = \bar{p}(A(z,t),z,t).
\end{equation}
Specifically, we use a linear elastic model, consistent with the CMM formulation discussed in the next section, given by
\begin{equation}
\bar{p}(A(z,t),z,t) = p_0(z) + \left(\frac{E\,h}{r_0(z)}\right)\left(\sqrt{\frac{A(z,t)}{A_0(z)}} - 1\right),
\end{equation}
where $p_0(z) = p(z,0)$ is the initial undeformed pressure, $E$ is the elastic modulus of the vascular tissue, $h$ the wall thickness, and $r_0(z)$ and $A_{0}(z)$ are the undeformed inner radius and area, respectively.

A weak (integral) formulation of these equations is solved on each segment using a stabilized finite element method in space and a discontinuous Galerkin method in time. The non-linear algebraic system of equations is finally solved using modified Newton iterations. Pressure and cross-sectional area continuity are enforced at the joints between two segments through Lagrange multipliers. Further details on the numerical implementation are provided in~\cite{Wan_2002}.

\subsection{Zero-dimensional models} \label{sec:Cardio Modeling 0D}

A lumped parameter network representation of blood flow is used as our lowest-fidelity model. This is a circuit layout formulated by hydrodynamic analogy in terms of flow rate (electrical current) and pressure differences (voltage), where each circuit element is associated with an algebraic or differential equation, i.e.
\begin{equation}
\text{resistor}\,\,\Delta P = R\,Q,\,\,\text{capacitor}\,\,Q = C\,\frac{dP}{dt},\,\,\text{inductor}\,\,\Delta P = L\,\frac{dQ}{dt}.
\end{equation}
These equations are assembled in a system of ODEs which can be efficiently solved, for example, with a $4^\mathrm{th}$--order Runge-Kutta scheme. 
Resistors are used to represent viscous dissipation on the vessel walls, capacitors are used to represent vascular tissue compliance, and inductors are used to represent the inertia of blood. A Poiseuille flow assumption is used to determine the model parameters for these circuit elements~\cite{milisic_quarteroni_2004}, i.e.
\begin{equation}
R = \frac{8\,\mu\,l}{\pi\,r^4},\,\,C = \frac{3\,l\,\pi\,r^3}{2\,E\,h},\,\,L = \frac{l\,\rho}{\pi\,r^2},
\end{equation}
where $\mu$ is the dynamic viscosity of blood, $l$ the vessel length, $r$ the vessel radius, $E$ the elastic modulus, $h$ the wall thickness and $\rho$ the density of blood. These models have been used extensively to model the heart and circulatory system, achieving remarkably realistic flow and pressure waveforms and allowing for the assessment of physiologic changes to treatments~\cite{Migliavacca2006,Schiavazzi_2017,Corsini2011,KUNG2013}.

\subsection{Model fidelity validation and generation} \label{sec:Cardio Modeling Validation}

To take full advantage of multiple fidelities in uncertainty propagation, generation of lower fidelity models from existing three-dimensional models should be accomplished with little effort. We therefore implemented a conversion tool which automatically extracts the one-dimensional model geometry from the centerlines or two-dimensional segmentations constructed as part of the three-dimensional modeling process, described in~\autoref{sec:Cardio Modeling 3D}. This allows one-dimensional models to be obtained at a minimal cost once a three-dimensional model has been built. The same inlet and outlet boundary conditions can be applied to the one- and three-dimensional models.

\begin{figure}[!ht]
\centering
\includegraphics[width=1.\textwidth]{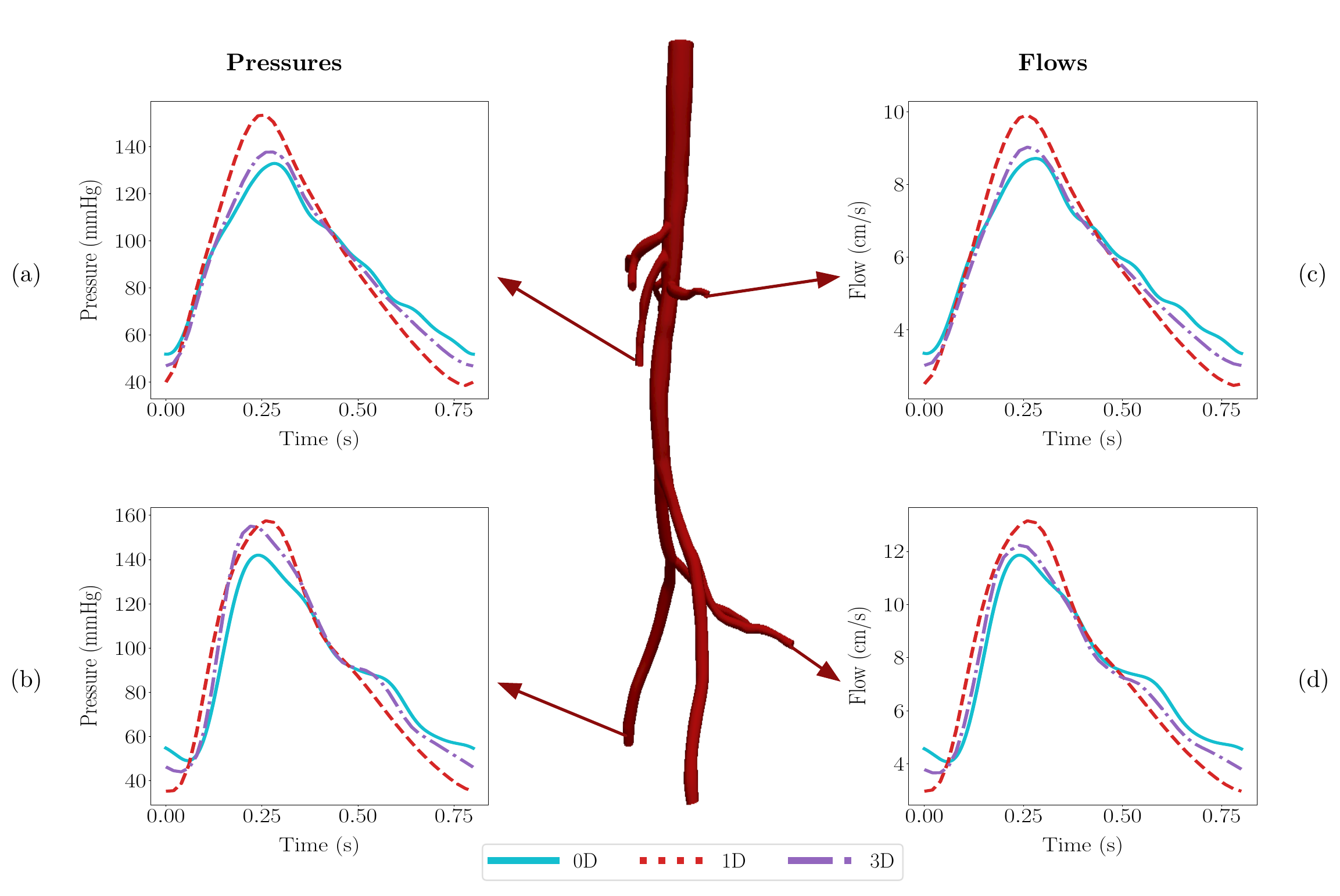}
\caption{Agreement for outlet flow and pressure quantities of interest is seen between 3D, 1D, and 0D models with resistance boundary conditions and vessel compliance built into the model. (a) Pressure at superior mesenteric artery. (b) Pressure at right iliac artery. (c) Flow at right renal artery. (d) Flow at left internal iliac artery.}
\label{fig:ModelValidate}
\end{figure}

To verify the agreement between low fidelity and three-dimensional model results, we performed a preliminary validation study on the aorto-femoral model. Results from all three model formulations were compared under steady and pulsatile flow conditions and with resistance and RCR boundary conditions at the model outlets. One- and zero-dimensional models produce very similar results as they both can be formulated to include vessel compliance. Good agreement is observed between the one-dimensional results and three-dimensional simulations with deformable (compliant) vessel walls. Agreement in selected flow and pressure quantities of interest for pulsatile inlet flow and resistance boundary conditions is acceptable for all three model fidelities (\autoref{fig:ModelValidate}). The agreement between models leads to high correlations between the model fidelities, which is desirable for a multifidelity approach.

\section{Uncertainty Quantification} \label{sec:UQ Methods}

\begin{figure}[!ht]
\centering
\includegraphics[width=0.7\textwidth]{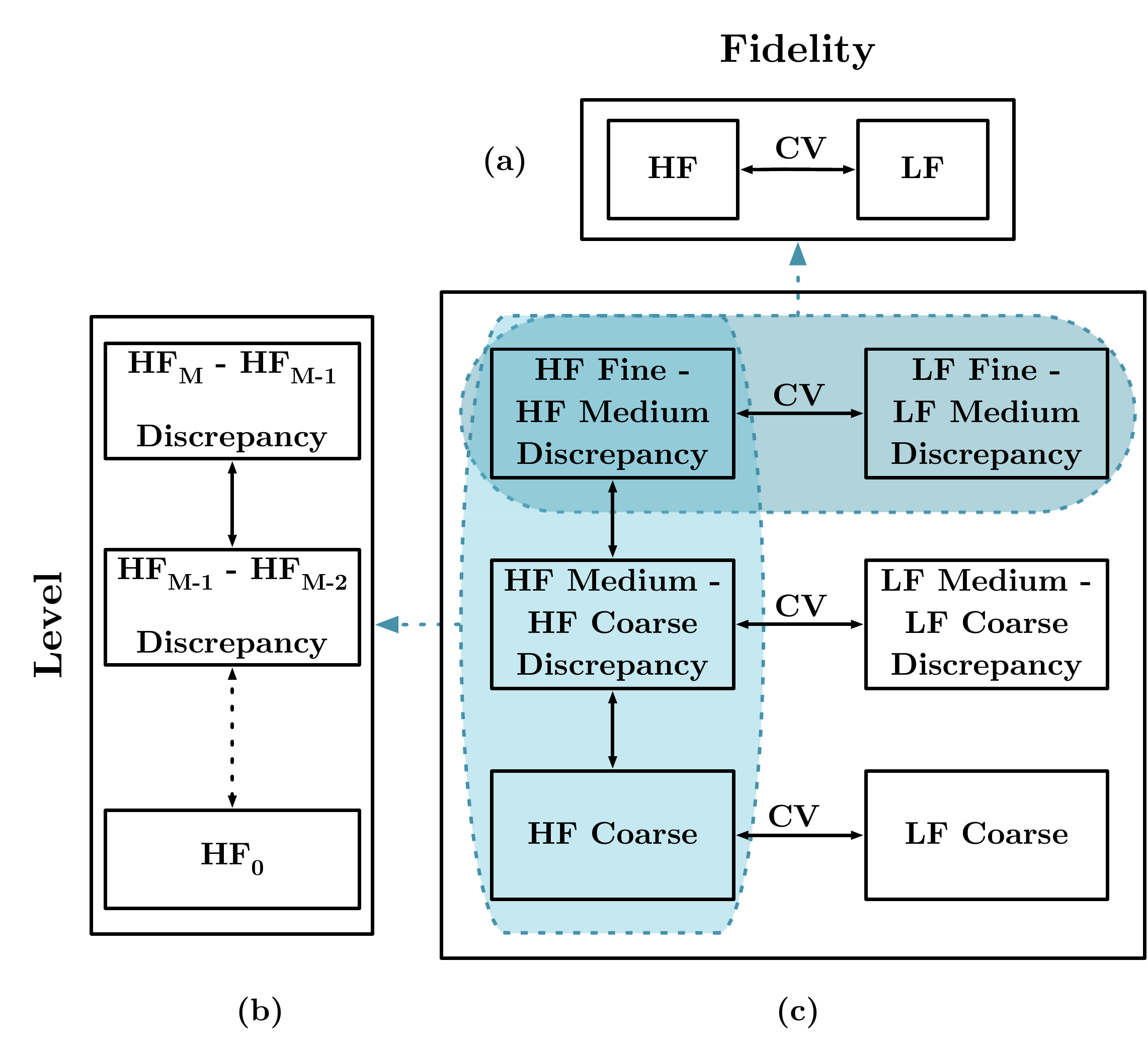}
\caption{Multilevel multifidelity simulations are comprised of (a) a multifidelity control variate approach coupling high and low fidelity models combined with (b) a multilevel approach incorporating varying mesh levels of the high fidelity model in a telescopic sum of discrepancies between each level. (c) In this study, the specific models we employ are three mesh levels of the 3D model coupled by control variates with either three mesh levels of the 1D model or a fine mesh resolution 1D model and two different zero-dimensional models, a ``full'' model with multiple circuit elements and a ``simple'' model with only resistor elements.}
\label{fig:Schematics}
\end{figure}

In this section we briefly explain multilevel, multifidelity control variate, and multilevel multifidelity Monte Carlo methods for uncertainty quantification (schematics in~\autoref{fig:Schematics}). The main goal of these methods is to leverage computationally inexpensive, but potentially less accurate, models to decrease the variance of our estimators for a variety of quantities of interest (QoIs) while using a limited number of highest-fidelity model evaluations. 

\subsection{Monte Carlo estimation} \label{sec:UQ MC Background}

Consider a complete probability space $(\Omega,\mathcal{F},P)$ where $\Omega$ is a set of elementary events, $\mathcal{F}$ is a Borel $\sigma$-algebra of subsets in $\Omega$, and $P$ a probability measure with values in $[0,1]$ over events in $\mathcal{F}$.
A vector of random variables $\bm{\xi} = (\xi_1,\xi_2,\dots,\xi_d)\in\mathbb{R}^{d}$ has components $\xi_i:\Omega\rightarrow\Sigma_{\xi_i},\,i=1,\dots,d$ with marginals $\xi_i\sim\rho_{i}(\xi_i)$ and joint probability density $\rho(\bm{\xi})$. These variables represent the sources of uncertainty in the model, while $\bm{x} \in D\subset\mathbb{R}^{3}$ and $t \in \mathbb{R}^+$ are the spatial and temporal variables, respectively.

We are interested in the statistical characterization of the quantity of interest $Q = Q(\bm{\xi})$, which is obtained as a functional of the numerical solution in time and space. Specifically we look at the first two moments, the expected value $\mathbb{E}[Q]$ or the variance $\mathbb{V}[Q]$, where for a given realization $\bm{\xi}^{(i)}$ of the random inputs, $Q$ is determined through the numerical solution of the incompressible Navier-Stokes equations on a discrete domain $D\subset\mathbb{R}^{3}$. While in general we assume our random inputs to be arbitrarily distributed and correlated, for this study we only consider independent inputs.

In the remainder of this work, we will always use an estimator for the mean value of our quantity of interest $Q$, with various methods employed to reduce its variance, but the proposed methodology can be extended to the estimation of higher-order statistical moments. In general, variance reduction techniques can be applied to estimators for other statistics pertaining to the quantity of interest $Q$.

Due to the discrete nature of the solution scheme adopted, the accuracy of the generic output $Q$ is affected by the \emph{discretization level} $M$, i.e., the number of degrees of freedom in the spatial or temporal meshes.
Therefore, a choice of discretization levels for which $Q_M \to Q$ as $M \to \infty$ guarantees that $\mathbb{E}[Q_M] \to \mathbb{E}[Q]$ as $M \to \infty$.
The Monte Carlo estimator based on $N$ realizations for the expected value of $Q_M$ is defined as
\begin{align}
\mathbb{E}[Q_M] = \int_\Omega Q_M(\bm{\xi})\,\rho(\bm{\xi})\,\text{d}\bm{\xi} \simeq \hat{Q}_{M,N}^{\MC} = \frac{1}{N}\sum_{i=1}^N Q_M^{(i)},
\end{align}
where the $i$-th realization is denoted as $Q_M^{(i)} = Q_M(\bm{\xi}^{(i)})$. The variance of this estimator can be computed as
\begin{align}
\mathbb{V}[\hat{Q}_{M,N}^{\MC}] &= \frac{1}{N}\mathbb{V}[Q_M],
\end{align}
whereas the mean squared error (MSE) is
\begin{equation}\label{equ:mcMSE}
\begin{split}
\mathbb{E}[(\hat{Q}_{M,N}^{\MC} - \mathbb{E}[Q])^2] &= \mathbb{V}[\hat{Q}_{M,N}^{\MC}] + (\mathbb{E}[Q_M - Q])^2.
\end{split}
\end{equation}
We see from $\eqref{equ:mcMSE}$ that the MSE contains two contributions, from the estimator variance $\mathbb{V}[\hat{Q}_{M,N}^{\MC}]$ and deterministic bias $(\mathbb{E}[Q_M - Q])^2$.
The deterministic bias can be reduced with a finer discretization for $Q_{M}$, i.e., a larger value of $M$, while the estimator variance is reduced by increasing the number of realizations $N$. 
As a consequence, obtaining a highly accurate MC estimator (i.e. an estimator with a small MSE) would potentially require a large number $N$ of highly accurate (large $M$) numerical simulations, making this approach impractical for high-fidelity models.
In this work, we make the following assumptions. We performed convergence studies (not reported here for brevity) to select the three-dimensional resolution level guaranteed to obtain an acceptable discretization error for the QoIs we considered. Following this convergence study, a fixed set of resolution levels was constructed via a comprehensive mesh convergence study (details in~\autoref{sec:Mesh Convergence}). 
Our goal in this work is to build an unbiased estimator with respect to the high-fidelity model. Hence, we only target the variance contribution of the MSE. We assume the bias of the high-fidelity model to already be satisfactory; that is, we do not extend the hierarchy of models to those with finer discretizations. However, it is also possible to extend this approach in order to target the full MSE as suggested in the MLMC literature (e.g. see \cite{Giles_2015}).   

Due to the large computational cost of evaluating the output $Q_{M}^{(i)}$ of a single hemodynamic simulation, increasing the number of samples is impractical, and multilevel, multifidelity, and multilevel multifidelity approaches (\hyperref[sec:UQ ML Background]{subsections~\ref*{sec:UQ ML Background}}, \ref{sec:UQ MF Background}, and \ref{sec:UQ MLMF Background}, respectively) offer a better alternative. 
These approaches allow us to combine models of multiple fidelities and computational costs into the same stochastic framework, thus providing a means to efficiently manage the available computational resources.

\subsection{A multilevel approach} \label{sec:UQ ML Background}

The main idea of multilevel MC (MLMC) approaches is to replace the quantity of interest $Q$ with a telescoping sum of the differences between the next coarsest resolution level (\hyperref[fig:Schematics]{Figure~\ref*{fig:Schematics}b}). For an extensive review on MLMC estimators we refer the reader to \cite{Giles_2015}.

Consider $L$ discretization (or resolution) levels, i.e., $\{M_\ell:\ell=0,\dots,L\}$, where $M_0 < M_1 < \dots < M_L := M$. 
Using linearity of expectation and a telescopic sum,  we re-write the expected value of our QoI after introducing the \emph{discrepancy} between QoIs at successive resolutions $Y_{\ell}$,
\begin{equation}
Y_\ell = \begin{cases}
Q_{M_0} &\text{ if } \ell = 0 \\
Q_{M_\ell} - Q_{M_{\ell -1}} &\text{ if } 0 < \ell \leq L
\end{cases}\quad\Rightarrow\quad
\mathbb{E}[Q_M] = \sum_{\ell=0}^L \mathbb{E}[Y_{\ell}].
\end{equation}
The multilevel Monte Carlo (MLMC) estimator for $\mathbb{E}[Q_M]$ is assembled from the Monte Carlo independent estimators of $\mathbb{E}[Y_{\ell}]$. 
At level $\ell$, $N_\ell$ realizations are used to estimate $Q^{\ML}_M$, and so the expected value of our QoI is
\begin{equation}\label{equ:ML}
\hat{Q}_{M,N}^{\ML} = \sum_{\ell=0}^L \hat{Y}_{\ell,N_\ell} = \sum_{\ell=0}^L \frac{1}{N_\ell} \sum_{i=1}^{N_\ell}Y_{\ell}^{(i)}.
\end{equation}

As before, $Y_\ell^{(i)}$ is evaluated at the $i$-th, $i=1,\dots,N_\ell$,  realization of the stochastic vector $\boldsymbol{\xi}$ drawn from the distribution $\rho(\boldsymbol{\xi})$. 
The MLMC estimator has variance equal to
\begin{align}
\mathbb{V}[\hat{Q}_{M,N}^{\ML}] = \sum_{\ell=0}^L \frac{1}{N_\ell} \mathbb{V}\left[Y_{\ell}\right].
\end{align}

The advantages of the MLMC method come from the hierarchical nature of $Y_{\ell}$, the difference in the quantity of interest between successive resolutions. As $Q_M \to Q$ as $M \to \infty$, $Y_{\ell} \to 0$ as $\ell$ increases and therefore the contribution to the overall variance from different resolution levels decreases with $\ell$. This allows us to shift the computational burden to the coarser levels, which are computationally cheaper. 

\emph{Extrapolation} can be performed to approximate the samples needed to improve estimator variance by a factor $\epsilon$.
The optimal sample allocation at each level can be determined by minimizing the computational cost of MLMC subject to a fixed target $\epsilon$ improvement factor~\cite{Giles_2015}. The minimum total computational cost is $\mathcal{C}[\hat{Q}^{\ML}_{M,N}] = \sum_{\ell=0}^L N_{\ell}\,\mathcal{C}_\ell,$
where $\mathcal{C}_\ell$ is the computational cost of one evaluation of $Y_\ell$ (note that this is the combined cost of two evaluations on successive resolution levels for $\ell\geq1$). 
This cost can be minimized using a Lagrange multiplier under a variance constraint resulting in an optimal number of samples at level $\ell$ given by
\begin{equation}\label{equ:MLOptSample}
\begin{split}
N_\ell = \frac{1}{\epsilon^2}\left(\sum_{k=0}^L \sqrt{\mathbb{V}[Y_k]\,\mathcal{C}_k} \right) \sqrt{\frac{\mathbb{V}[Y_\ell]}{\mathcal{C}_\ell}}.
\end{split}
\end{equation}

\subsection{A multifidelity approach} \label{sec:UQ MF Background}

Multifidelity (MFMC) approaches represent a flavor of the better known \emph{control variate} (CV) variance reduction technique in Monte Carlo estimation (\hyperref[fig:Schematics]{Figure~\ref*{fig:Schematics}a}). Two models are now defined, i.e., a low-fidelity (LF) and a high-fidelity (HF) model, at discretization levels $M_{\LF}$ and $M$, respectively. 
Without loss of generality, in the following we only use the level $M$, where it is implicitly assumed that the discretization levels of the HF and LF model are in general independent.  
In this approach, the generic quantity of interest $Q_M^{\HF}$ is replaced by $Q_M^{\mathrm{CV},\HF}$ which embeds a correction term based on the LF model. For full details of the approximate CV estimators used here, we refer the reader to \cite{Pasupathy_2014,NG_2014}. For more details of the general multivariate CV approach, we refer the reader to \cite{Rubinstein1985}.

The final form of the selected approximate CV MFMC estimator is obtained by combining the MC estimators as
\begin{equation}
\hat{Q}_{M,N^{\HF}}^{\mathrm{CV},\HF} = \hat{Q}_{M,N^{\HF}}^{\HF} + \alpha \left(\hat{Q}_{M,N^{\HF}}^{\LF} - \hat{Q}_{M,N^{\LF}}^{\LF}\right),
\end{equation}
where $N^{\LF} = N^{\HF} + \Delta_{\LF} = N^{\HF}(1+r)$ and the additional LF realizations $\Delta^{\LF} = r N^{\HF}$ are drawn independently of the initial set of $N^{\HF}$.

The regression coefficient $\alpha$ and the value of the parameter $r >0$ are obtained by minimizing the variance of the CV MFMC estimator. After optimization, the optimal $\alpha$ value
\begin{equation}
\alpha = -\rho\sqrt{\frac{\mathbb{V}[\hat{Q}_{M,N^{\HF}}^{\HF}]}{\mathbb{V}[\hat{Q}_{M,N^{\HF}}^{\LF}]}}
\end{equation}
yields a minimal variance
\begin{equation}\label{equ:MFwithRho2}
\mathbb{V}[\hat{Q}_{M,N^{\HF}}^{\mathrm{CV},\HF}] = \mathbb{V}[\hat{Q}_{M,N^{\HF}}^{\HF}] \left(1- \frac{r}{1+r}\rho^2\right),
\end{equation}
where $\rho$ is Pearson's correlation coefficient between the LF and HF estimators. Note that this multifidelity estimator always guarantees variance reduction, since $\rho^{2}\in(0,1)$. 

In practice, we first compute ratio $r^{\star}$
\begin{equation}
r^{\star} = -1 + \sqrt{w \dfrac{\rho^2}{1 - \rho^2}}\,,
\end{equation}
where $w = \mathcal{C}^{\HF} / \mathcal{C}^{\LF}$ is the cost ratio between the two fidelities. 
With this, we can compute the optimal number of high-fidelity samples by extrapolation targeting an estimator variance improvement factor $\epsilon$:
\begin{equation}\label{equ:MFOptSample}
\begin{split}
N^\HF = \frac{\mathbb{V}[Q_M^\HF]}{\epsilon^2}\left(1 - \frac{r}{r+1}\rho^2 \right).
\end{split}
\end{equation}
This follows the extrapolation for MLMC estimators in~\autoref{sec:UQ ML Background}. For an extension of the CV MFMC estimators to multiple fidelity models we refer the reader to \cite{Peher_2016,Gorodetsky_2020}.

\subsection{A multilevel multifidelity approach} \label{sec:UQ MLMF Background}

We combine ideas from \hyperref[sec:UQ ML Background]{subsections~\ref*{sec:UQ ML Background}} and~\ref{sec:UQ MF Background} to further reduce the variance of our estimators in a MLMF approach (\hyperref[fig:Schematics]{Figure~\ref*{fig:Schematics}c}). 
We apply one control variate for each (high-fidelity) resolution level $\ell$, or equivalently apply the multifidelity control variate approach to the Monte Carlo estimators associated with the difference $Y_\ell$. MLMF schemes can be implemented for either the same or different numbers of high-fidelity and low-fidelity model levels. For further details, the reader is referred to~\cite{Nobile2015, Geraci_A_2017,Fairbanks2016}.

The MLMF estimator which assumes the same number of high-fidelity and low-fidelity model levels is given by
\begin{equation}\label{equ:MLMF}
\begin{split}
\mathbb{E}[Q_M^{\HF}] \approx \hat{Q}^{\mathrm{MLMF}}_M = \sum_{\ell =0}^L \left(\hat{Y}_{M_\ell, N_\ell^{\HF}}^{\HF} + \alpha_\ell\left(\hat{Y}_{M_\ell, N_\ell^{\HF}}^{\LF} - \mathbb{E}[Y_{M_\ell}^{\LF}] \right)\right).
\end{split}
\end{equation}

As before, we determine the optimal sampling distribution (by level) $N_\ell^{\HF}$ to compute $\hat{Y}_{M_\ell, N_\ell^{\HF}}^{\HF}$ and $\hat{Y}_{M_\ell, N_\ell^{\HF}}^{\LF}$ by extrapolation.
We then calculate the redistribution of computational burden to the more inexpensive LF model as the set of (independent) samples $\Delta_\ell^{\LF} = r_\ell\,N^{HF}_\ell$ needed to compute $\hat{\mathbb{E}}[Y_{M_\ell,N_\ell^{LF}}^{\LF}]$ as in~\autoref{sec:UQ ML Background}. The solution for the optimal number of samples per level $\ell$ is given by
\begin{equation}\label{equ:MLMFOptSample}
N_{\ell}^{\HF} = \frac{2}{\epsilon^2}\left(\sum_{k=0}^{L^{HF}} \left(\frac{\mathbb{V}[Y_k^{\HF}]\mathcal{C}_k^{\HF}}{1 - \rho_{\ell}^2}\right)^{\frac12} \Lambda_k(r_k)\right) \sqrt{(1-\rho_\ell^2)\frac{\mathbb{V}[Y_\ell^{\HF}]}{\mathcal{C}_\ell^{\HF}}}\,,
\end{equation}
with the optimal redistribution of samples given by $\Delta^{\LF}_\ell = r_\ell^{\star} N^{\HF}_\ell$ where 
\begin{equation}
r_\ell^{\star} = -1 + \sqrt{w_\ell \dfrac{\rho_\ell^2}{1 - \rho_\ell^2}}
\end{equation}
and the quantity 
\begin{equation}
\Lambda_\ell(r_\ell) = 1 - \frac{r_\ell}{1 + r_\ell}\rho_\ell^2
\end{equation}
measures the variance reduction on each level $\ell$ by accounting for the CV effect. 

At each level, the allocation of samples between LF and HF is controlled by $r_\ell$, which is a function of the correlation $\rho_\ell^2$ and the computational cost ratio $w_\ell = \mathcal{C}_\ell^{\HF}/\mathcal{C}_\ell^{\LF}$. 
The optimal variance reduction therefore corresponds to $1 - \Lambda(r_\ell^\star)$. 
Intuitively this is correct since for an increase in correlation and/or in the cost ratio, more low-fidelity simulations (i.e., larger $r_{\ell}$) are required. To ensure our model correlations are sufficiently high for all fidelity and discretization levels, we employ the version of the MLMF algorithm detailed in~\cite{Geraci_A_2017}.

\section{Methods} \label{sec:Methods}

\subsection{Workflow} \label{sec:Workflow}
In this study, we developed an interface coupling the three-, one-, and zero-dimensional hemodynamic solvers with the functionality of Sandia National Laboratories' Dakota toolkit. 
Dakota is an extensive open-source framework providing a library of computational tools for optimization, uncertainty quantification, parameter estimation and sensitivity analysis ~\cite{Dakota_Adams_Version}. It allowed us to assign distributions to uncertain parameters, to automatically generate parameter samples, to manage the simultaneous execution of multiple simulations, and to characterize statistically the resulting QoIs. The MLMF estimators within Dakota have been successfully applied to aerospace applications~\cite{Geraci_A_2017}, wind power plants~\cite{Maniaci_2018}, and other applications. Initial work by these authors for the application of cardiovascular modeling~\cite{Fleeter_2018,Schiavazzi_2018} is extended and improved in this paper.

The Dakota cardiovascular UQ interface was created to be easily adaptable to different UQ scenarios. Most files that connected Dakota to the hemodynamic solvers remained unchanged as the cardiovascular models were exchanged. Similarly, changing uncertain parameter distributions and QoIs was achieved with an update to a small number of interface files. 
The power of Dakota's behind-the-scenes simulation management came at runtime. We developed the interface such that after providing the model-specific files and all cardiovascular solver executables, only a single executable was called to launch an uncertainty quantification study on a high performance computing system. Dakota then managed the allocation of each model fidelity and resolution level to achieve a specified convergence tolerance. Upon completion, all statistics for the QoIs were automatically generated. As such, the developed framework was user-friendly and did not require constant management of parameter realizations or solution files. We also note that the framework was not limited to MLMF estimators, but could interface with any of Dakota's capabilities simply by providing a different input file.

\begin{figure}[!ht]
\centering
\includegraphics[width=\textwidth]{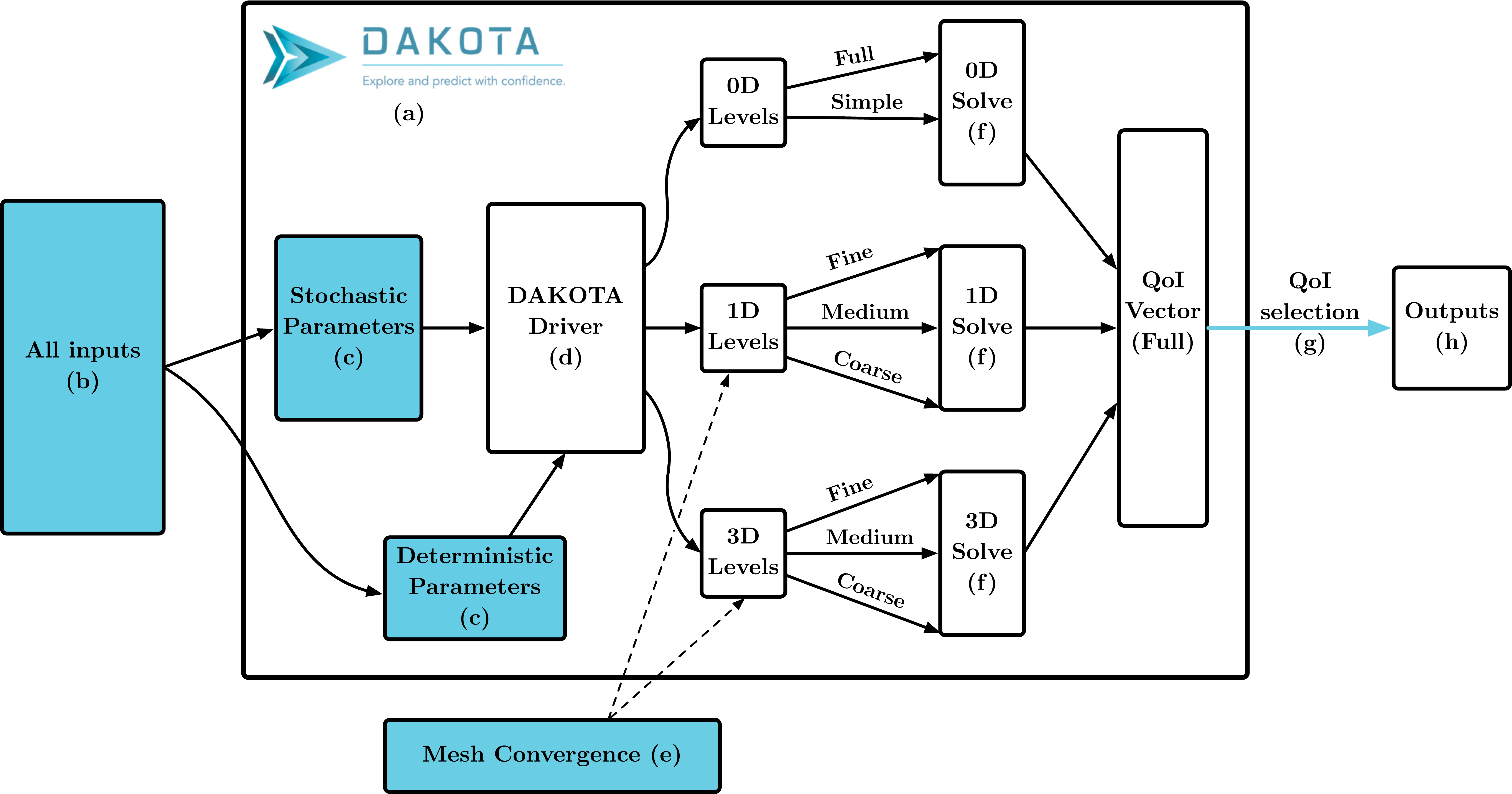}
\caption{The developed workflow utilizes Sandia National Laboratories' Dakota toolkit to automate the uncertainty quantification process. (a) Dakota (official logo from~\cite{Dakota_Adams_Version}) automatically manages all steps within the outer box after the user provides in a single input file (b) with all problem-specific inputs (such as stochastic parameters and their distributions, deterministic parameters, the desired number of simulations, and convergence criterion). The user can easily change (c) the parameters from stochastic to deterministic and vice versa. Dakota automates (d) the selection of models of different fidelities and mesh levels, with the relevant simulation files provided by the user after (e) a mesh convergence study is conducted to select the appropriate models. Dakota also (f) automatically invokes the 0D, 1D or 3D solvers, including all pre- and post-processing, and stores the complete set of model outputs. Finally, Dakota extracts (g) the QoIs specified by the user and delivers (h) the moments (mean, standard deviation) of the estimators for the QoIs.}
\label{fig:Workflow}
\end{figure}

Specific details of the models used in this paper are discussed in \hyperref[sec:Models]{subsections~\ref*{sec:Models}},~\ref{sec:Mesh Convergence}, and~\ref{sec:Params}, while the QoIs are discussed in~\autoref{sec:QoIs}. A repository detailing the necessary user-provided files, implementation, and usage of the Dakota cardiovascular UQ interface is publicly available on GitHub~\cite{FleeterCode2019}. The repository includes more detailed information on the simulation process. A schematic representation of the UQ workflow is provided as a general overview (\autoref{fig:Workflow}).

\subsection{Model construction} \label{sec:Models}

\begin{figure}[!ht]
\centering
\begin{subfigure}[t]{0.24\textwidth}
	\centering
	\includegraphics[width=\textwidth,height=1.3\textwidth]{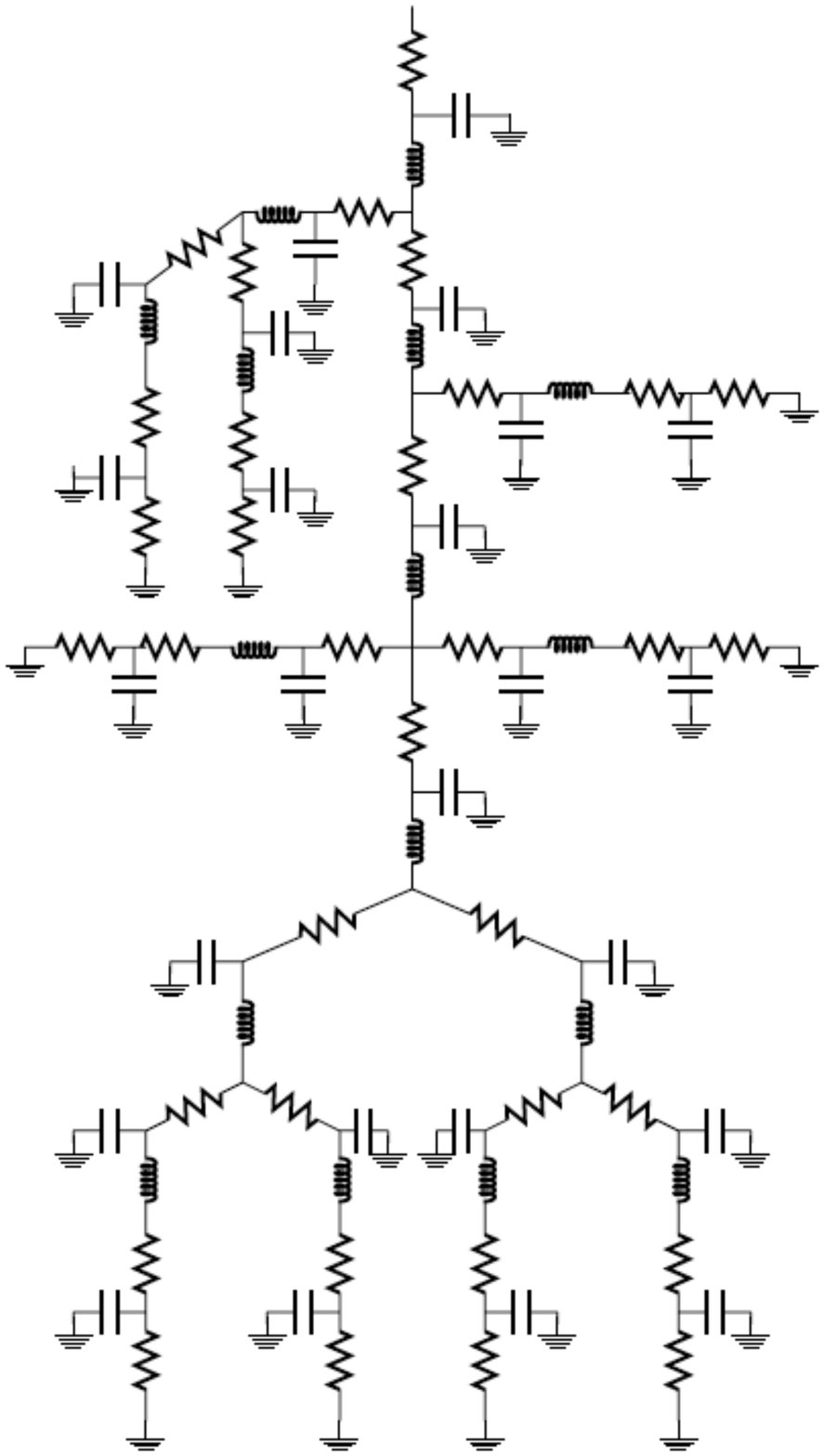}
	\caption{}
	\label{fig:Aorta0D}
\end{subfigure}
\hfill
\begin{subfigure}[t]{0.24\textwidth}
	\centering
	\includegraphics[width=\textwidth]{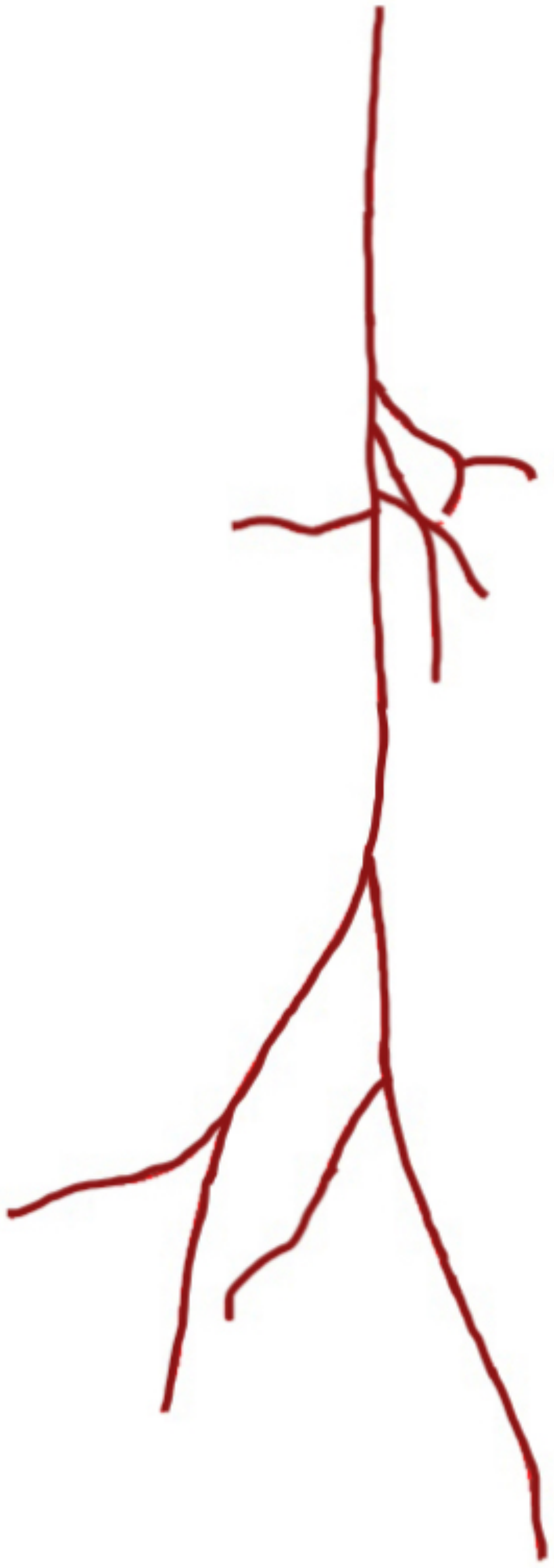}
	\caption{}
	\label{fig:Aorta1D}
\end{subfigure}
\hfill
\begin{subfigure}[t]{0.24\textwidth}
	\centering
	\includegraphics[width=\textwidth]{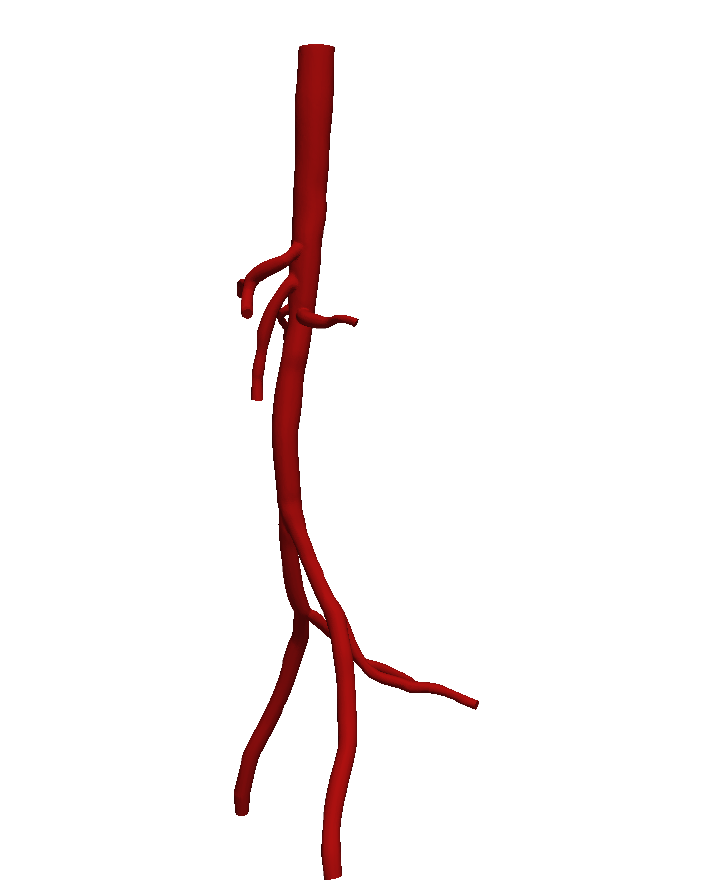}
	\caption{}
	\label{fig:AortaHealthy}
\end{subfigure}
\hfill
\begin{subfigure}[t]{0.24\textwidth}
	\centering
	\includegraphics[width=\textwidth]{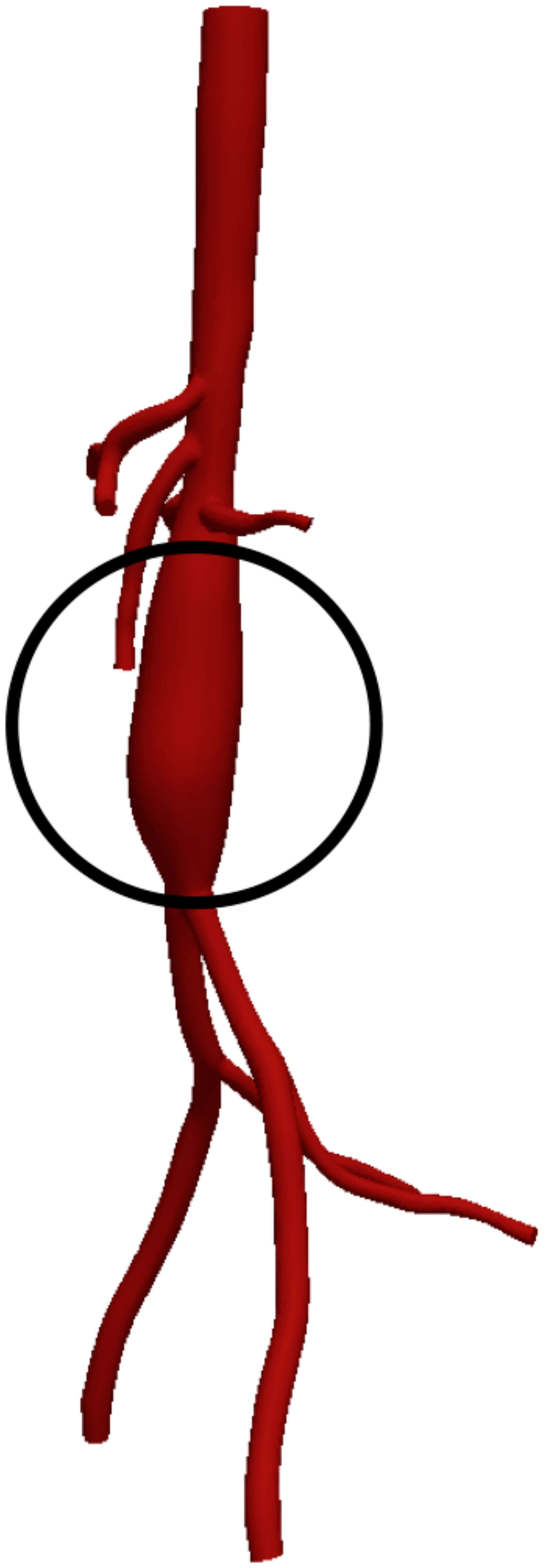}
	\caption{}
	\label{fig:AortaDiseased} 
\end{subfigure} 
\\
\begin{subfigure}[t]{0.26\textwidth}
	\centering
	\includegraphics[width=\textwidth,height=0.95\textwidth]{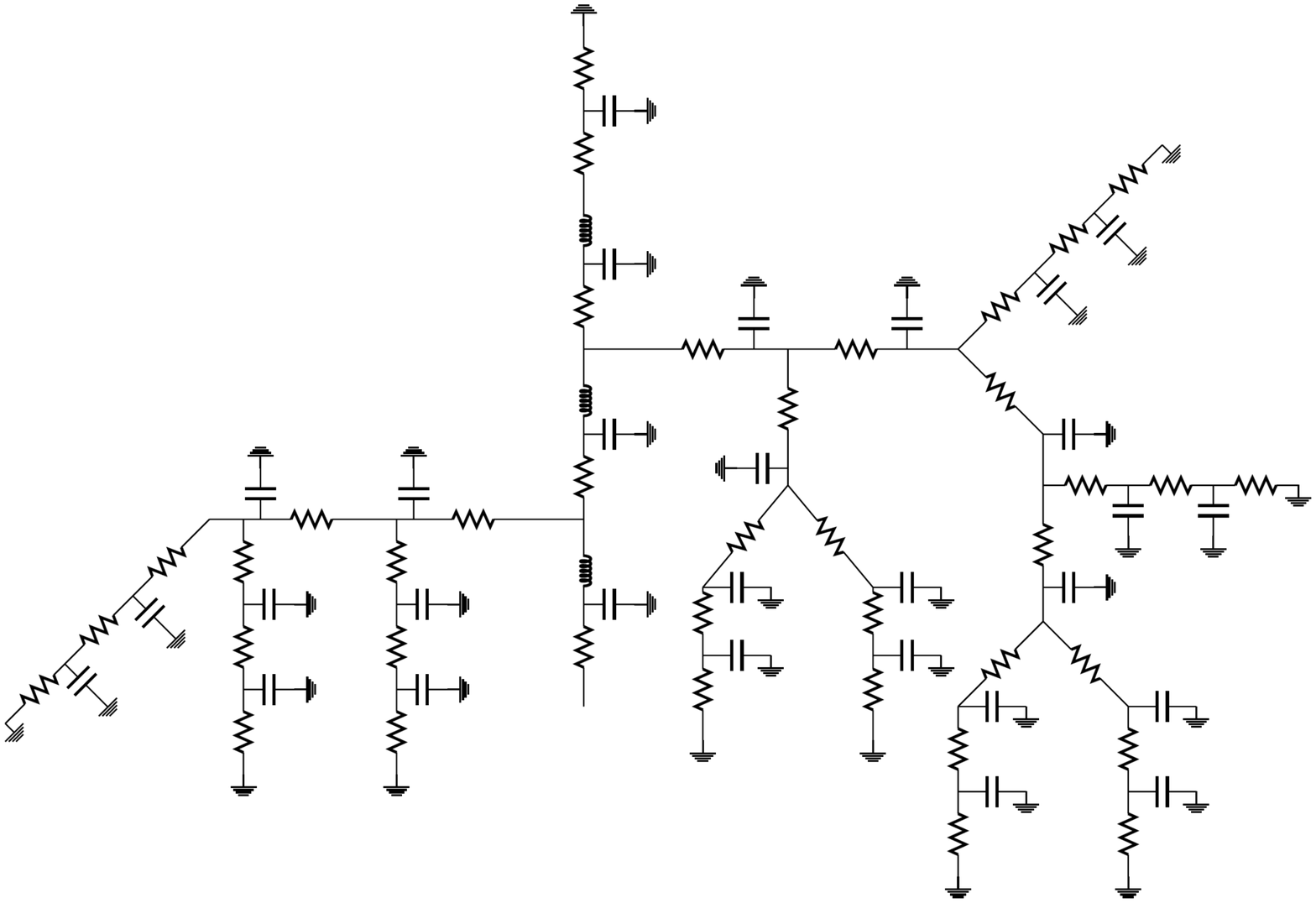}
	\caption{}
	\label{fig:Coronary0D}
\end{subfigure}
\hfill
\begin{subfigure}[t]{0.22\textwidth}
	\centering
	\includegraphics[width=0.88\textwidth]{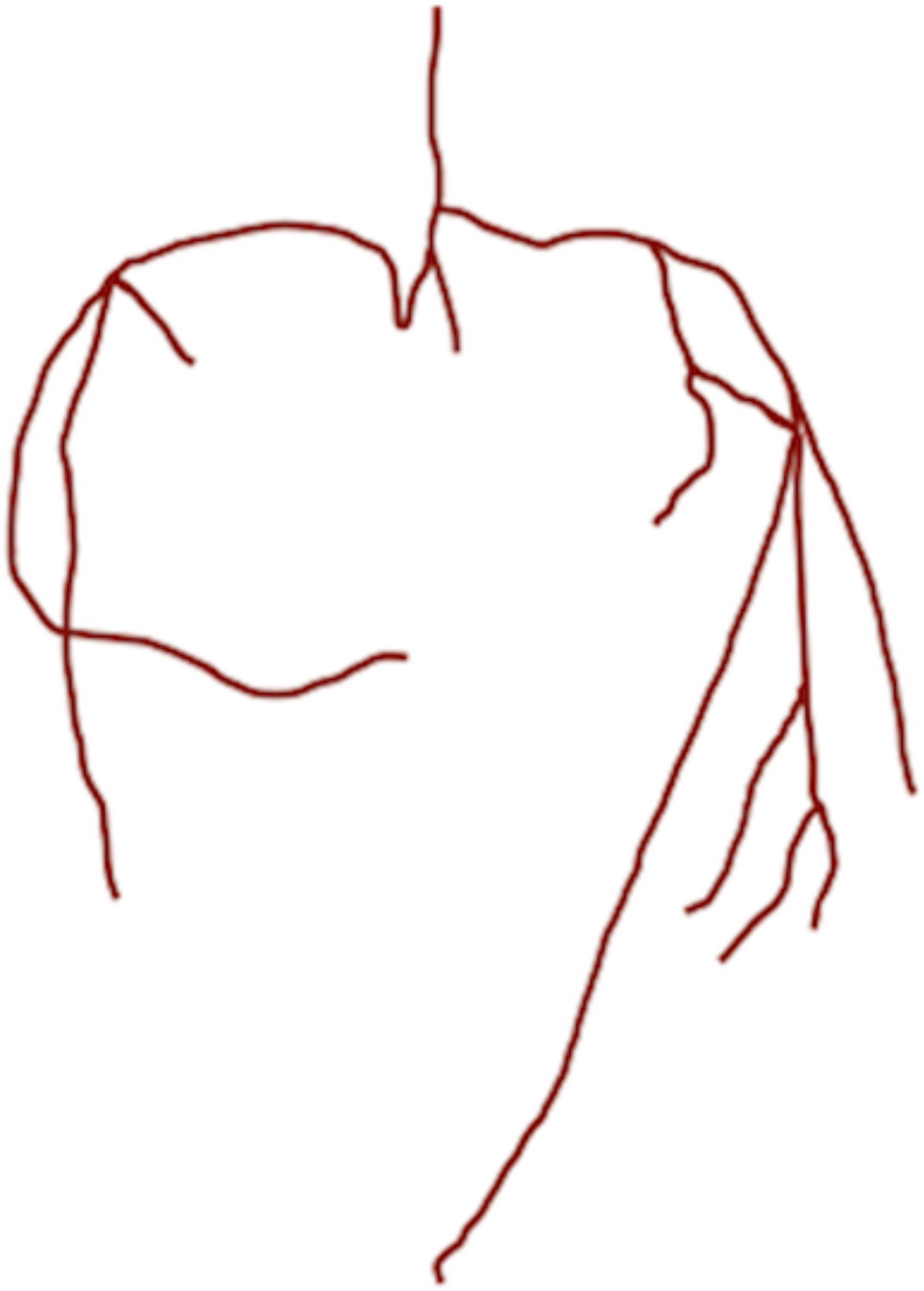}
	\caption{}
	\label{fig:Coronary1D}
\end{subfigure}
\hfill
\begin{subfigure}[t]{0.24\textwidth} 
	\centering
	\includegraphics[width=\textwidth]{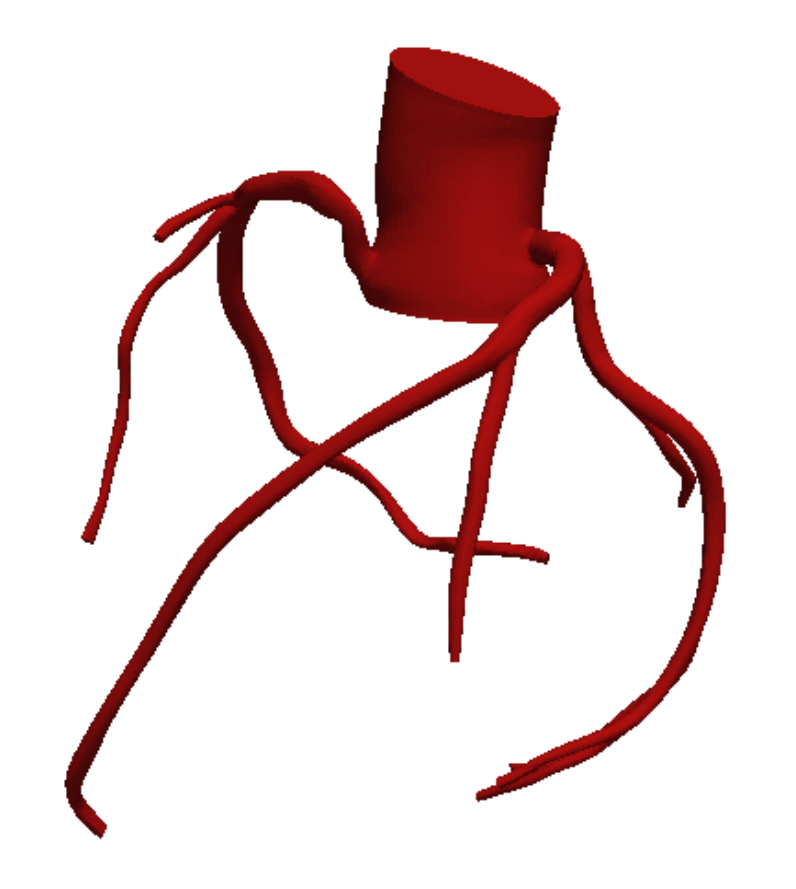}
	\caption{}
	\label{fig:CoronaryHealthy}  
\end{subfigure}	 
\hfill
\begin{subfigure}[t]{0.24\textwidth} 
	\centering
	\includegraphics[width=\textwidth]{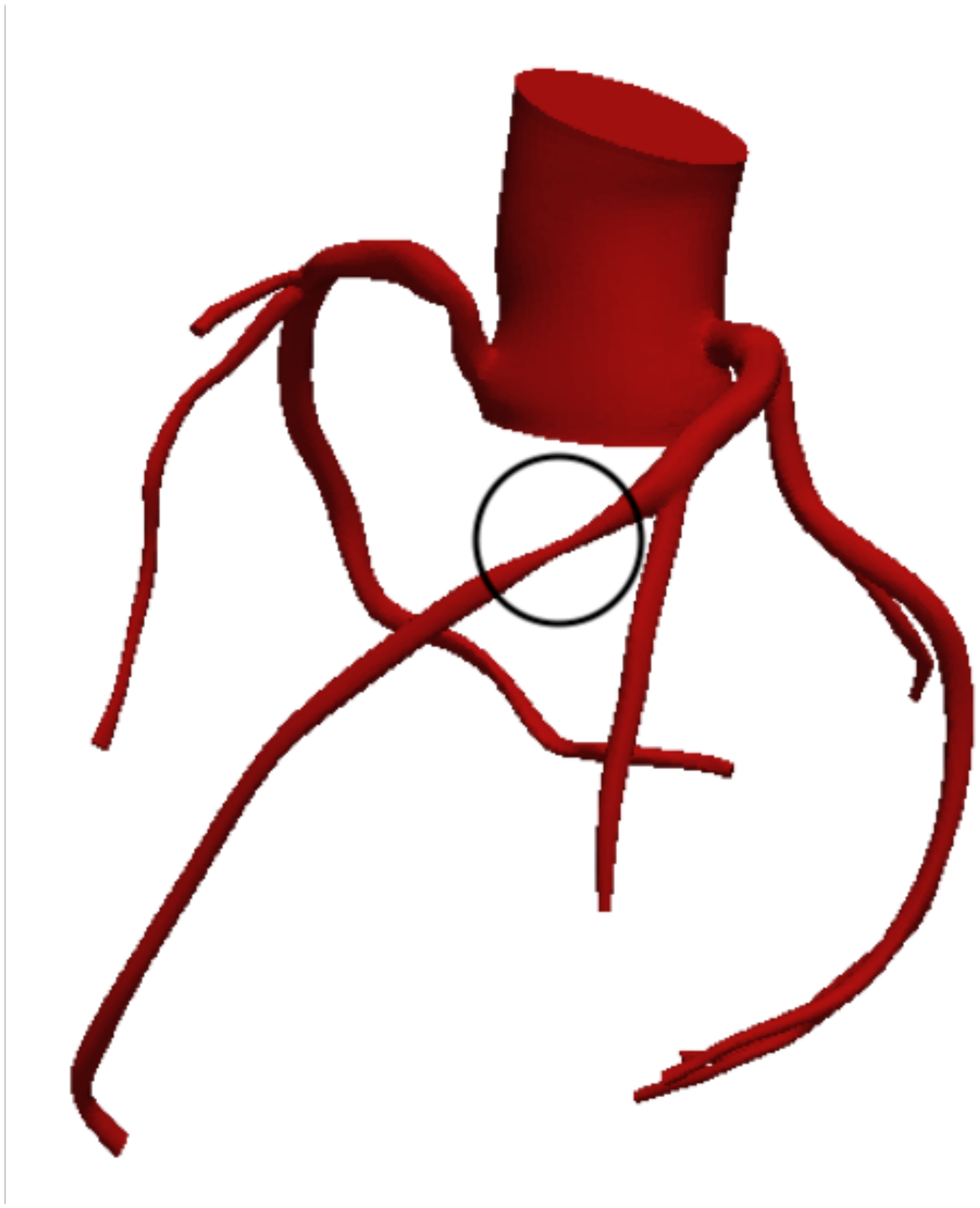}
	\caption{}
	\label{fig:CoronaryDiseased}  
\end{subfigure}	 
\caption{Healthy and diseased aorto-femoral and coronary models of three varying fidelities are used in this study. Specifically, we utilize (a) 0D, (b) 1D, (c) 3D healthy, and (d) 3D diseased (AAA) aorto-femoral models. We also utilize (e) 0D, (f) 1D, (g) 3D healthy and (h) 3D diseased (75\% stenosis) coronary models. The diseased model regions are circled in (d) and (h).}
\label{fig:Models}
\end{figure}

Four models, healthy and diseased geometry of aorto-femoral and coronary anatomy, each with three fidelities (3D, 1D, and 0D) were used in this study, with original image data and models provided by the Vascular Model Repository (\url{http://www.vascularmodel.org}). The realistic diseased models were constructed by adapting the geometry of the healthy models to introduce an abdominal aortic aneurysm (AAA) and coronary vessel stenosis (\autoref{fig:Models}).

The healthy aorto-femoral model is a patient-specific model of a healthy abdominal aorta with iliac and femoral arteries, with nine outlet branches.
The three model fidelities were generated as discussed in~\autoref{sec:Cardio Modeling Background}, illustrated in \hyperref[fig:Aorta0D]{Figures~\ref*{fig:Aorta0D}},~\ref{fig:Aorta1D},~\ref{fig:AortaHealthy}, and~\ref{fig:AortaDiseased}.
The diseased aorto-femoral model was adapted in SimVascular to include an AAA at a location typically observed in the clinic. The aneurysm was induced by increasing the aorta radius by 50\%, which is the accepted threshold for classification as an AAA~\cite{Kent_2014}.

The healthy coronary model includes the aorta, the left and right coronary arteries and ten outlet coronary branches. The fluid dynamics in the coronaries are more complex due to large area differences between coronary and aortic inlets and outlets and the offset in time between pressure and flow waveforms. 
The three model fidelities were generated as discussed in~\autoref{sec:Cardio Modeling Background} above, illustrated in \hyperref[fig:Coronary0D]{Figures~\ref*{fig:Coronary0D}},~\ref{fig:Coronary1D},~\ref{fig:CoronaryHealthy}, and~\ref{fig:CoronaryDiseased}.
The diseased coronary model was adapted in SimVascular to include a stenosis in the LAD, a common location of coronary stenoses~\cite{Giannoglou2010}. Significant vessel stenosis is often defined as a reduction in vessel diameter of at least 50\%; we used a reduction in diameter of 75\%. 

Two zero-dimensional models were constructed for each anatomy. The first, more complex zero-dimensional model uses both RCL and RC blocks for each vessel segment between bifurcations. Including the inductance and capacitance elements allows these models to incorporate inertia and wall elasticity. The second model uses only resistance elements for each vessel segment, creating a model simplification with the same uncertain parameters. While this is not the only possible simplified zero-dimensional model, the resistor-only model remains highly correlated with the higher fidelity models. It also performs well as an even less computationally expensive zero-dimensional model, allowing for the two reduced models to be considered as coarse and medium ``resolution levels'' in the framework. In the future, additional reduced-order models could be added to the hierarchy. Many such examples for one- and zero-dimensional models are enumerated in~\cite{Peiró2009}.

\subsection{Mesh convergence} \label{sec:Mesh Convergence}

After creating three-, one-, and zero-dimensional models of all four types (aorto-femoral healthy and diseased, coronary healthy and diseased), a mesh convergence study was performed to select coarse, medium, and fine resolution levels for each model which demonstrated increasing accuracy compared to a reference mesh for our quantities of interest. These QoIs included time-averaged outlet flow, time-averaged outlet pressure, time-averaged pressure spatially-averaged throughout various regions of the model, and time-averaged wall shear stress (TAWSS) values also spatially-averaged on various regions of the model wall, as described in~\autoref{sec:QoIs}.  For three- and one-dimensional models, we performed a spatial mesh convergence. As mentioned in ~\autoref{sec:Models}, in lieu of a mesh convergence for the zero-dimensional models lacking in traditional spatial resolution, a simplified LPN was constructed from only resistor elements to serve the purpose of a coarse model representation. 

For each of the four model geometries, our three-dimensional mesh convergence study compared the performance of seven mesh resolutions, ranging from about 50,000 to 2 million elements, against a reference mesh of approximately 4 million tetrahedral elements. At this reference mesh resolution, the QoI values are no longer changing by more than 1\% as the mesh is further refined.

Deformable wall simulations were used, requiring first a rigid simulation for each mesh level followed by a deformable wall simulation. 
Steady inlet flow and resistance boundary conditions were used. The total resistance was equivalent to that of the mean value of the resistance for the boundary conditions used in the UQ study (see~\autoref{sec:Params}). Non-linear finite element iterations were regarded as converged when the residual norm was below a 0.0001 threshold. Each simulation was run for four cardiac cycles, and the convergence of each was confirmed to ensure a fair comparison of the mesh levels. 

\begin{figure}
\begin{floatrow}
\ffigbox[\FBwidth][\FBheight][t]{%
	\includegraphics[width=0.3\textwidth]{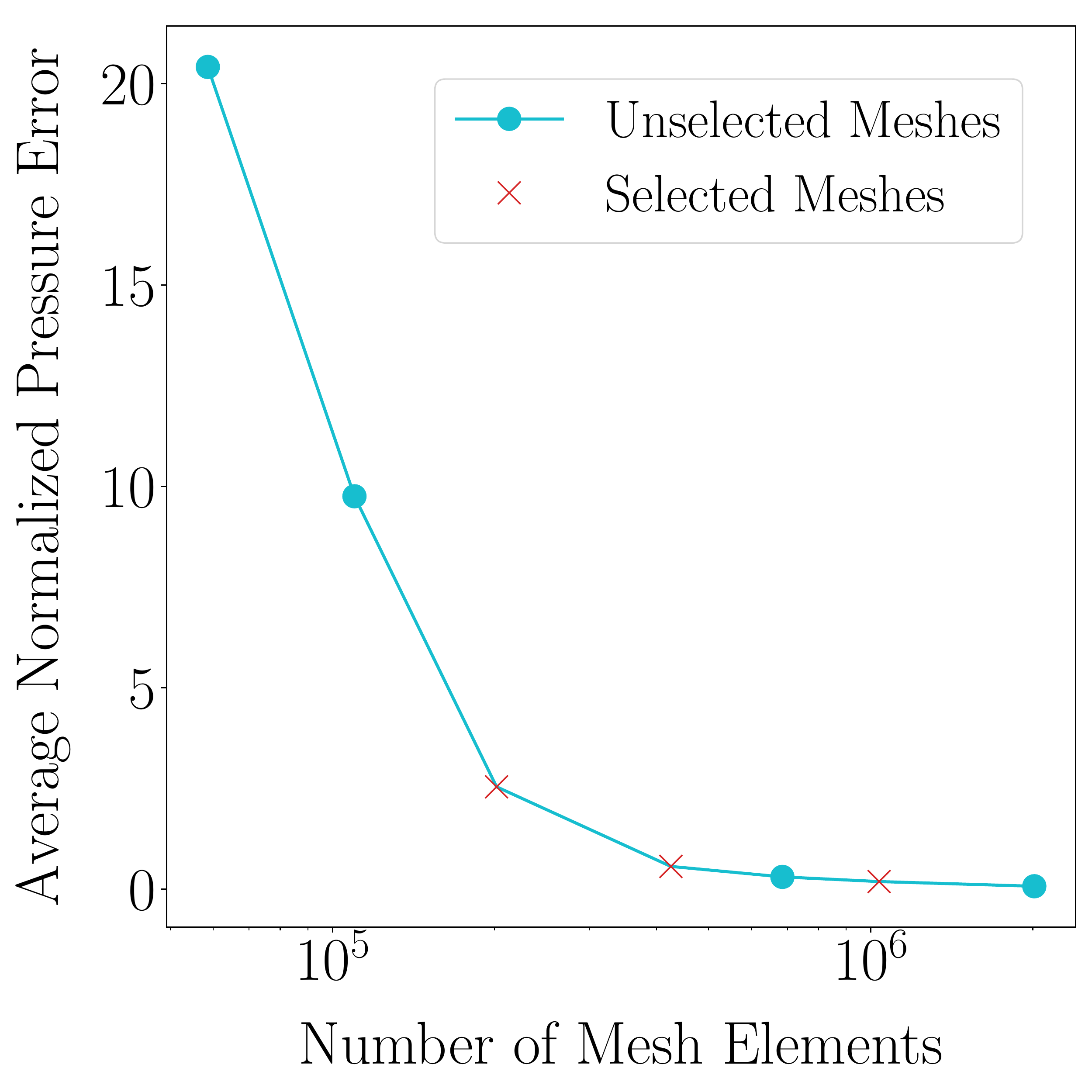}
}{%
	\caption{Sample convergence plot for mesh selection. Selected meshes for aorto-femoral diseased model demarcated with red xs. (For interpretation  of  the  references  to  color  in  this  figure  legend,  the  reader  is  referred  to  the  web  version  of  this  article.)}
	\label{fig:MeshConverge}
}
\capbtabbox{%
	\resizebox{0.6\textwidth}{!}{
	\begin{tabular}{l | r | r |  r |  r } 
	\toprule 
	 & Aorto-Femoral & Aorto-Femoral & Coronary  & Coronary \\
	 & Healthy & Diseased & Healthy  & Diseased \\
	\midrule
	\midrule
	3D Fine Mesh 	  & 1\,219\,672 & 1,036\,483 & 1\,026,675   & 1\,069\,328	\\
	3D Medium Mesh    & 446\,806	& 425\,416 	 & 411\,958 	& 401\,170  	\\
	3D Coarse Mesh    & 223\,927 	& 200\,610 	 & 106\,681 	& 204\,762 	\\
	\midrule
	1D Fine Mesh      &  50 		& 50 		 &  25 			& 25 \\
	1D Medium Mesh    &  10 		& 10 		 &  10 			& 10 \\
	1D Coarse Mesh    &   5 		& 5			 &	5 			& 5 \\
	\bottomrule 
	\end{tabular}
	}
}{%
	\caption{Total number of elements in each of the meshes selected for the UQ study for the 3D and 1D models.}
	\label{tab:Meshes}
}
\end{floatrow}
\end{figure}

Convergence plots for the quantities of interest were used to select the mesh levels used for the multi-level framework for the UQ study. We confirmed that the finest mesh produced QoI values close to those of the reference mesh in the linear scaling range (\autoref{fig:MeshConverge}). After selecting the fine mesh, medium and coarse meshes were chosen from among the remaining meshes in the linear scaling range.

Mesh convergence was also performed for the one-dimensional models. One-dimensional models are constructed automatically in SimVascular after the lumen segmentation phase of three-dimensional model building through the ``1D-Plugin'' feature, which is based on the method developed and discussed in~\autoref{sec:Cardio Modeling Validation}. When the one-dimensional models were extracted from the three-dimensional geometry, linear segments with different diameters were automatically generated to approximate changes in vessel diameter throughout the model. Different mesh resolutions for these one-dimensional models were generated by varying the number of finite elements used to discretize each of these model segments along the $z$-axis (the vessel centerlines). The coarsest level used 1 element per segment, the finest level used 250 elements, and a reference mesh used 500 elements.
Details of the selected meshes are shown in \hyperref[tab:Meshes]{Table~\ref*{tab:Meshes}}. Mesh quality values for the aspect ratio and Jacobian were checked for all 3D meshes to ensure they were of high quality for our UQ study.

\subsection{Computational cost assessment} \label{sec:Computational Cost}

After selecting the mesh resolution levels for the three-, one-, and zero-dimensional models of all four types (aorto-femoral healthy and diseased, coronary healthy and diseased), an initial computational cost of each type was determined by running each model with the mean value realizations of all stochastic parameters (see~\autoref{sec:Params}), along with the deterministic parameters, including transient inlet flow. The significant cost difference between the model fidelities and resolution levels lends itself to the MLMF framework. These costs were updated to reflect the averaged cost from the 25 pilot runs for each model after the pilot phase (details in ~\autoref{sec:ResMCMLMFnCI}) of our UQ study (\autoref{tab:ModelCosts}).

\begin{table}[!ht]
\centering
\resizebox{\textwidth}{!}{
\begin{tabular}{l | r | l | r | l | r | l | r | l} 
\toprule 
& \multicolumn{2}{|c|}{Aorto-Femoral Healthy} &\multicolumn{2}{c|}{Aorto-Femoral Diseased} &\multicolumn{2}{c|}{Coronary Healthy} &\multicolumn{2}{c}{Coronary Diseased}\\
\midrule
Fidelity \& Level & \multicolumn{1}{|l|}{Cost} & Effective Cost & \multicolumn{1}{l|}{Cost} & Effective Cost & \multicolumn{1}{l|}{Cost} & Effective Cost & \multicolumn{1}{l|}{Cost} & Effective Cost \\ 
\midrule
\midrule
3D Fine Mesh 	& 870.80 h & 1 			    & 667.23 h & 1 			    & 2\,164.61 h & 1			     & 1\,198.48 h & 1 \\
3D Medium Mesh  & 228.44 h & \num{2.62e-01} & 157.05 h & \num{2.35e-01} &    497.23 h & \num{2.30e-01} &      286.88 h & \num{2.39e-01} \\
3D Coarse Mesh  &  98.02 h & \num{1.13e-01} &  56.21 h & \num{8.42e-02} &     78.65 h & \num{3.63e-02} &      120.63 h & \num{1.01e-01} \\
\midrule
1D Fine Mesh    &  11.60 m & \num{2.22e-04} &  11.87 m & \num{2.96e-04} &      4.33 m & \num{3.34e-05} &        4.78 m & \num{6.65e-05} \\
1D Medium Mesh  &   2.95 m & \num{5.65e-05} &   2.62 m & \num{6.54e-05} & 	   1.90 m & \num{1.46e-05} & 	    2.00 m & \num{2.78e-05} \\
1D Coarse Mesh  &   1.90 m & \num{3.64e-05} &   1.52 m & \num{3.79e-05} & 	   1.08 m & \num{8.34e-06} & 		1.13 m & \num{1.58e-05} \\
\midrule
0D Full Model   &   0.49 m & \num{3.64e-06} &   0.50 m & \num{1.25e-05} &      0.17 m & \num{7.66e-05} & 		0.16 m & \num{1.36e-04} \\
0D Simple Model &   0.03 m & \num{6.60e-07} &   0.03 m & \num{7.60e-07} & 	   0.03 m & \num{2.51e-07} & 		0.03 m & \num{4.72e-07} \\ 
\bottomrule 
\end{tabular}}
\caption{Costs needed to generate simulation results using various hemodynamic solvers and resolution levels. Cost refers to the cost of one simulation while the effective cost is scaled relative to the cost of the corresponding 3D fine mesh simulation.}
\label{tab:ModelCosts}
\end{table}

\subsection{Stochastic and deterministic inputs} \label{sec:Params}

Both stochastic and deterministic parameters were used for all hemodynamic models in this study. When considering healthy and diseased models of the same anatomy, identical parameters were used to isolate the effect of adapting the model geometry. All models included eight independent uncertain inputs with values drawn from a uniform distribution with limits equal to $\pm 30\%$ of the literature values for each parameter (\autoref{tab:UncertainParams}).

\begin{table}[!ht]
\centering
\resizebox{\textwidth}{!}{
\begin{tabular}{l | l | l | l | l} 
\toprule 
& \multicolumn{2}{|c|}{Aorto-Femoral Ranges} & \multicolumn{2}{|c}{Coronary Ranges}\\
\midrule
Uncertain Parameter& Min & Max & Min & Max \\ 
\midrule
\midrule
BC: Total $R$ & $\SI{1.0079e+03}{}$ & $\SI{1.8718e+03}{}$ & $\SI{1.0500e+03}{}$ & $\SI{1.9500e+03}{}$ \\ 

BC: Total $C$ & $\SI{7.0000e-04}{}$ & $\SI{1.3000e-03}{}$ & $\SI{7.0000e-04}{}$ & $\SI{1.3000e-03}{}$ \\

BC: Ratio of $R_p/R_{total}$ & $\SI{3.9200e-02}{}$ & $\SI{7.2800e-02}{}$ & $\SI{6.3000e-02}{}$ & $\SI{1.1700e-01}{}$ \\ 

BC: Ratio of $R_p/R_{total}$ (renal arteries) & $\SI{1.9600e-01}{}$ & $\SI{3.6400e-01}{}$ & --- & --- \\ 

Young's Modulus & $\SI{4.9700e+05}{}$ & $\SI{9.2300e+05}{}$ & $\SI{4.9700e+05}{}$ & $\SI{9.2300e+05}{}$ \\

Young's Modulus (coronary arteries) & --- & --- & $\SI{8.0500e+05}{}$ & $\SI{1.4950e+06}{}$ \\

Inlet waveform total flow & $\SI{5.8333e+01}{}$ & $\SI{1.0833e+02}{}$ & $\SI{6.3490e+01}{}$  & $\SI{1.1791e+02}{}$ \\

Blood Density & $\SI{7.4200e-01}{}$  & $\SI{1.3780e+00}{}$ & $\SI{7.4200e-01}{}$  & $\SI{1.3780e+00}{}$ \\

Blood Viscosity & $\SI{2.8000e-02}{}$  & $\SI{5.2000e-02}{}$ & $\SI{2.8000e-02}{}$ & $\SI{5.2000e-02}{}$ \\
\bottomrule 
\end{tabular}}
\caption{Uncertain parameters and ranges for uniform distributions of these parameters for aorto-femoral and coronary models. Resistances have units of $\SI{}{\text{dyn }\second\per\centi\meter^5}$. Capacitances have units of $\SI{}{\centi\meter^5\per\text{dyn}}$. Young's moduli have units of $\SI{}{\pascal}$. Flow has units of $\SI{}{\centi\meter\per\second}$. Density has units of $\SI{}{\gram\per\centi\meter^3}$. Viscosity has units of $\SI{}{\pascal\times\second}$.}
\label{tab:UncertainParams}
\end{table}

All parameters were assigned reference (mean) values chosen to match targets from clinical literature. RCR boundary conditions were applied to all outlets of both models to represent the downstream vasculature. The total resistance and capacitance in the model were tuned using a three-element Windkessel model to produce a physiologic pressure waveform, then distributed proportional to the vessel outlet areas \cite{zhou1999design}. The mean values for the proximal ($R_p$) and distal ($R_d$) resistance split was consistent with those used in~\cite{Les2010}. As in that study, a different ratio was used for the renal arteries to account for their unique flow features in the aorto-femoral models. Further details on boundary condition assignment can be found in~\cite{Les2010}.
Mean values chosen for the Young's moduli were corroborated by other studies~\cite{Coogan_2013,Roccabianca_2014,Ramachandra_2016}.
A physiologic inlet waveform was applied to the aortic inlet of both models. As the aorto-femoral model inlet is in the abdominal aorta and the coronary model inlet is in the thoracic aorta, these waveforms differed slightly, but both were consistent with typical clinical assessments for their respective anatomy. All analysis presented in this paper was also carried out with steady inlet waveforms in a preliminary study~\cite{Fleeter_2018}.

Though all stochastic parameters used in this study were uniformly distributed and independent, an advantage of the MLMF method is the flexibility it affords. Generally, sampling based methods such as this allow for heterogeneous sources of uncertainty. Parameters can be a combination of independent parameters with distributions assumed from literature data and parameters assimilated or sampled from clinical data, for example using a Markov Chain Monte Carlo approach as in~\cite{TRAN_2017128}. Allowing for a spectrum of methods for defining uncertainties is an asset of sampling-based approaches such as the MLMF method.

\subsection{Quantities of Interest} \label{sec:QoIs}
Both local and global quantities of interest were included in this study. These were divided into four main categories: time-averaged flow and pressure values at model outlets, and time-averaged blood pressure and wall shear stress values spatially-averaged over individual branches, strips of interest, and the entire model. The strips of interest are the highlighted regions seen in \hyperref[fig:AortaHD]{Figures~\ref*{fig:AortaHD}} and~\ref{fig:CoronaryHD}. A total of 132 and 148 QoIs were computed for the aorta and coronary models, respectively. The same quantities of interest measured at the same regions of interest were used for the healthy and diseased geometries to facilitate comparisons.

While all QoIs used in this study are single values computed by temporally averaging the results over one cardiac cycle, the MLMF method is capable in principle of handling QoI estimators at multiple time points, for example to generate flow and pressure waveforms with confidence intervals over a cardiac cycle. Estimators for a single time step were used here to demonstrate the performance of the MLMF method.

As outlined in~\autoref{sec:UQ Methods}, high correlations between the model fidelities at each resolution level are necessary for the success of our MLMF methods. We find high ($>0.6$) averaged values for Pearson's correlation coefficient across all global and local QoI categories (~\autoref{tab:ModelCorr}). As expected, the global QoIs are more highly correlated than the local QoIs. Within the local QoIs, higher 3D-3D mesh level correlations are observed for the coarser resolution levels than at the fine resolution level, due to the fact that the most spatially resolved models are less well approximated by the low-fidelity models. We look at the correlation between both low- and high-fidelity models as well as within the mesh resolution levels of the high-fidelity model.

\begin{table}
\resizebox{0.98\textwidth}{!}{
	\begin{tabular}{l l | c c | c c | c c | c c } 
	\toprule 
	& & \multicolumn{2}{c|}{Outlet Flow} & \multicolumn{2}{c|}{Outlet Pressure} & \multicolumn{2}{c|}{Model Pressure} & \multicolumn{2}{c}{Model TAWSS} \\
	\multicolumn{2}{c|}{Model and Mesh} & 3D-1D & {3D-0D} & {3D-1D} & {3D-0D} & {3D-1D} & {3D-0D} & {3D-1D} & {3D-0D} \\
	\midrule
	\midrule
	\multirow{3}{*}{\rotatebox[origin=c]{90}{AFH}} 
	& Coarse 	& 0.997	 & 0.995   & 0.995   & 0.995   & 0.995   & 0.992   & 0.825   & 0.647 \\
	& Medium 	& 0.997	 & 0.996   & 0.995   & 0.989   & 0.995   & 0.990   & 0.843   & 0.782  \\
	& Fine  	& 0.997	 & ---     & 0.994   & ---     & 0.995   & ---     & 0.792   & ---   \\
	\midrule
	\multirow{3}{*}{\rotatebox[origin=c]{90}{AFD}} 
	& Coarse 	& 0.998  & 0.995   & 0.996  & 0.996   & 0.996  & 0.992   & 0.834  & 0.612  \\ 
	& Medium	& 0.993  & 0.992   & 0.995  & 0.989   & 0.994  & 0.990   & 0.880  & 0.812  \\
	& Fine 		& 0.998  & ---     & 0.994  & ---     & 0.995  & ---     & 0.818  & ---  \\
	\midrule
	\multirow{3}{*}{\rotatebox[origin=c]{90}{CH}} 
	& Coarse 	& 0.999   & 0.999   & 0.998   & 0.999   & 0.998   & 0.999   & 0.854   & 0.775  \\
	& Medium	& 0.999   & 0.999   & 0.998   & 0.995   & 0.998   & 0.995   & 0.851   & 0.850   \\
	& Fine 		& 0.999   & ---     & 0.998   & ---     & 0.998   & ---     & 0.811   & ---   \\
	\midrule
	\multirow{3}{*}{\rotatebox[origin=c]{90}{CD}} 
	& Coarse	& 0.999   & 0.999   & 0.997   & 0.999   & 0.997   & 0.999   & 0.847   & 0.744  \\
	& Medium	& 0.999   & 0.999   & 0.997   & 0.993   & 0.997   & 0.994   & 0.853   & 0.798  \\
	& Fine 		& 0.999   & ---     & 0.996   & ---     & 0.996   & ---     & 0.799   & --- \\
	\bottomrule
	\vspace{0.05em} \\
	\multicolumn{10}{c}{(a) HF-LF Model Correlations} \\
	\end{tabular}
}
\\
\vspace{2em}
\resizebox{0.98\textwidth}{!}{
	\begin{tabular}{l l | c | c | c | c } 
	\toprule 
	& & Outlet Flow & Outlet Pressure & Model Pressure & Model TAWSS \\
	\multicolumn{2}{c|}{Model and Mesh} & {3D-3D} & {3D-3D} & {3D-3D} & {3D-3D} \\
	\midrule
	\midrule
	\multirow{2}{*}{\rotatebox[origin=c]{90}{AFH}} 
	& Coarse--Medium 	& 0.999	 & 0.999   & 0.999   & 0.999  \\
	& Medium--Fine	    & 0.999	 & 0.999   & 0.999   & 0.998  \\
	\midrule
	\multirow{2}{*}{\rotatebox[origin=c]{90}{AFD}} 
	& Coarse--Medium 	& 0.996	 & 0.999   & 0.999   & 0.939   \\
	& Medium--Fine	    & 0.995	 & 0.998   & 0.998   & 0.924   \\
	\midrule
	\multirow{2}{*}{\rotatebox[origin=c]{90}{CH}} 
	& Coarse--Medium 	& 0.999	 & 0.999   & 0.999   & 0.999  \\
	& Medium--Fine	    & 0.999	 & 0.999   & 0.997   & 0.998  \\
	\midrule
	\multirow{2}{*}{\rotatebox[origin=c]{90}{CD}} 
	& Coarse--Medium 	& 0.999	 & 0.999   & 0.999   & 0.999  \\
	& Medium--Fine	    & 0.999	 & 0.999   & 0.998   & 0.998 \\
	\bottomrule
	\vspace{0.05em} \\
	\multicolumn{6}{c}{(b) HF Mesh Resolution Level Correlations} \\
	\end{tabular}
}
\caption{Averaged Pearson correlation coefficients for each QoI category for all models between (a) the 3D and 1D models (3D-1D) and the 3D and 0D models (3D-0D) of the same resolution level and between (b) the 3D models of different mesh resolutions levels. The former correlations are needed for the control variate portion and the latter are needed for the MLMC portion of our approach. The correlations are very high for the global (flow and pressure) QoIs, and slightly lower for the local (TAWSS) QoIs. However, even for the local QoIs, the correlations are large enough to produce good performance from the MLMF UQ methods. No 0D model fine resolution level was used in our study, so no correlation is reported. Abbreviations: aorto-femoral healthy (AFH), aorto-femoral diseased (AFD), coronary healthy (CH), and coronary diseased (CD).}
\label{tab:ModelCorr}
\end{table}

\section{Results} \label{sec:Results}

\subsection{Comparing Monte Carlo, multilevel, and multilevel multifidelity estimators} \label{sec:ResMCMLMLF}

To justify the additional overhead of introducing lower-fidelity models for the MLMF estimators, we compare the performance of the MLMF estimators to Monte Carlo (MC) and multilevel Monte Carlo (MLMC) estimators, which rely only on three-dimensional models. We see that across a variety of metrics, the MLMF estimators outperform the MC and MLMC estimators, demonstrating the power of the method when subject to constrained computational budgets. MLMF schemes using only one-dimensional low fidelity models and including both one- and zero-dimensional models were evaluated for all four healthy and diseased model geometries, as discussed in~\autoref{sec:UQ MLMF Background}. We will refer to the setup including zero-dimensional models as the \emph{3D-1D-0D} scheme and the setup excluding zero-dimensional models as the \emph{3D-1D} scheme. For the {3D-1D} scheme, there are three spatial mesh resolutions of both the HF and LF (1D) models, leading to three discrepancy levels $Y_\ell$. For the 3D-1D-0D scheme, the same three HF mesh resolutions are employed. Now, the two lower resolutions of the 1D LF model are replaced with 0D models. The lowest level is the 0D model comprised of only resistor elements, while the middle level is a 0D model incorporating resistance, capacitance, and inductance elements (see~\autoref{sec:Cardio Modeling 0D} for additional details). The highest LF resolution level remains unchanged from the 3D-1D scheme, and is still the finest 1D resolution.

\subsubsection{Bias in Monte Carlo estimators of low-fidelity models} \label{sec:ResBias}

As we observed in~\autoref{tab:ModelCorr}, the correlations between the low- and high-fidelity models are high across our various QoIs. For the global QoIs, we see very high ($> 0.9$) correlations between both the 0D and 3D and the 1D and 3D models. These observed correlations, combined with the inexpensive nature of the low-fidelity models, led us to investigate the bias in the low-fidelity models, in order to justify the use of our multifidelity estimators instead of simply relying on MC estimators from the low-fidelity models.

\begin{figure}[!ht]
\centering
\begin{subfigure}[t]{0.48\textwidth}
	\centering
	\includegraphics[width=\textwidth]{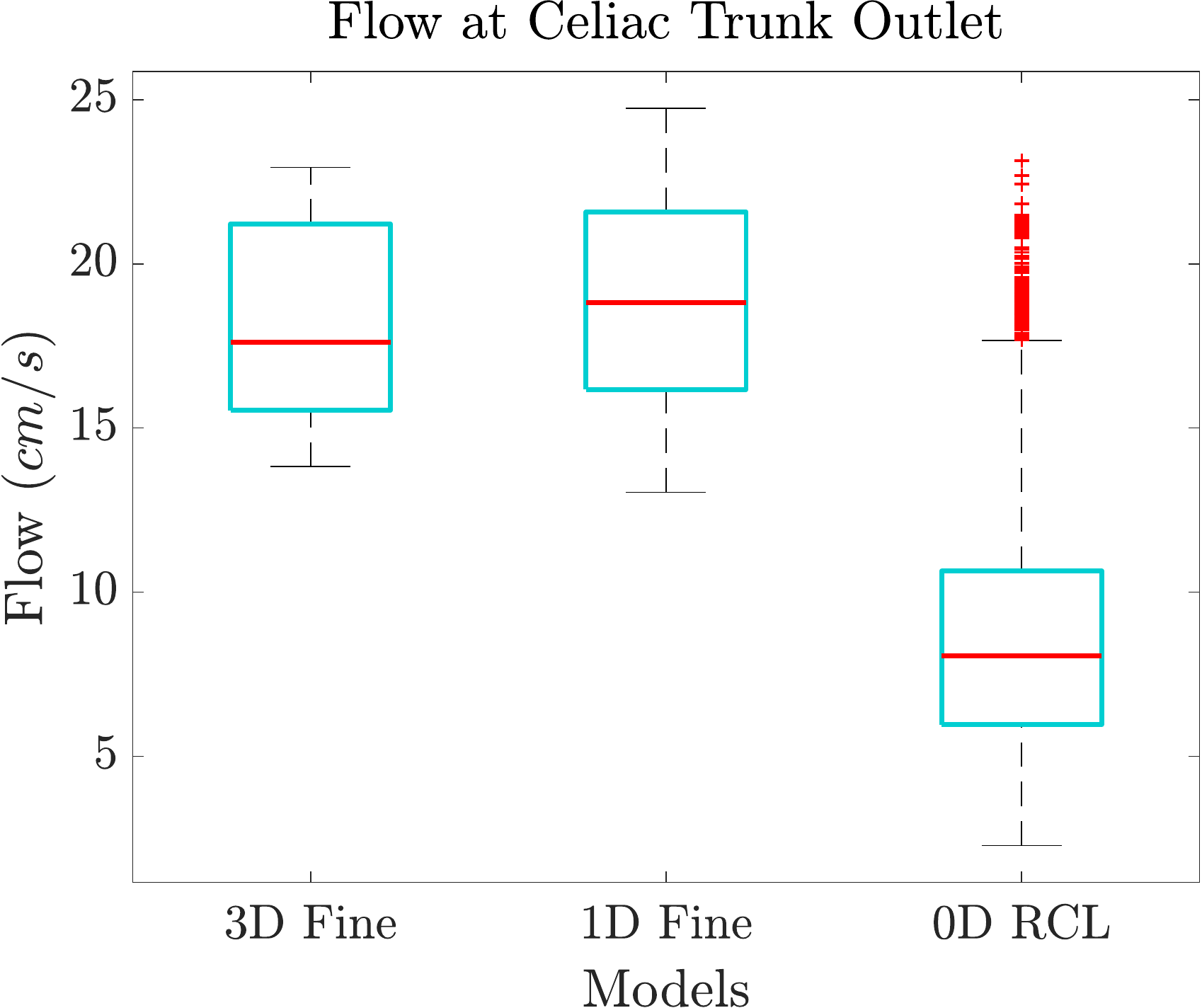}
	\caption{}
	\label{fig:Bias-A1}
\end{subfigure}
\hfill
\begin{subfigure}[t]{0.48\textwidth}
	\centering
	\includegraphics[width=\textwidth]{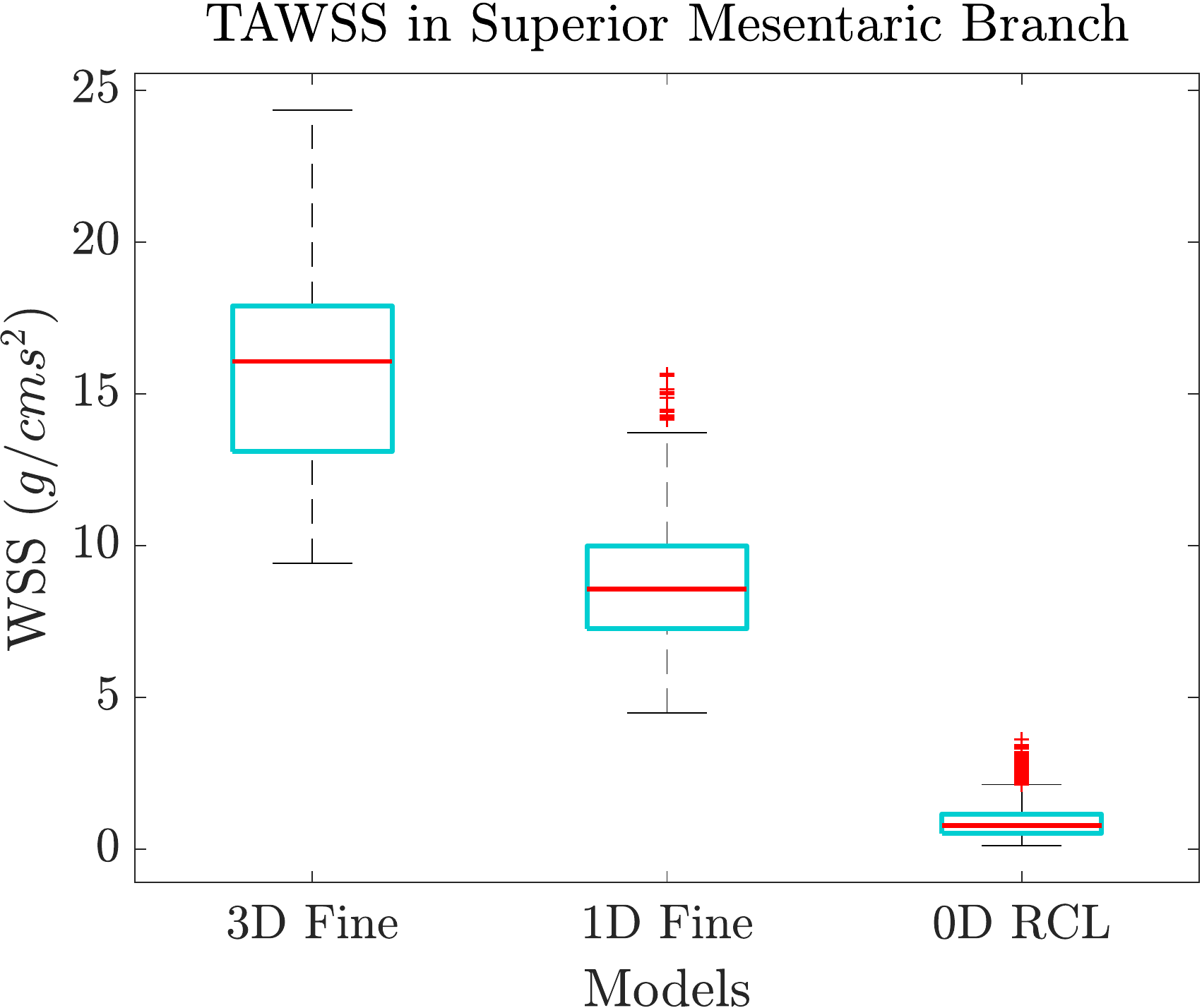}
	\caption{}
	\label{fig:Bias-A2}
\end{subfigure} 
\caption{Examining the bias between MC estimators for different model fidelities for (a) a global flow QoI and (b) a local TAWSS QoI in the aorto-femoral model. MC estimators were built individually from simulations of the 3D Fine model, the 1D Fine model, and 0D RCL model.}
\label{fig:BiasValidate}
\end{figure}

We utilized a Monte Carlo method with only the 0D RCL model simulation results or only the 1D fine model results for various quantities of interest, and compared these to the 3D fine model results. We found varying degrees of bias for different QoIs between the LF model MC estimators and the HF MC estimator. Box plots for representative global and local QoIs from the aorto-femoral model show that even for the global QoI (flow), the 0D RCL MC estimator has a significantly different mean value than the 3D or 1D MC estimators do (\autoref{fig:BiasValidate}). The 3D and 1D model MC estimators do agree in this case. However, for the local QoI (TAWSS), all three MC estimators exhibit different mean values. In this case, the 0D model fails to generate any simulations with QoI values near the 3D MC estimator mean value.

The bias exhibited by the 0D and 1D MC estimators is not a simple correction value, as the bias value differs across QoIs. This prevents solely the 0D or 1D model from being used without any 3D simulations. For this, and other, reasons, the general applicability of the MLMF method appeals to us. Even with a small number of high-fidelity evaluations we can rely heavily on a biased 0D model without having to explicitly calculate or consider this bias. Due to the way the MLMF estimators are constructed, any bias in the LF model is automatically compensated for.

\subsubsection{Estimator cost under a fixed normalized confidence interval} \label{sec:ResMCMLMFnCI}

In our uncertainty quantification workflow, we first complete a \emph{pilot run}, an a priori prescribed number of simulations of each model level and fidelity, chosen here to be 25 samples of each fidelity discrepancy level $Y_{\ell}^{\HF}$ and $Y_{\ell}^{\LF}$ for $\ell = 0,1,2$, for a total of 250 simulations. Recall that for all $\ell > 0$, $Y_\ell$ is the discrepancy between a lower-resolution and higher-resolution simulation, amounting to a total of 50 simulations for each $Y_\ell, \ell >0$. This educated sample size was informed as a reasonable portion of the overall available computational budget. The purpose of the pilot run is threefold: to estimate (i) the variances of the QoIs on each level, (ii) the correlations between high and low fidelity models, and, possibly, (iii) the average model evaluation cost per level. From the estimator variances, we compute the normalized confidence interval of each QoI after the pilot run. After assessing the initial normalized confidence intervals, we estimate how many additional simulations are needed to converge the solution to the desired normalized confidence interval value for the QoIs.

\emph{The normalized confidence interval (nCI)} of a QoI estimator $Q$ is defined here in accordance with previous literature~\cite{Geraci_A_2015,Geraci_A_2017} as six times the coefficient of variation, or the size of three standard deviations corresponding to a 99.7\% confidence interval normalized by the expected value, 
\begin{equation}
nCI[Q] = \frac{6\sqrt{\mathbb{V}[Q]}}{\mathbb{E}[Q]} = \frac{6\,\sigma}{\mu},
\end{equation}
for each QoI. As such, smaller values of nCI are preferred as they indicate a higher level of confidence, i.e. tighter confidence intervals, in the estimator $Q$. For this study, a value of $nCI[Q] = 0.01$ was arbitrarily chosen as the target, though this may be more accurate than needed for different applications. A more extensive study of the proper nCI target value for clinical applications will be the subject of future work.

As explained in~\autoref{sec:UQ MLMF Background}, \emph{extrapolation} can be performed to estimate the additional number of simulations of each model and fidelity needed to improve the estimator variance by a factor $\epsilon$, with $\epsilon \in (0,1]$. This factor $\epsilon$ can be changed to target different variances and, by extension, different nCIs for the QoIs, as nCI is a function of variance. The formulas for the optimal number of samples needed for extrapolation are given by $\eqref{equ:MLOptSample}$, $\eqref{equ:MFOptSample}$, and $\eqref{equ:MLMFOptSample}$ for the MLMC, MFMC, and MLMF estimators, respectively. 

\begin{figure}[!ht]
\centering
\hspace{\baselineskip}
\columnname{Aorto-Femoral Healthy (Estimator Standard Deviation)}\\
\begin{subfigure}[t]{0.24\textwidth}
	\centering
	\includegraphics[width=\textwidth]{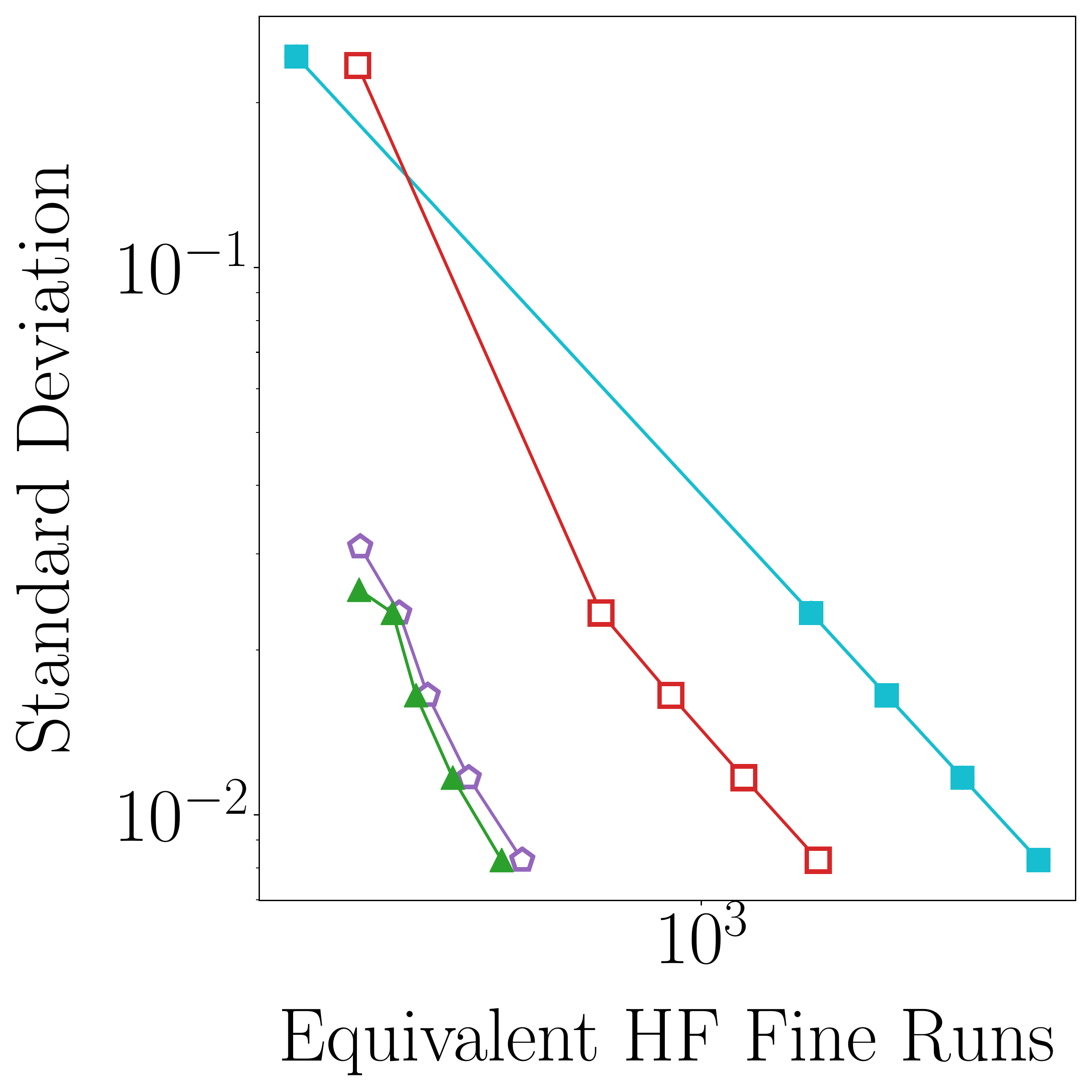}
	\caption{}
	\label{fig:FlowAcc-A}
\end{subfigure}
\begin{subfigure}[t]{0.24\textwidth}
	\centering
	\includegraphics[width=\textwidth]{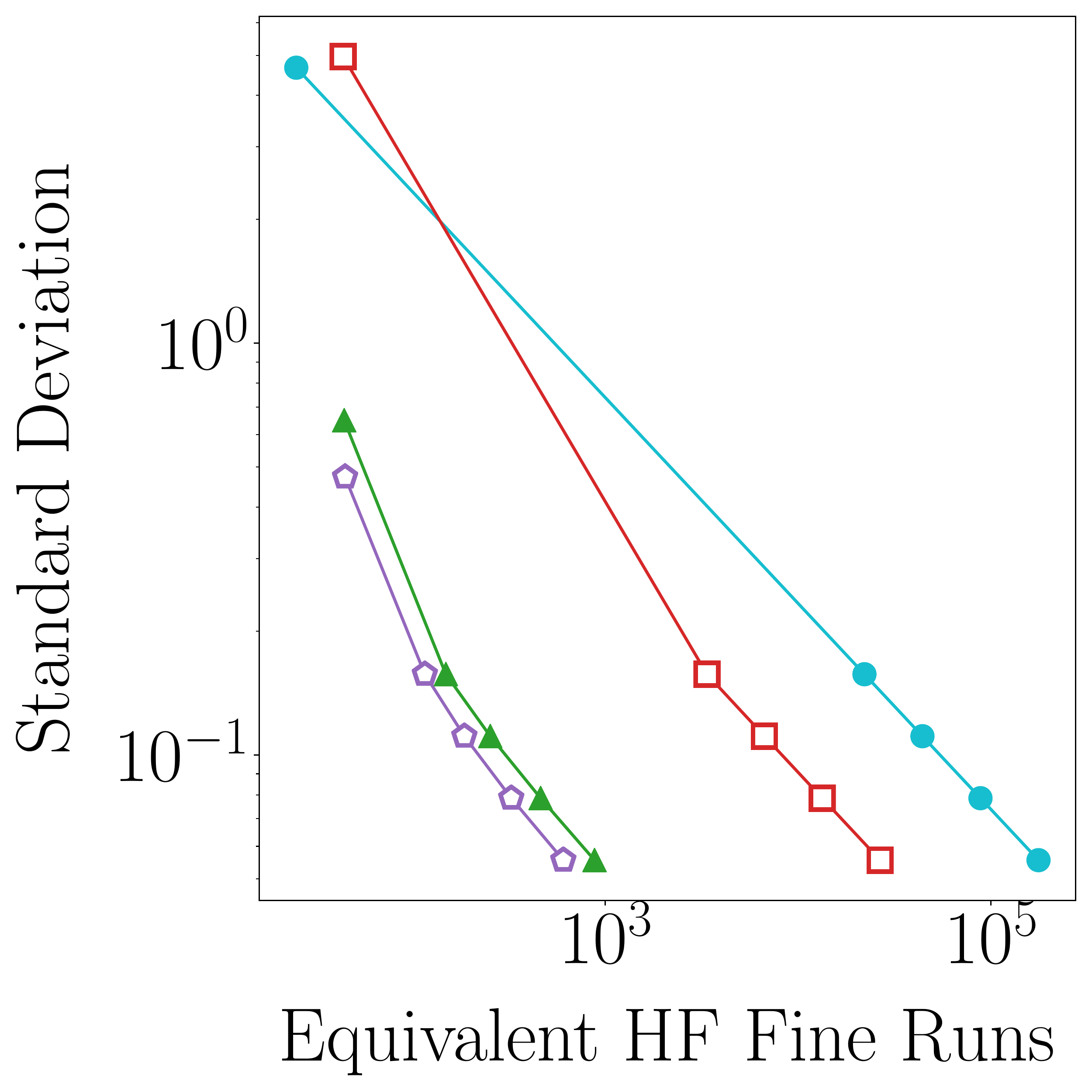}
	\caption{}
	\label{fig:PressAcc-A} 
\end{subfigure} 
\begin{subfigure}[t]{0.24\textwidth}
	\centering
	\includegraphics[width=\textwidth]{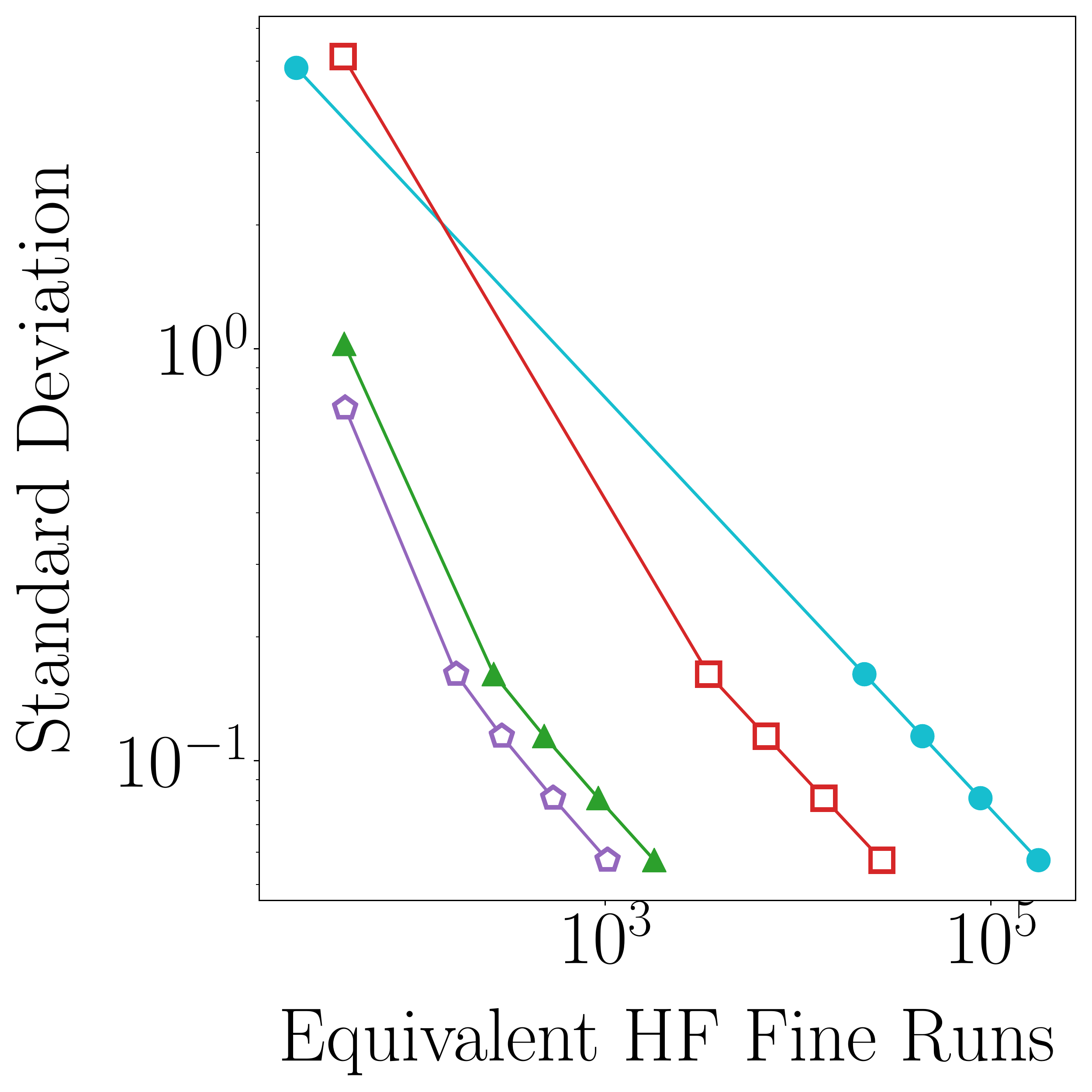}
	\caption{}
	\label{fig:TAPAcc-A} 
\end{subfigure} 
\begin{subfigure}[t]{0.24\textwidth}
	\centering
	\includegraphics[width=\textwidth]{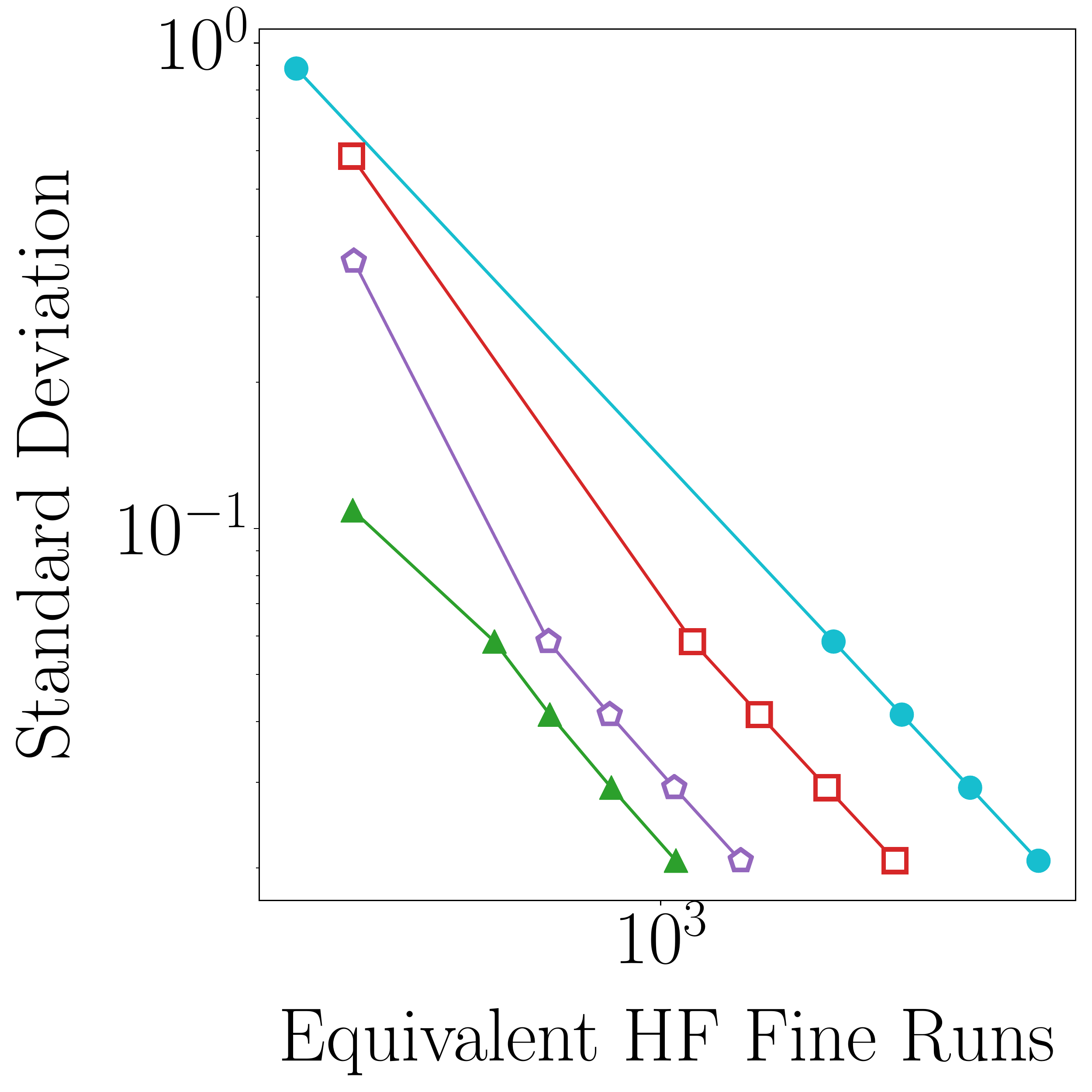}
	\caption{}
	\label{fig:TAWSSAcc-A} 
\end{subfigure} 
\\
\vspace{1em}
\hspace{\baselineskip}
\columnname{Coronary Healthy (Estimator Standard Deviation)}\\
\begin{subfigure}[t]{0.24\textwidth}
	\centering
	\includegraphics[width=\textwidth]{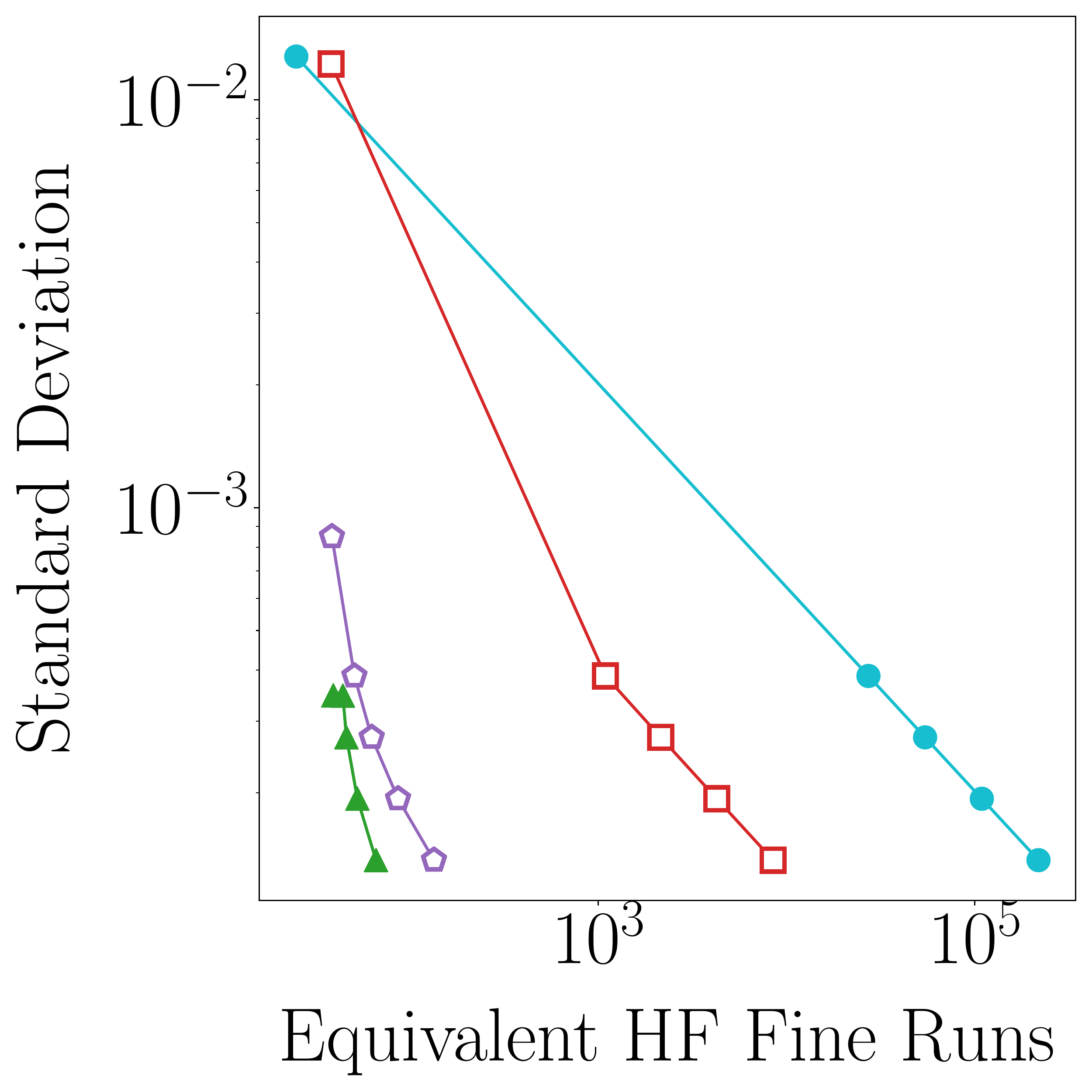}
	\caption{}
	\label{fig:FlowAcc-C}
\end{subfigure}
\begin{subfigure}[t]{0.24\textwidth}
	\centering
	\includegraphics[width=\textwidth]{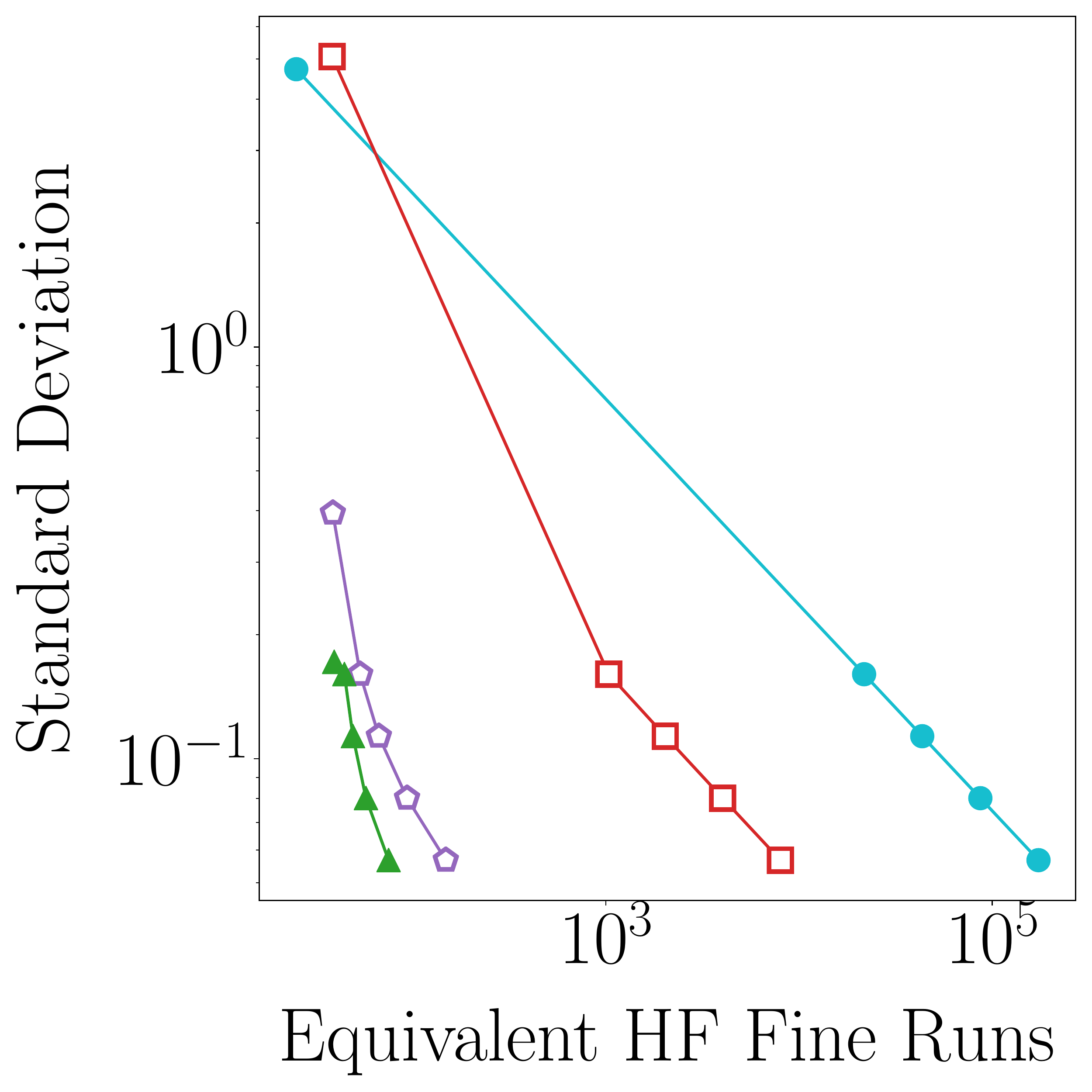}
	\caption{}
	\label{fig:PressAcc-C} 
\end{subfigure} 
\begin{subfigure}[t]{0.24\textwidth}
	\centering
	\includegraphics[width=\textwidth]{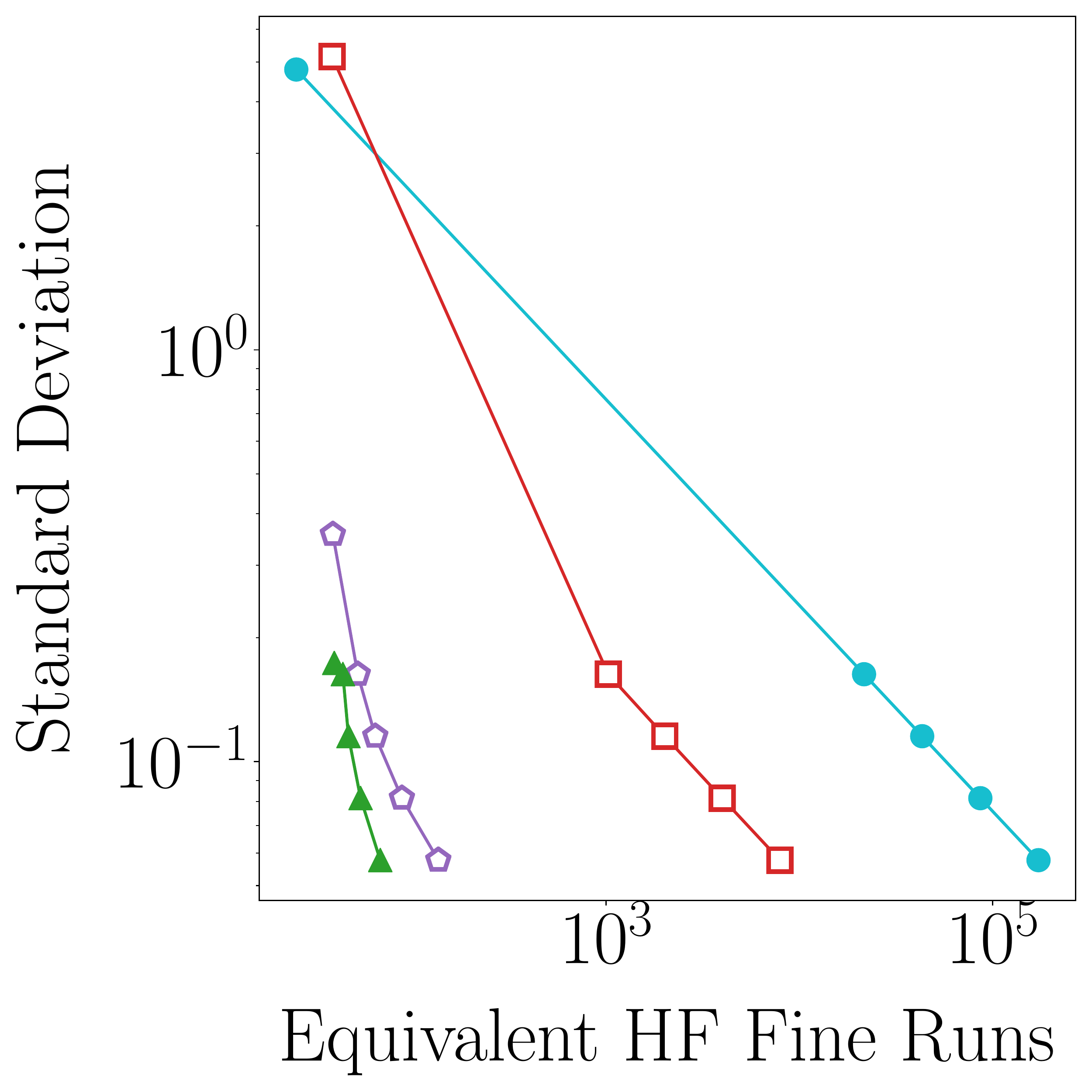}
	\caption{}
	\label{fig:TAPAcc-C} 
\end{subfigure} 
\begin{subfigure}[t]{0.24\textwidth}
	\centering
	\includegraphics[width=\textwidth]{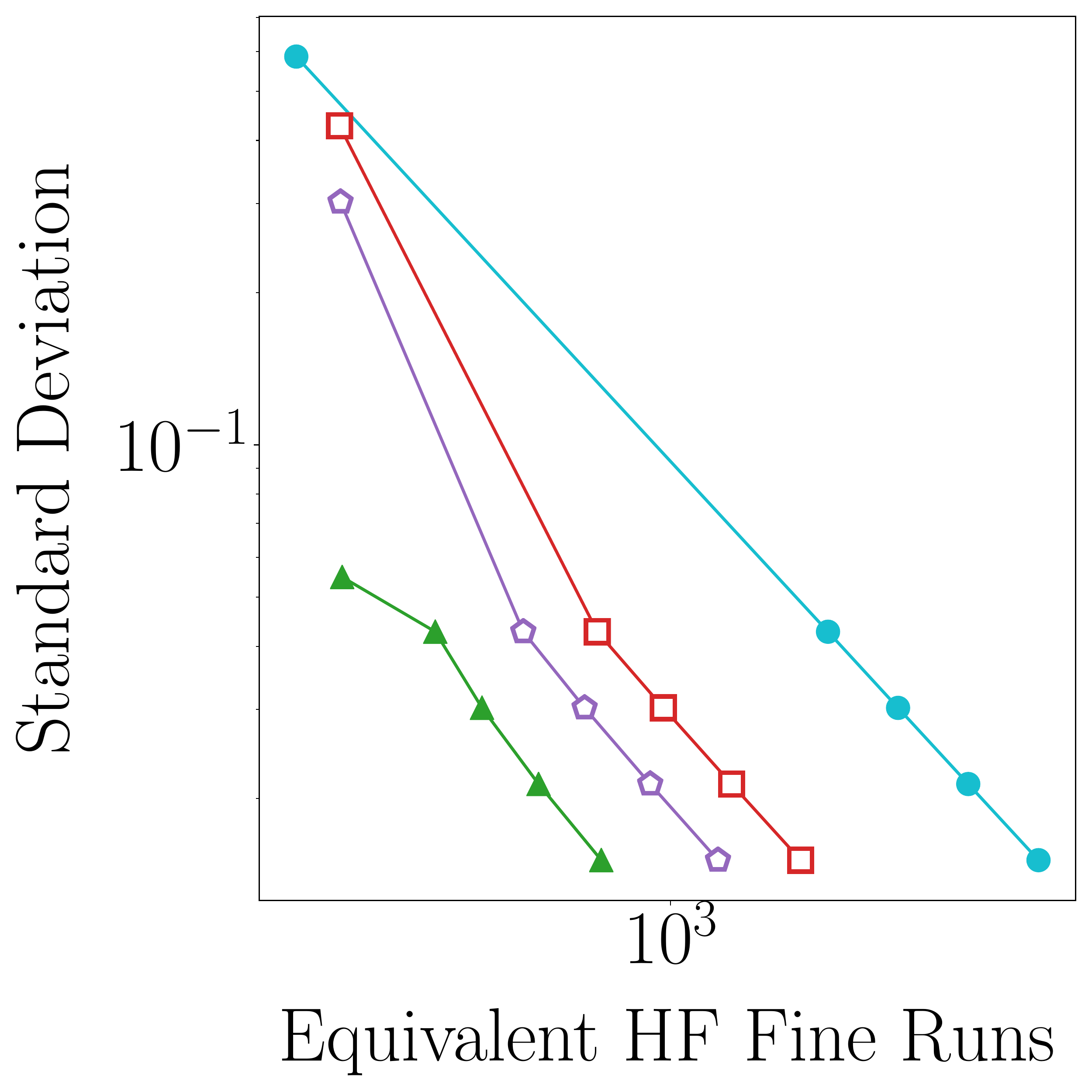}
	\caption{}
	\label{fig:TAWSSAcc-C} 
\end{subfigure} 
\\
\begin{subfigure}[t]{0.7\textwidth} 
	\centering
	\includegraphics[width=\textwidth]{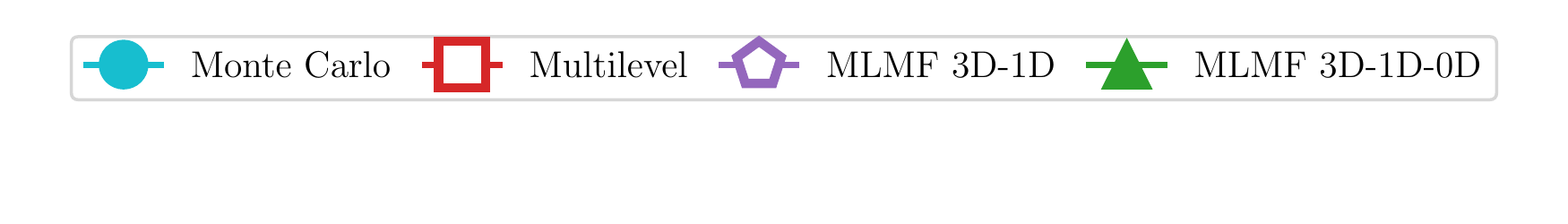}
	\label{fig:3d1dlegend}  
\end{subfigure}	
\vspace{-3em}
\caption{Comparing the performance of estimators from MC, MLMC, and MLMF UQ schemes for (top) aorto-femoral and (bottom) coronary healthy models. The values shown are extrapolated expected costs (in comparable units of the equivalent number of 3D fine simulations) to obtain various estimator standard deviation values for representative QoIs from each category. Standard deviation cost curves are shown for representative outlet flow QoIs in figures (a) and (e), outlet pressure QoIs in figures (b) and (f), model time-averaged pressure QoIs in figures (c) and (g), and model TAWSS QoIs in figures (d) and (h) for the aorto-femoral and coronary models, respectively. These quantities are spatially and temporally averaged over the outlet face for the flow and pressure QoIs, throughout regions of the model volume for the model pressure QoIs, or over regions of the model surface for model TAWSS QoIs respectively.}
\label{fig:AccCompare}
\end{figure}

MLMF estimators outperformed MC and MLMC methods when considering extrapolation with different target variances (i.e. different $\epsilon$ factors) for a specific QoI. In this context, outperforming means that the computational cost to obtain a given variance is lower for the MLMF schemes. The superior MLMF performance holds for both healthy and diseased models and both global and local QoIs (\autoref{fig:AccCompare}).

When extrapolating to a nCI of 0.01, the exact breakdown of projected costs for the same QoIs as in~\autoref{fig:AccCompare} shows that the MLMF methods offer one to two orders of magnitude improvement in computational cost over the MLMC method and one to three orders of magnitude improvement over the MC method (\hyperref[tab:MethodsCompare]{Tables~\ref*{tab:MethodsCompare}a} and \hyperref[tab:MethodsCompare]{\ref*{tab:MethodsCompare}b} for the aorto-femoral and coronary models, respectively). Lower improvement factors are seen for local QoIs, as the local QoIs rely heavily on the detailed spatial information from the HF model evaluations. This means that the extrapolated costs will be more similar across UQ methods for these QoIs. The extrapolated cost reported includes the initial cost of the pilot run. Extrapolated costs are reported in comparable units of the \emph{equivalent number 3D fine simulations}. This is calculated as 
\begin{align} 
\begin{split}
N_{Eq, Fine}^{3D} = &\bigl(N_{Fine}^{3D} C_{Fine}^{3D} + N_{Med}^{3D} C_{Med}^{3D} + N_{Coa}^{3D} C_{Coa}^{3D} \\[5pt]
&+ N_{Fine}^{1D} C_{Fine}^{1D} + N_{Med}^{1D} C_{Med}^{1D} + N_{Coa}^{1D} C_{Coa}^{1D}\bigr) / C_{Fine}^{3D}, 
\end{split}
\end{align} 
where $C_{Level}^{Fidelity}$ is the cost of one simulation of that level and fidelity and $N_{Level}^{Fidelity}$ includes the total number of all pilot run simulations and all additional simulations needed for the extrapolation to $nCI[Q] = 0.01$.

\begin{table}
\resizebox{0.48\textwidth}{!}{
	\begin{tabular}{l l | r  | r} 
	\toprule 
	& & \multicolumn{1}{c|}{Healthy Model} & \multicolumn{1}{c}{Diseased Model} \\
	\multicolumn{2}{c|}{QoI and Method} & \multicolumn{1}{c|}{Effective Cost} & \multicolumn{1}{c}{Effective Cost} \\
	\midrule
	\midrule
	\multirow{4}{*}{\rotatebox[origin=c]{90}{Outlet}}
	\multirow{4}{*}{\rotatebox[origin=c]{90}{Flow}} 
	& MC 				& 9\,758.0 	 & 9\,471.0   \\
	& MLMC 				& 1\,293.6   & 2\,593.8   \\
	& MLMF (3D-1D) 		&     77.9   &    468.3   \\
	& MLMF (3D-1D-0D) 	&     66.5   &    440.2   \\
	\midrule
	\multirow{4}{*}{\rotatebox[origin=c]{90}{Outlet}}
	\multirow{4}{*}{\rotatebox[origin=c]{90}{Pressure}} 
	& MC 				& 19\,340.0  & 18\,374.0  \\
	& MLMC 				&  2\,915.4  &  4\,588.1  \\
	& MLMF (3D-1D) 		&      82.2  &     282.2  \\
	& MLMF (3D-1D-0D)	&     107.0  &     282.2  \\
	\midrule
	\multirow{4}{*}{\rotatebox[origin=c]{90}{Model}}
	\multirow{4}{*}{\rotatebox[origin=c]{90}{Pressure}} 
	& MC 				& 19\,598.0  & 19\,050.0  \\
	& MLMC 				&  3\,016.3  &  4\,430.4  \\
	& MLMF (3D-1D) 		&     119.4  &     321.9  \\
	& MLMF (3D-1D-0D) 	&     210.5  &     369.2  \\
	\midrule
	\multirow{4}{*}{\rotatebox[origin=c]{90}{Model}}
	\multirow{4}{*}{\rotatebox[origin=c]{90}{TAWSS}} 
	& MC 				& 23\,94.0   & 22\,581.0  \\
	& MLMC 				&  5\,578.7  &  4\,695.4  \\
	& MLMF (3D-1D) 		&  1\,151.4  &  1\,922.4  \\
	& MLMF (3D-1D-0D) 	&     581.2  &     847.0  \\
	\bottomrule 
	\vspace{0.05em} \\
	\multicolumn{4}{c}{(a) Aorto-Femoral Model} \\
	\end{tabular}
}
\quad
\resizebox{0.48\textwidth}{!}{
	\begin{tabular}{l l | r  | r } 
	\toprule 
	& & \multicolumn{1}{c|}{Healthy Model} & \multicolumn{1}{c}{Diseased Model} \\
	\multicolumn{2}{c|}{QoI and Method} & \multicolumn{1}{c|}{Effective Cost} & \multicolumn{1}{c}{Effective Cost}\\
	\midrule
	\midrule
	\multirow{4}{*}{\rotatebox[origin=c]{90}{Outlet}}
	\multirow{4}{*}{\rotatebox[origin=c]{90}{Flow}} 
	& MC 				& 10\,656.0  & 10\,619.0 \\
	& MLMC 				&     435.1  &  1\,095.5  \\
	& MLMF (3D-1D) 		&      41.0  &      49.4  \\
	& MLMF (3D-1D-0D)	&      39.5  &      43.7  \\
	\midrule
	\multirow{4}{*}{\rotatebox[origin=c]{90}{Outlet}}
	\multirow{4}{*}{\rotatebox[origin=c]{90}{Pressure}} 
	& MC 				& 20\,467.0  & 20\,587.0  \\
	& MLMC				&     948.1  &  2\,480.6   \\
	& MLMF (3D-1D) 		&      46.6  &      65.1   \\
	& MLMF (3D-1D-0D) 	&      40.4  &      44.7   \\
	\midrule
	\multirow{4}{*}{\rotatebox[origin=c]{90}{Model}}
	\multirow{4}{*}{\rotatebox[origin=c]{90}{Pressure}} 
	& MC 				& 20\,213.0   & 20\,330.0  \\
	& MLMC 				&     934.1   &  2\,445.8   \\
	& MLMF (3D-1D) 		&      45.1   &      65.0   \\
	& MLMF (3D-1D-0D)	&      40.4   &      46.4   \\
	\midrule
	\multirow{4}{*}{\rotatebox[origin=c]{90}{Model}}
	\multirow{4}{*}{\rotatebox[origin=c]{90}{TAWSS}} 
	& MC 				& 21\,748.0   & 16\,391.0  \\
	& MLMC 				&  2\,066.7   &  2\,391.4   \\
	& MLMF (3D-1D) 		&     901.9   &  1\,303.8  \\
	& MLMF (3D-1D-0D) 	&     290.6   &     269.7   \\
	\bottomrule 
	\vspace{0.05em} \\
	\multicolumn{4}{c}{(b) Coronary Model} \\
	\end{tabular}
}
\caption{Comparison of the performance of different UQ methods. Each of the four UQ methods compared (Monte Carlo, multilevel, MLMF with 3D, 1D, and 0D models, and MLMF with only 3D and 1D models) requires differing costs to achieve a nCI value of 0.01. For four representative outlet flow, outlet pressure, time-averaged model pressure, and TAWSS QoIs, the cost of each method (in units of the equivalent number of 3D fine simulations) is shown. The same representative QoI is used for both healthy and diseased models. These QoIs are spatially and temporally averaged over the outlet face, throughout regions of the model volume, or over regions of the model surface for the flow and pressure, model pressure, and model TAWSS QoIs respectively.}
\label{tab:MethodsCompare}	
\end{table}

We observe several trends across all UQ methods which we examine in depth in subsequent sections. Local QoIs are more expensive to converge than global QoIs. QoIs for the diseased model can be more expensive to converge than those same QoIs in the healthy model. While the exact extrapolated values differ between QoIs of the same category, the specific QoIs presented here for extrapolation are representative of the general behavior we see.

\subsubsection{MLMF estimator cost breakdown under a fixed normalized confidence interval} 

To demonstrate how the cost burden is shifted to the least expensive models with our estimators, a level-by-level breakdown of the extrapolated simulations is shown for one global and one local QoI. From the aorto-femoral model we have chosen the flow and TAWSS QoIs (\hyperref[tab:MethodsBreakdown]{Table~\ref*{tab:MethodsBreakdown}a} and \hyperref[tab:MethodsBreakdown2]{Table~\ref*{tab:MethodsBreakdown2}a}). From the coronary model we have chosen the pressure and TAWSS QoIs (\hyperref[tab:MethodsBreakdown]{Table~\ref*{tab:MethodsBreakdown}b} and \hyperref[tab:MethodsBreakdown2]{Table~\ref*{tab:MethodsBreakdown2}b}). We look at both the breakdown in terms of pure number of simulations of each model level (\autoref{tab:MethodsBreakdown}), where a single LF and HF model are considered equivalent, and the breakdown in terms of contribution to the total cost of the extrapolation (\autoref{tab:MethodsBreakdown2}).

\begin{table}[!ht]
\centering
\resizebox{0.8\textwidth}{!}{
	\begin{tabular}{l l | l l l l | l l l l } 
	\toprule 
	& & \multicolumn{4}{c|}{Outlet Flow } & \multicolumn{4}{c}{Model TAWSS} \\
	& & MC & MLMC & MLMF & MLMF & MC & MLMC & MLMF & MLMF \\
	\multicolumn{2}{c|}{} &    &    & 3D-1D & 3D-1D-0D &  &  & 3D-1D & 3D-1D-0D \\
	\midrule
	\midrule
	\multirow{6}{*}{\rotatebox[origin=c]{90}{Healthy}} 
	& LF Coarse	& --- 	& ---  		& 96.6\% 	& 98.9\% 	& --- 		& ---	 	& 77.9\% 	& 94.7\% \\
	& LF Medium 	& --- 	& ---  		& 1.7\%    	& 0.9\% 	& --- 		& ---	 	& 10.8\%    & 4.1\%    \\
	& LF Fine 	& --- 	& ---  		& 1.4\%    	& 0.2\% 	& --- 		& ---  		& 10.1\%	& 1.1\%    \\
	& HF Coarse 	& --- 	& 98.1\%  	& 0.2\%    	& 0.0\% 	& --- 		& 84.4\%  	& 1.1\%	  	& 0.0\%    \\
	& HF Medium	& --- 	& 1.1\%   	& 0.0\%     & 0.0\% 	& --- 		& 8.2\%   	& 0.0\%	  	& 0.0\%	   \\
	& HF Fine 	& 100\% & 0.8\% 	& 0.0\%     & 0.0\% 	& 100\% 	& 7.4\%   	& 0.1\%	  	& 0.0\%	   \\
	\multicolumn{2}{l|}{\textbf{Total No. Sims.}}& 
	\textbf{9\,758} & 
	\textbf{10\,386} & 
	\textbf{133\,843}   & 
	\textbf{883\,690} & 

	\textbf{23\,940} & 
	\textbf{25\,469}   & 
	\textbf{543\,309} & 
	\textbf{3\,060\,780}  \\

	\midrule
	\multirow{6}{*}{\rotatebox[origin=c]{90}{Diseased}} 
	& LF Coarse	& --- 	& ---	 	& 92.2\% 	& 97.1\%	& --- 	& ---  		& 78.6\% 	& 94.0\% \\
	& LF Medium 	& --- 	& ---	  	& 5.3\% 	& 2.4\%		& --- 	& ---  		& 14.1\%    & 5.3\%  \\
	& LF Fine 	& --- 	& ---   	& 2.3\%	   	& 0.4\%		& --- 	& ---  		& 5.6\%     & 0.7\%  \\
	& HF Coarse 	& --- 	& 89.9\%   	& 0.1\%	   	& 0.0\%		& --- 	& 84.3\%  	& 1.4\%    	& 0.0\%  \\
	& HF Medium	& --- 	& 6.7\%    	& 0.1\%	   	& 0.0\%		& --- 	& 10.3\%   	& 0.1\%     & 0.0\%   \\
	& HF Fine 	& 100\% & 3.4\%     & 0.1\%	   	& 0.0\%		& 100\% & 5.4\%    	& 0.1\%     & 0.0\%    \\
	\multicolumn{2}{l|}{\textbf{Total No. Sims.}}& 
	\textbf{9\,471} & 
	\textbf{18\,626} & 
	\textbf{363\,442} & 
	\textbf{2\,240\,202} & 

	\textbf{22\,581} & 
	\textbf{27\,537}   & 
	\textbf{673\,023} & 
	\textbf{3\,407\,848}  \\
	
	\bottomrule 
	\vspace{0.05em} \\
	\multicolumn{10}{c}{(a) Aorto-Femoral Model} \\
	\end{tabular}
}
\vspace{1.5em}\\
\resizebox{0.8\textwidth}{!}{
	\begin{tabular}{l l | l l l l | l l l l } 
	\toprule 
	& & \multicolumn{4}{c|}{Outlet Pressure} & \multicolumn{4}{c}{Model TAWSS} \\
	& & MC & MLMC & MLMF & MLMF & MC & MLMC & MLMF & MLMF \\
	\multicolumn{2}{c|}{} &    &    & 3D-1D & 3D-1D-0D &  &  & 3D-1D & 3D-1D-0D \\
	\midrule
	\midrule
	\multirow{6}{*}{\rotatebox[origin=c]{90}{Healthy}} 
	& LF Coarse	& --- 	& ---     & 96.4\%  & 98.4\% & ---   & ---	    & 82.4\%  & 98.6\% \\
	& LF Medium 	& --- 	& ---     & 2.0\%   & 0.9\%  & ---   & ---	    & 11.6\%  & 0.9\%    \\
	& LF Fine 	& --- 	& ---     & 1.4\%   & 0.7\%  & ---   & ---     	& 4.6\%   & 0.4\%    \\
	& HF Coarse 	& --- 	& 99.6\%  & 0.1\%   & 0.0\%  & ---   & 91.6\% 	& 1.3\%   & 0.0\%  \\
	& HF Medium	& --- 	& 0.3\%   & 0.0\%   & 0.0\%  & ---   & 6.3\%  	& 0.1\%   & 0.0\%	   \\
	& HF Fine 	& 100\% & 0.1\%   & 0.0\%   & 0.0\%  & 100\% & 2.1\%   	& 0.0\%   & 0.0\%	   \\
	\multicolumn{2}{l|}{\textbf{Total No. Sims.}}& 
	\textbf{20\,467} & 
	\textbf{24\,799} & 
	\textbf{197\,712}   & 
	\textbf{653\,682} & 
	
	\textbf{21\,748} & 
	\textbf{27\,298} & 
	\textbf{932\,759}   & 
	\textbf{3\,717\,084} \\

	\midrule
	\multirow{6}{*}{\rotatebox[origin=c]{90}{Diseased}} 
	& LF Coarse	& ---  	& ---     & 96.7\%  & 98.5\% 	& ---   & ---	    & 91.2\%  & 99.0\% \\
	& LF Medium 	& --- 	& ---     & 1.3\%   & 0.6\% 	& ---   & ---	    & 3.5\%   & 0.4\%    \\
	& LF Fine 	& --- 	& ---     & 1.9\%   & 0.9\% 	& ---   & ---     	& 4.0\%   & 0.6\%    \\
	& HF Coarse 	& --- 	& 99.6\%  & 0.1\%   & 0.0\% 	& ---   & 94.0\%  	& 1.2\%   & 1.2\%  \\
	& HF Medium	& --- 	& 0.3\%   & 0.0\%   & 0.0\% 	& ---   & 2.8\%   	& 0.0\%   & 0.0\%  \\
	& HF Fine 	& 100\% & 0.1\%   & 0.0\%   & 0.0\% 	& 100\% & 3.2\%   	& 0.0\%   & 0.0\%  \\
	\multicolumn{2}{l|}{\textbf{Total No. Sims.}}&
	\textbf{20\,587} & 
	\textbf{24\,188} & 
	\textbf{214\,491}   & 
	\textbf{563\,280} & 
	
	\textbf{16\,391} & 
	\textbf{16\,615} & 
	\textbf{699\,517}   & 
	\textbf{2\,894\,913}   \\

	\bottomrule 
	\vspace{0.05em} \\
	\multicolumn{10}{c}{(b) Coronary Model} \\
	\end{tabular}
}
\caption{Comparison of the extrapolated number of simulations to obtain $nCI[Q] = 0.01$ for different UQ methods. The distribution of number of simulations by model level and fidelity is shown as a percentage of the total number of simulations needed for extrapolation. The total number of simulations is shown on the final row for the healthy and diseased models. The same representative QoI is used for both healthy and diseased models. Values marked as 0.0\% contribute negligibly to the overall percentage of simulations.}
\label{tab:MethodsBreakdown}
\end{table}

\begin{table}[!ht]
\centering
\resizebox{0.8\textwidth}{!}{
	\begin{tabular}{l l | l l l l | l l l l } 
	\toprule 
	& & \multicolumn{4}{c|}{Outlet Flow } & \multicolumn{4}{c}{Model TAWSS} \\
	& & MC & MLMC & MLMF & MLMF & MC & MLMC & MLMF & MLMF \\
	\multicolumn{2}{c|}{} &    &    & 3D-1D & 3D-1D-0D &  &  & 3D-1D & 3D-1D-0D \\
	\midrule
	\midrule
	\multirow{6}{*}{\rotatebox[origin=c]{90}{Healthy}} 
	& LF Coarse	& --- 	& ---  		& 6.0\% 	& 0.8\% 	& --- 		& ---	 	& 1.3\% 	& 0.3\% \\
	& LF Medium 	& --- 	& ---  		& 0.3\%    	& 0.1\% 	& --- 		& ---	 	& 0.5\%     & 0.2\%    \\
	& LF Fine 	& --- 	& ---  		& 0.7\%    	& 0.6\% 	& --- 		& ---  		& 1.3\%		& 1.4\%    \\
	& HF Coarse 	& --- 	& 88.7\%  	& 40.5\%    & 36.9\% 	& --- 		& 43.4\%  	& 59.8\% 	& 5.9\%    \\
	& HF Medium	& --- 	& 3.2\%   	& 12.0\%    & 14.1\% 	& --- 		& 14.0\%   	& 7.0\%	  	& 8.4\%	   \\
	& HF Fine 	& 100\% & 8.1\% 	& 40.5\%    & 47.4\% 	& 100\% 	& 42.6\%   	& 30.0\%	& 83.8\%	   \\
	\multicolumn{2}{l|}{\textbf{Total Cost (Hours)}}& 
	\textbf{8\,497\,300} & 
	\textbf{1\,126\,500} & 
	\textbf{67\,833}   & 
	\textbf{57\,927} & 

	\textbf{20\,847\,000} & 
	\textbf{4\,857\,900}   & 
	\textbf{1\,002\,700} & 
	\textbf{506\,120}  \\
	
	\midrule
	\multirow{6}{*}{\rotatebox[origin=c]{90}{Diseased}} 
	& LF Coarse	& --- 	& ---	  	& 2.7\% 	& 0.4\%			& --- 	& ---  		& 1.0\% 	 & 0.3\% \\
	& LF Medium 	& --- 	& ---	  	& 0.4\% 	& 0.2\%			& --- 	& ---  		& 0.5\%      & 0.3\%  \\
	& LF Fine 	& --- 	& ---   	& 0.6\%	   	& 0.6\%			& --- 	& ---  		& 0.7\%      & 0.9\%  \\
	& HF Coarse 	& --- 	& 54.4\%   	& 7.9\%	   	& 10.3\%		& --- 	& 41.7\%  	& 42.7\%     & 7.7\%  \\
	& HF Medium	& --- 	& 15.5\%    & 27.9\%	& 27.6\%		& --- 	& 19.3\%   	& 10.4\%     & 16.3\%   \\
	& HF Fine 	& 100\% & 30.1\%    & 60.4\%	& 60.9\%		& 100\% & 39.0\%    & 44.7\%     & 74.5\%    \\
	\multicolumn{2}{l|}{\textbf{Total Cost (Hours)}}& 
	\textbf{6\,319\,300} & 
	\textbf{1\,730\,600} & 
	\textbf{312\,490} & 
	\textbf{293\,710} & 

	\textbf{15\,067\,00} & 
	\textbf{3\,132\,900}   & 
	\textbf{1\,282\,700} & 
	\textbf{565\,160}  \\

	\bottomrule 
	\vspace{0.05em} \\
	\multicolumn{10}{c}{(a) Aorto-Femoral Model} \\
	\end{tabular}
}
\vspace{1.5em}\\
\resizebox{0.8\textwidth}{!}{
	\begin{tabular}{l l | l l l l | l l l l } 
	\toprule 
	& & \multicolumn{4}{c|}{Outlet Pressure} & \multicolumn{4}{c}{Model TAWSS} \\
	& & MC & MLMC & MLMF & MLMF & MC & MLMC & MLMF & MLMF \\
	\multicolumn{2}{c|}{} &    &    & 3D-1D & 3D-1D-0D &  &  & 3D-1D & 3D-1D-0D \\
	\midrule
	\midrule
	\multirow{6}{*}{\rotatebox[origin=c]{90}{Healthy}} 
	& LF Coarse	& --- 	& ---     & 3.4\%  & 0.4\%   & ---   & ---	    	& 0.7\%  & 0.3\% \\
	& LF Medium 	& --- 	& ---     & 0.2\%   & 1.1\%   & ---   & ---	    	& 0.3\%  & 0.\%   \\
	& LF Fine 	& --- 	& ---     & 0.3\%   & 1.2\%   & ---   & ---     	& 0.2\%   & 0.6\%    \\
	& HF Coarse 	& --- 	& 94.7\%  & 15.9\%   & 4.9\%  & ---   & 44.0\% 	    & 47.1\%   & 4.1\%  \\
	& HF Medium	& --- 	& 2.1\%   & 14.3\%   & 16.4\%  & ---   & 22.1\%  	& 15.4\%   & 27.6\%	   \\
	& HF Fine 	& 100\% & 3.2\%   & 65.9\%   & 76.0\%  & 100\% & 33.9\%   	& 36.3\%   & 66.4\%   \\
	\multicolumn{2}{l|}{\textbf{Total Cost (Hours)}}& 
	\textbf{44\,303\,000} & 
	\textbf{3\,052\,200} & 
	\textbf{100\,830}   & 
	\textbf{87\,551} & 
	
	\textbf{47\,076\,000} & 
	\textbf{4\,473\,700} & 
	\textbf{1\,952\,300}   & 
	\textbf{628\,990} \\
	
	\midrule
	\multirow{6}{*}{\rotatebox[origin=c]{90}{Diseased}} 
	& LF Coarse	& --- 	& ---     & 5.0\%    & 0.6\% 	& ---   & ---	    & 0.8\%  & 0.5\% \\
	& LF Medium 	& --- 	& ---     & 0.2\%    & 0.3\% 	& ---   & ---    & 0.1\%   & 0.2\%    \\
	& LF Fine 	& --- 	& ---     & 0.6\%    & 1.2\% 	& ---   & ---     	& 0.2\%   & 0.7\%    \\
	& HF Coarse 	& --- 	& 97.8\%  & 33.6\%   & 9.5\% 	& ---   & 65.7\%  	& 64.2\%   & 16.8\%   \\
	& HF Medium	& --- 	& 0.9\%   & 13.1\%   & 19.0\% 	& ---   & 6.5\%   	& 5.0\%   & 14.2\%	   \\
	& HF Fine 	& 100\% & 1.3\%   & 47.6\%   & 69.4\% 	& 100\% & 27.7\%   	& 29.8\%   & 67.6\%   \\
	\multicolumn{2}{l|}{\textbf{Total Cost (Hours)}}&
	\textbf{24\,673\,000} & 
	\textbf{2\,972\,900} & 
	\textbf{78\,018}   & 
	\textbf{53\,530} & 
	
	\textbf{19\,644\,000} & 
	\textbf{2\,866\,000} & 
	\textbf{1\,562\,600}   & 
	\textbf{323\,170}   \\

	\bottomrule 
	\vspace{0.05em} \\
	\multicolumn{10}{c}{(b) Coronary Model} \\
	\end{tabular}
}
\caption{Comparison of the cost of the extrapolated simulations to obtain $nCI[Q] = 0.01$ for different UQ methods. The cost distribution by model level and fidelity is shown as a percentage of the total cost of all simulations needed for extrapolation. The total cost of simulations, in hours, is shown on the final row for the healthy and diseased models. The same representative QoI is used for both healthy and diseased models.}
\label{tab:MethodsBreakdown2}
\end{table}

As the LF models are much cheaper to evaluate than the HF models (\autoref{tab:ModelCosts}), this allocation keeps the cost of the MLMF methods below the MC or MLMC methods, even though more than twice the additional simulations are needed for the MLMF methods than for the MC or MLMC methods. Both the MLMC and MLMF methods allocate the majority of their simulations to the least expensive model. For the MLMC method, the majority of additional simulations are HF coarse. For the MLMF method, the majority are LF coarse. However, in the MLMF cases the LF coarse simulations do not comprise to the majority of the total cost of the extrapolation.

\subsubsection{Normalized confidence intervals for global and local quantities of interest}

When relying on estimator values and their variances from only the pilot run, MLMF estimators outperform the MC and MLMC methods. In this context, outperforming means that the estimator variance is lower from the MLMF schemes than from the other methods. In fact, with the size of this pilot, we would consider the MLMF estimators to be converged after the pilot run. The size of the pilot run can be shaped such that the variance is larger and a specific variance or nCI can be targeted with extrapolation following the pilot phase. A computational budget constraint on, for instance, the number of most expensive model evaluations can also be enforced.

\begin{figure}[!ht]
\centering
\hspace{\baselineskip}
\columnname{Aorto-Femoral Healthy}\\
\begin{subfigure}[t]{0.24\textwidth}
	\centering
	\includegraphics[width=\textwidth]{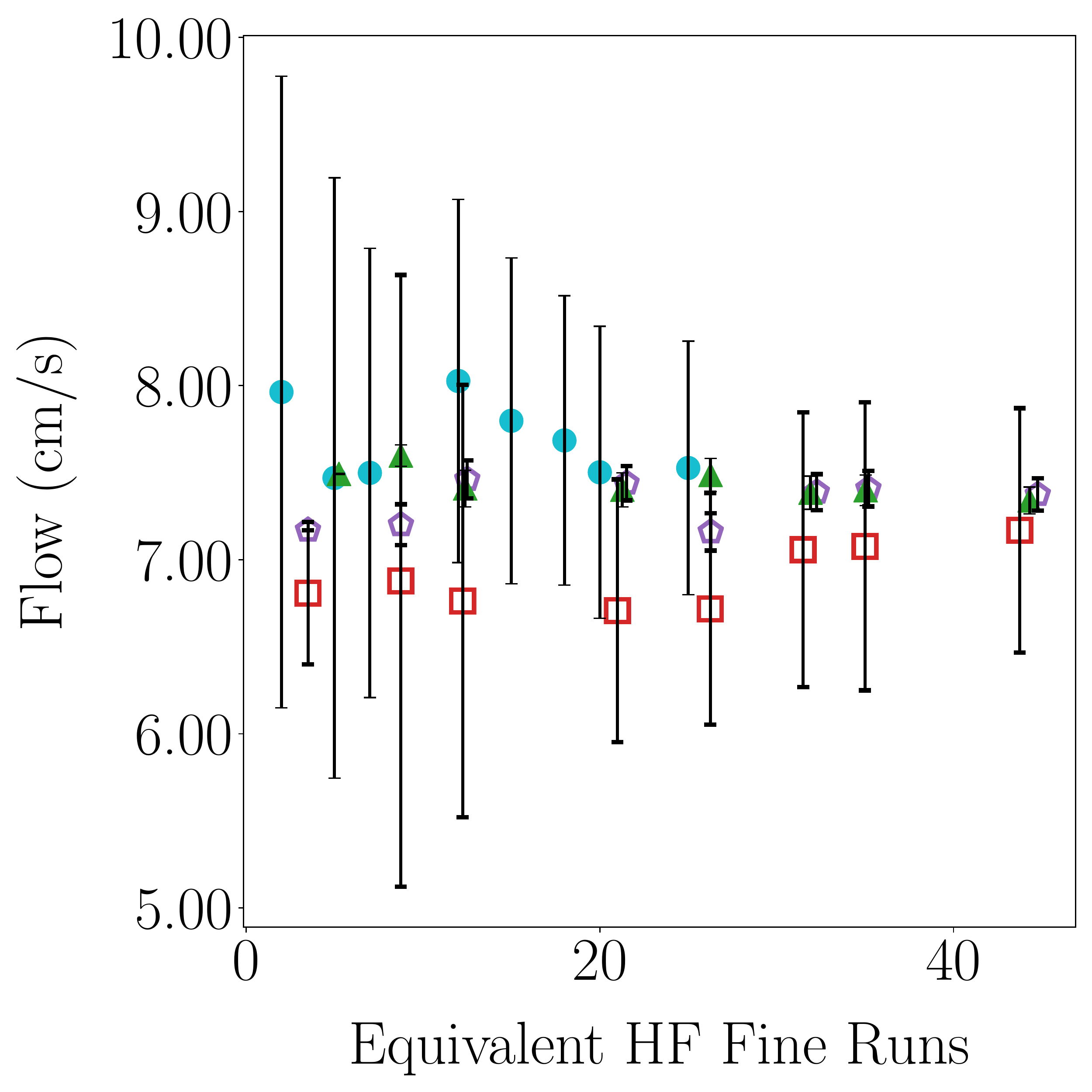}
	\caption{}
	\label{fig:methodsflow-A}
\end{subfigure}
\begin{subfigure}[t]{0.24\textwidth}
	\centering
	\includegraphics[width=\textwidth]{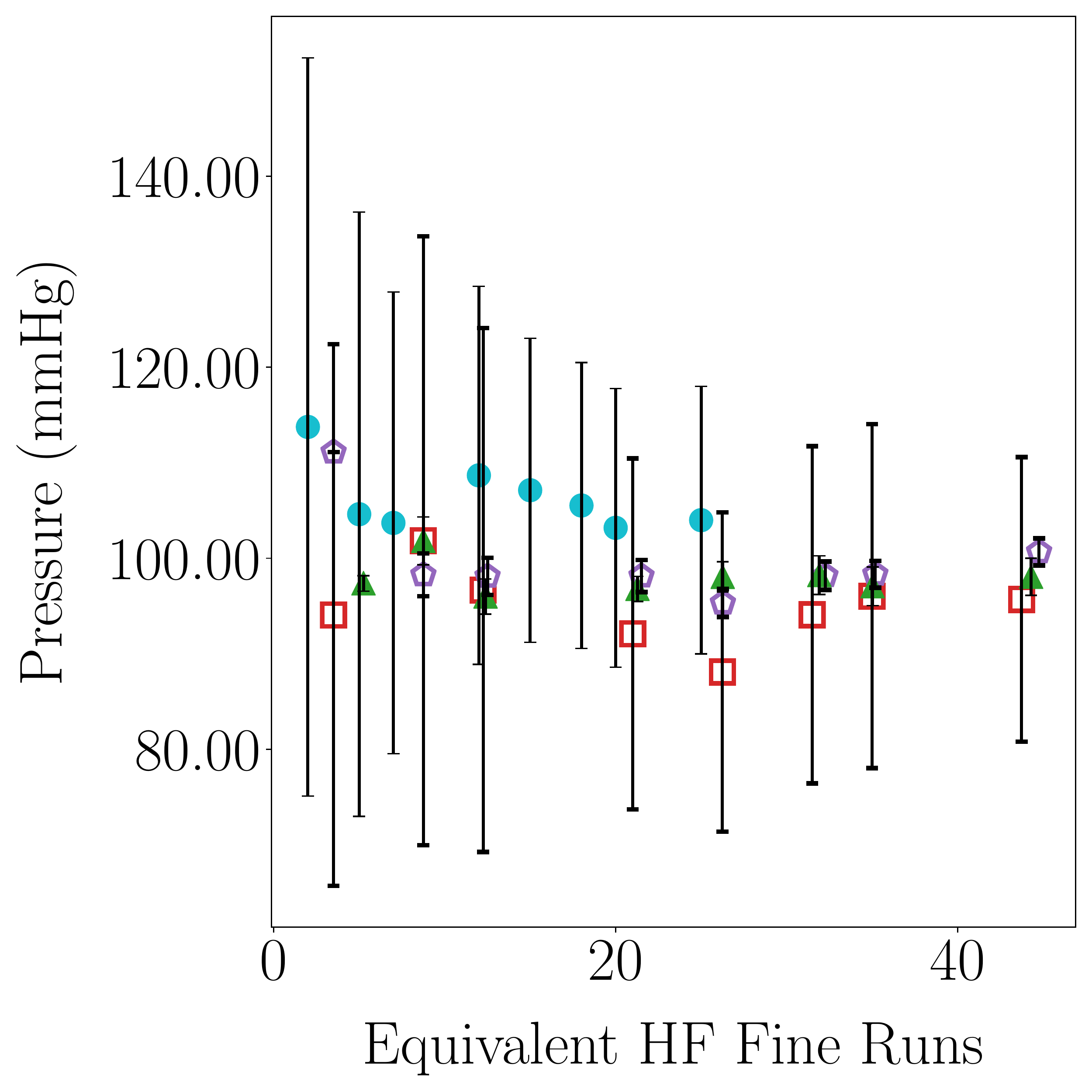}
	\caption{}
	\label{fig:methodspres-A} 
\end{subfigure} 
\begin{subfigure}[t]{0.24\textwidth}
	\centering
	\includegraphics[width=\textwidth]{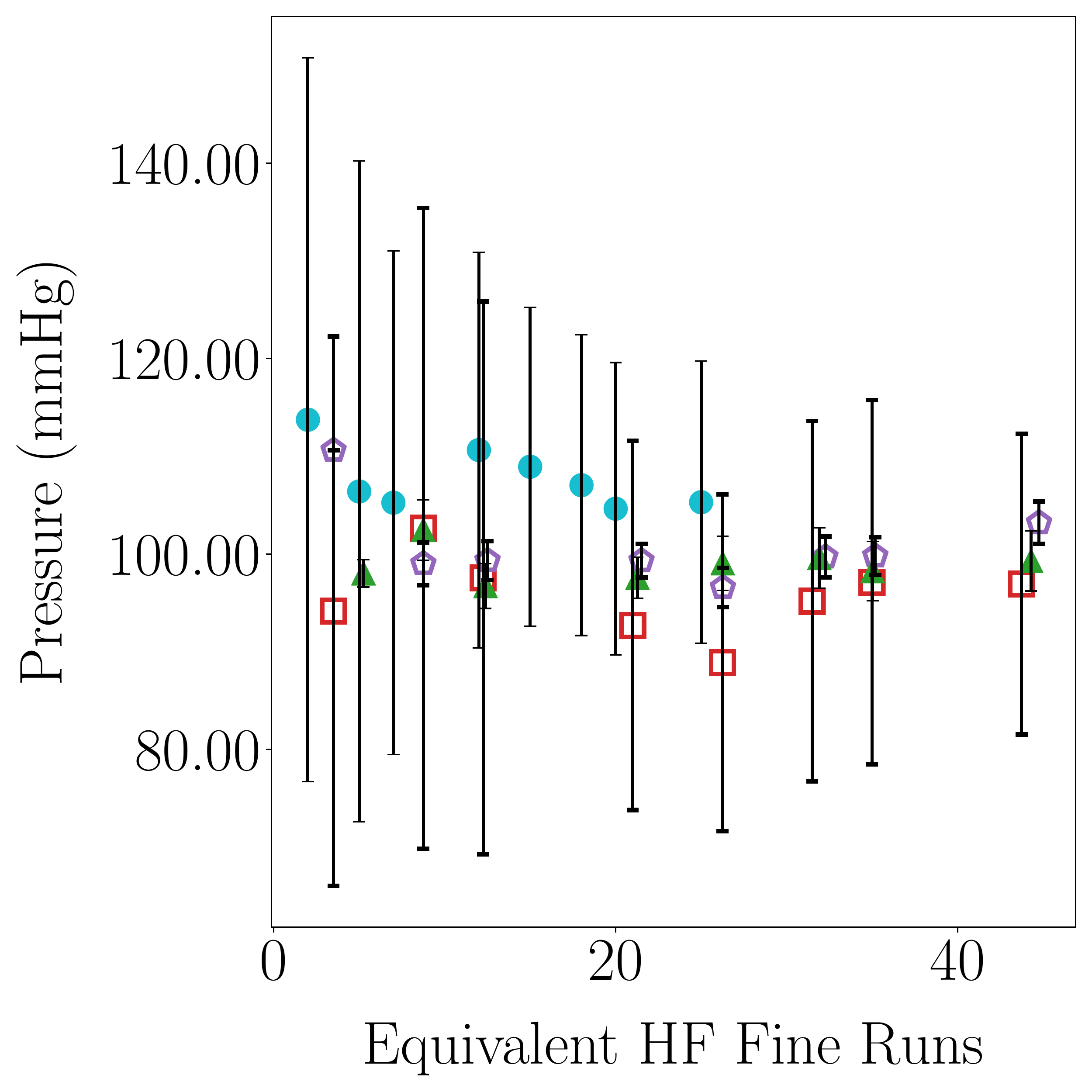}
	\caption{}
	\label{fig:methodsTAP-A} 
\end{subfigure} 
\begin{subfigure}[t]{0.24\textwidth}
	\centering
	\includegraphics[width=\textwidth]{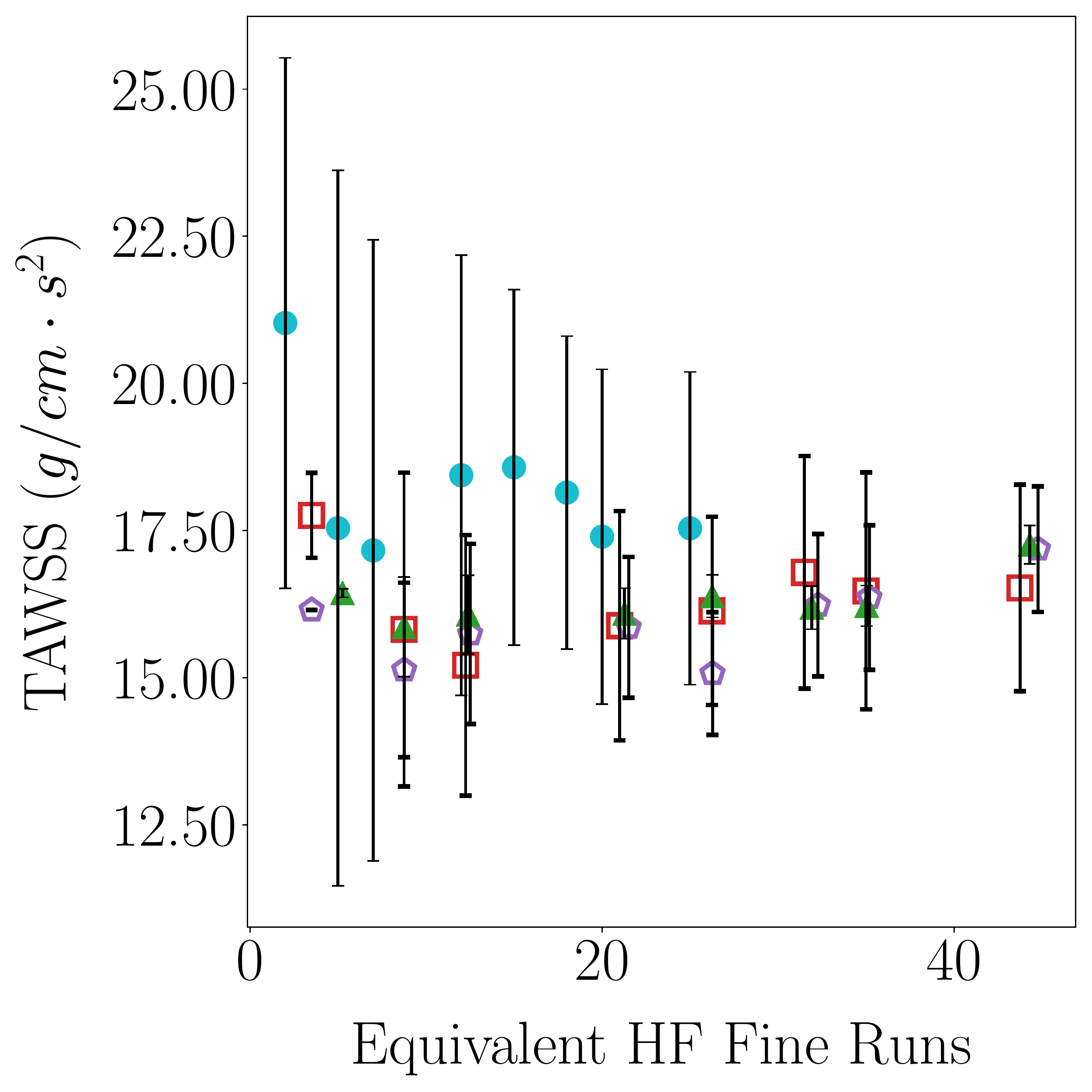}
	\caption{}
	\label{fig:methodstawss-A}
\end{subfigure}
\\
\vspace{1em}
\hspace{\baselineskip}
\columnname{Coronary Healthy}\\
\begin{subfigure}[t]{0.24\textwidth} 
	\centering
	\includegraphics[width=\textwidth]{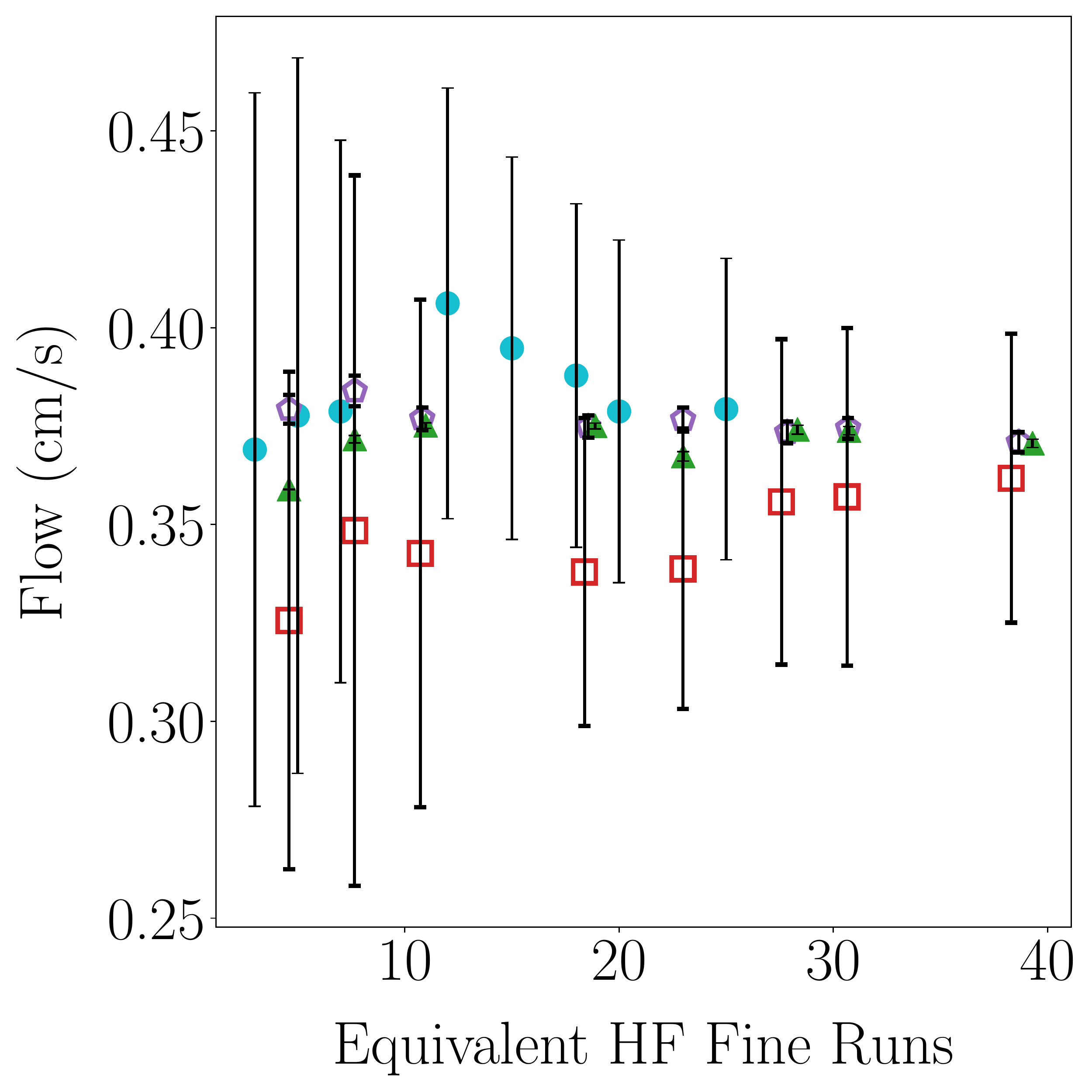}
	\caption{}
	\label{fig:methodsflow-C}  
\end{subfigure}	 
\begin{subfigure}[t]{0.24\textwidth} 
	\centering
	\includegraphics[width=\textwidth]{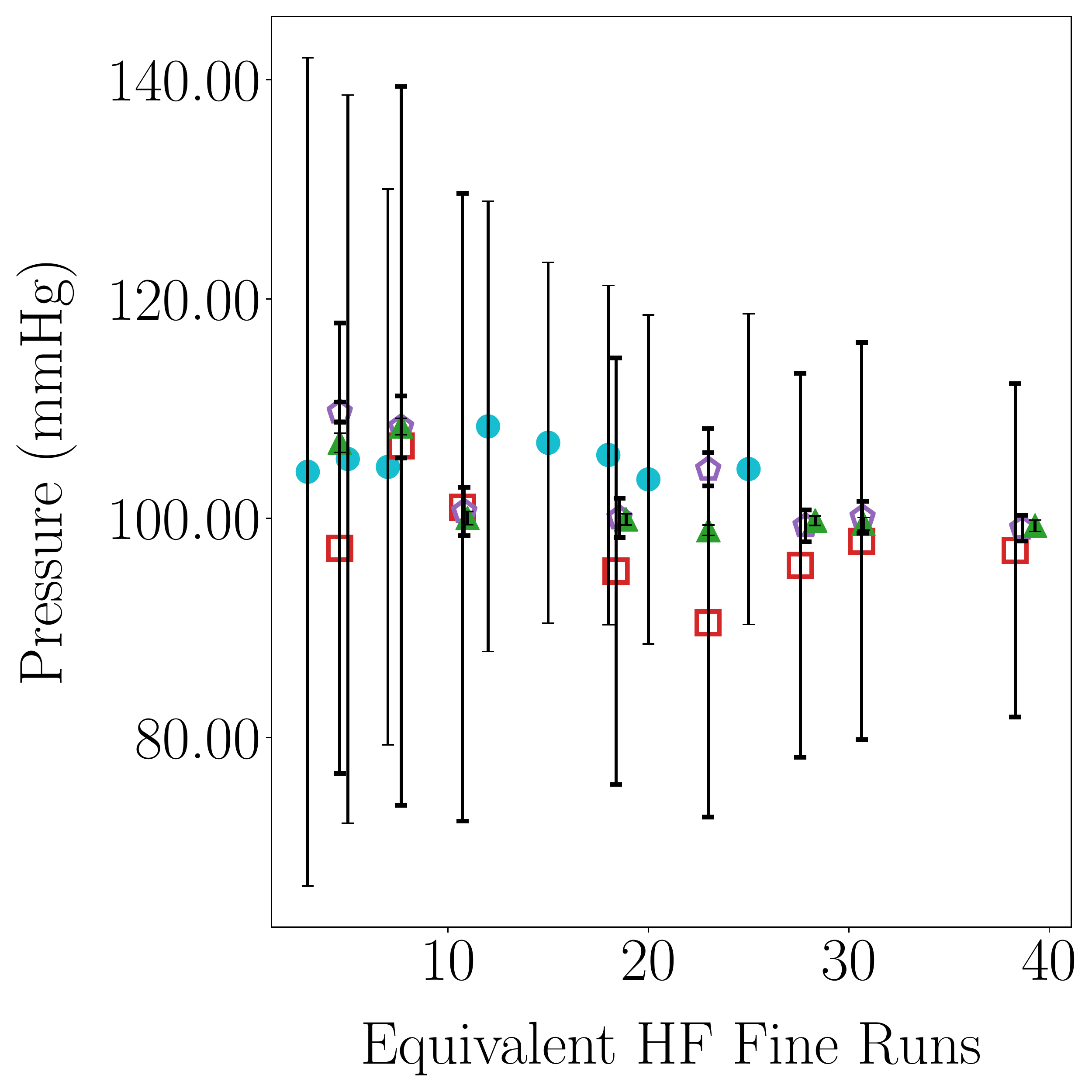}
	\caption{}
	\label{fig:methodspres-C}  
\end{subfigure}	 
\begin{subfigure}[t]{0.24\textwidth} 
	\centering
	\includegraphics[width=\textwidth]{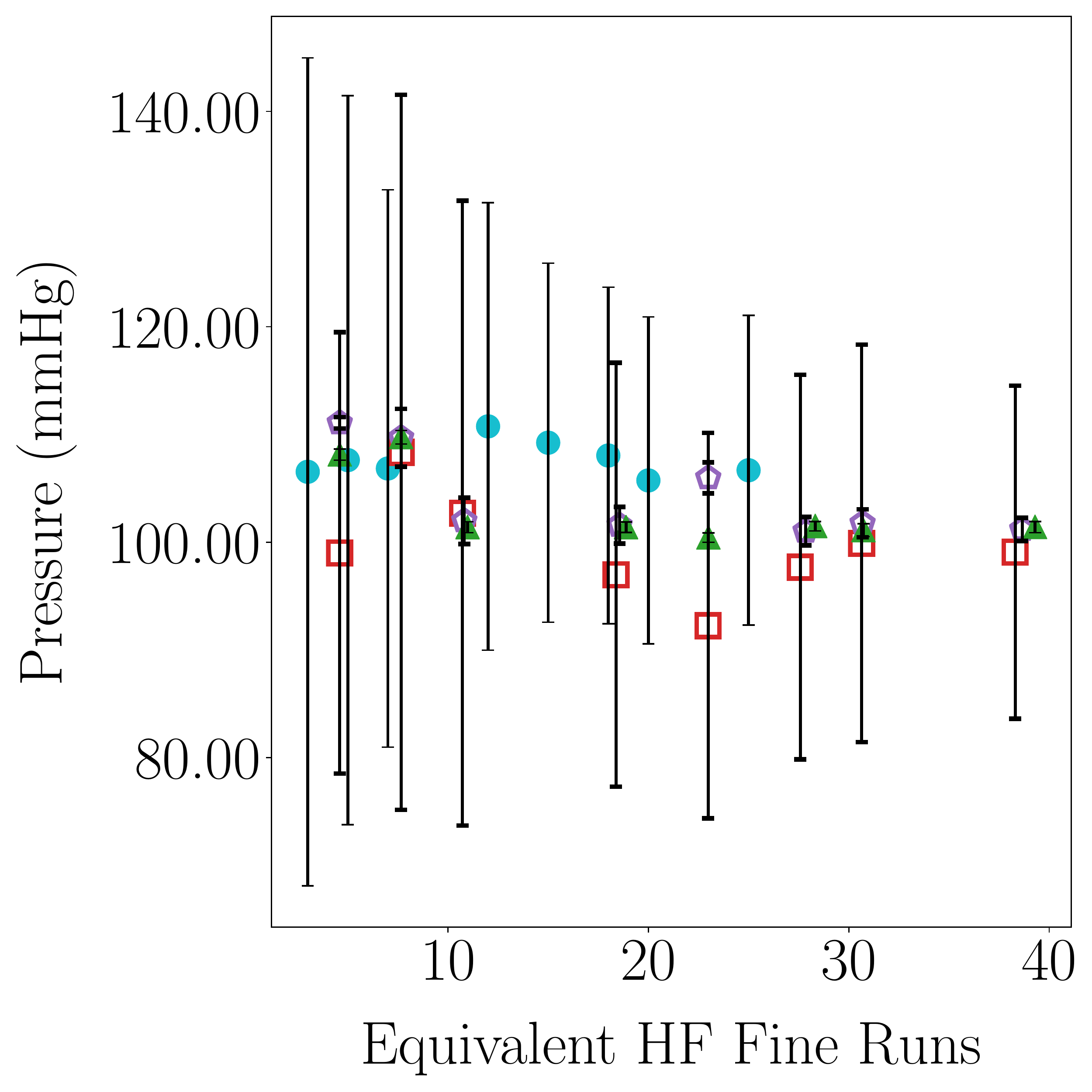}
	\caption{}
	\label{fig:methodsTAP-C}  
\end{subfigure}	
\begin{subfigure}[t]{0.24\textwidth} 
	\centering
	\includegraphics[width=\textwidth]{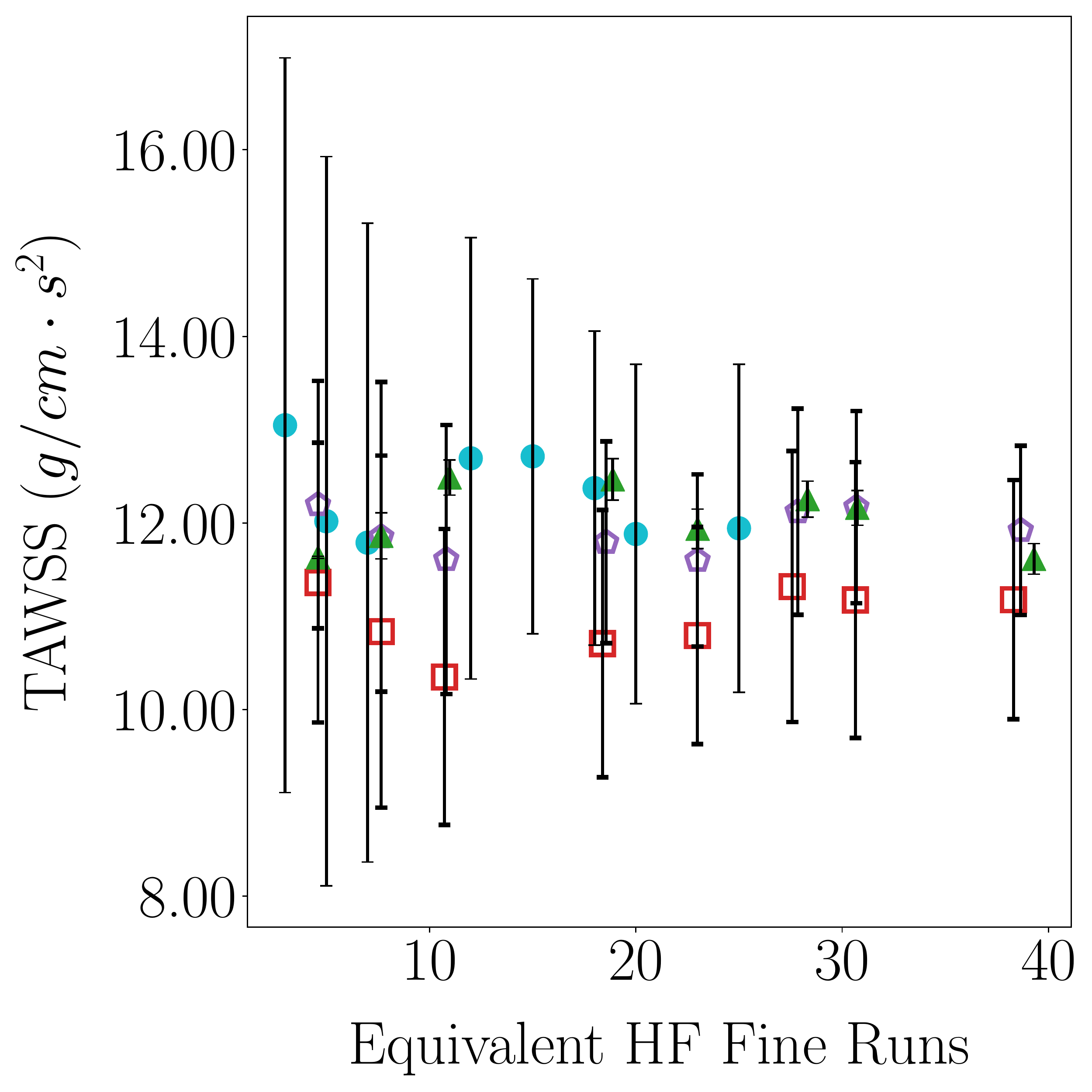}
	\caption{}
	\label{fig:methodstawss-C}  
\end{subfigure}
\begin{subfigure}[t]{0.7\textwidth} 
	\centering
	\includegraphics[width=\textwidth]{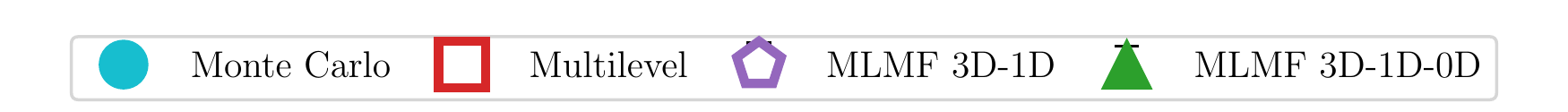}
	\label{fig:MethodsLeg}  
\end{subfigure}		 
\caption{Comparing the performance of UQ estimators from four methods for global and local QoIs. The four methods are Monte Carlo using 3D models, multilevel using 3D models, MLMF using 3D and 1D models, and MLMF using 3D, 1D, and 0D models. Estimators with their confidence intervals are shown converging to the values in the pilot run for representative outlet flow QoIs in figures (a) and (e), outlet pressure QoIs in figures (b) and (f), model time-averaged pressure QoIs in figures (c) and (g), and model TAWSS QoIs in figures (d) and (h) for the aorto-femoral and coronary models, respectively. These quantities are spatially and temporally averaged over the outlet face for the flow and pressure QoIs, throughout regions of the model volume for the model pressure QoIs, or over regions of the model surface for model TAWSS QoIs respectively.}
\label{fig:MethodsCompare}
\end{figure}

We show the convergence of four quantities of interest to the final estimated mean and variance obtained from the pilot run (\autoref{fig:MethodsCompare}). As we build up the full pilot sample, we can see the differences in variance between the methods arise. The MC, MLMC, and both MLMF estimators are shown at increasing subsamples of the full set of pilot run simulations. These estimators are plotted against the equivalent cost of the samples making up the estimators. The final, rightmost points in each plot are the estimators obtained from the pilot study. 

As the sample comprising the estimator grows, each method becomes more accurate. We can see that the MLMF methods obtain very accurate, i.e. small variance and nCI, estimators at a much lower cost than the MLMC or MC estimators.

Only the healthy model results are presented here, as the diseased models exhibit the same trends. Similarly, specific quantities of interest from each category were used. The conclusions drawn from these quantities are consistent with other quantities of interest of the same type (outlet flows and pressures, time-averaged model pressure, or TAWSS values).

\subsection{Extrapolated cost validation} \label{sec:ResExtrap}

We introduced the process of \emph{extrapolation} to estimate the additional number of simulations of each model and fidelity needed to decrease the estimator variance by a factor $\epsilon$. The formulas for the optimal number of samples needed for extrapolation are given by $\eqref{equ:MLOptSample}$, $\eqref{equ:MFOptSample}$, and $\eqref{equ:MLMFOptSample}$ for the MLMC, MFMC, and MLMF estimators, respectively. These formulas have been used successfully for applications such as engine nozzles, scramjets, and wind turbines~\cite{Geraci_A_2017,Maniaci_2018}. To validate these for cardiovascular modeling, we needed to ensure that the cost of converging to various target estimator variances is reflective of the projected extrapolated costs.

\begin{figure}[!ht]
\centering
\begin{subfigure}[t]{0.48\textwidth}
	\centering
	\includegraphics[width=\textwidth]{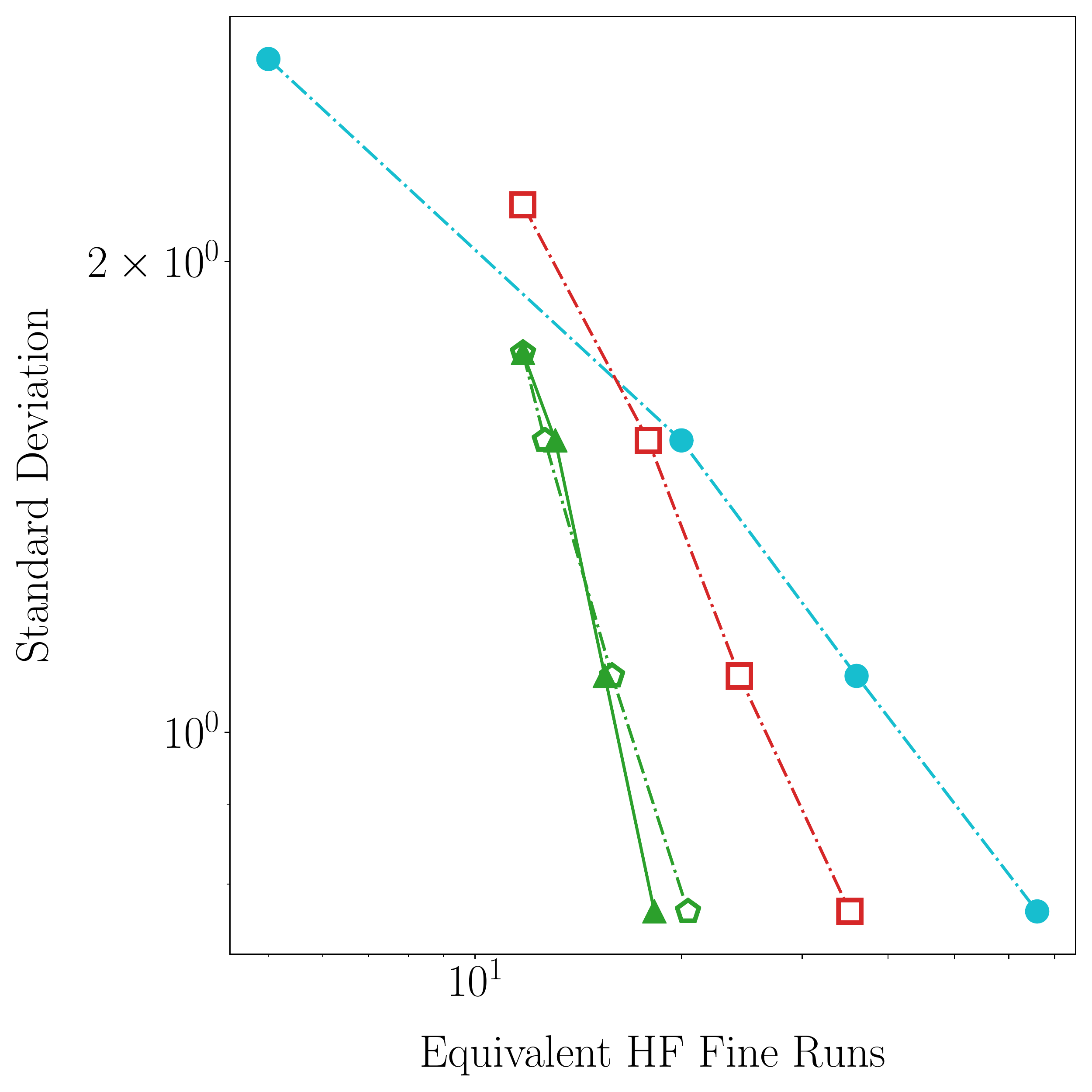}
	\caption{}
	\label{fig:FlowAcc-A2}
\end{subfigure}
\begin{subfigure}[t]{0.48\textwidth}
	\centering
	\includegraphics[width=\textwidth]{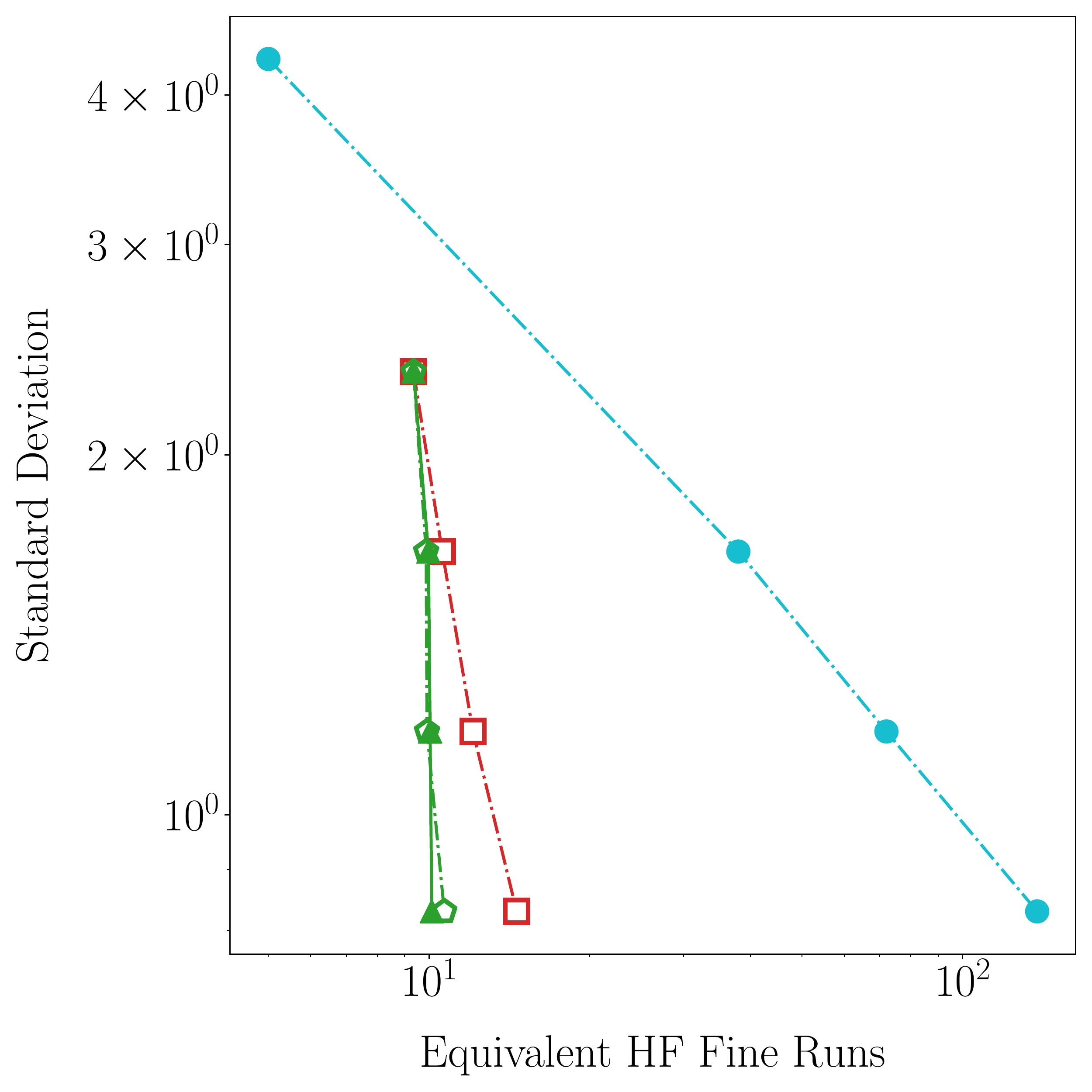}
	\caption{}
	\label{fig:FlowAcc-C2}
\end{subfigure} 
\\
\begin{subfigure}[t]{0.98\textwidth} 
	\centering
	\includegraphics[width=\textwidth]{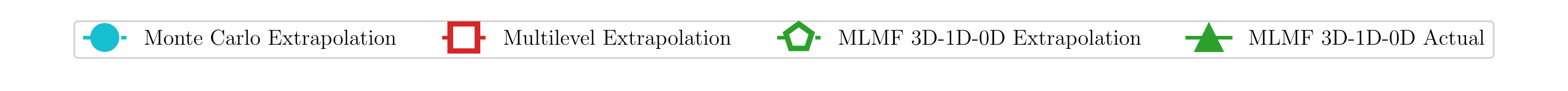}
	\label{fig:AccLegend2}  
\end{subfigure}	
\vspace{-2em}
\caption{Validating the extrapolation of estimators for the MLMF UQ schemes for (a) aorto-femoral and (b) coronary healthy models. The values shown in green pentagons and triangles, respectively, are the projected and actual extrapolated costs (in units of the equivalent number of 3D fine simulations) to obtain various estimator standard deviations for representative QoIs. The extrapolated costs for the MC and MLMC estimators are also shown for comparison. Representative time- and spatially-averaged (a) outlet and (b) model pressure QoIs are shown for the aorto-femoral and coronary model, respectively. (For interpretation  of  the  references  to  color  in  this  figure  legend,  the  reader  is  referred  to  the  web  version  of  this  article.)}
\label{fig:AccValidate}
\end{figure}

We utilized the following validation procedure. We first performed a pilot run, with 15 of the lowest discrepancy level $Y_\ell = Y_1$, 10 of the middle discrepancy level $Y_2$, and 5 of the highest discrepancy level $Y_3$. This is a smaller pilot than used elsewhere, producing larger estimator variances. With the QoI estimators and their variances computed from this data, we chose target estimator variances equal to an improvement factor of $\epsilon=[0.5, 0.25, 0.125]$, meaning
\begin{equation}
\mathbb{V}_{target}[\hat{Q}^{MLMF}] = \epsilon\hspace{0.15em}\mathbb{V}_{pilot}[\hat{Q}^{ML}],
\end{equation}
where $\hat{Q}^{ML}$ is the MLMC estimator defined in $\eqref{equ:ML}$ and $\hat{Q}^{MLMF}$ is the MLMF estimator defined in $\eqref{equ:MLMF}$. This equation is consistent with the convergence criterion within Dakota. After selecting the target variances, we then extrapolated to determine the optimal number of additional samples of each model level and fidelity. After then running the simulations to convergence in Dakota, the final costs were approximately those predicted by our extrapolation for the MLMF method (\autoref{fig:AccValidate}). 

\subsection{Comparing multilevel multifidelity estimators with and without zero-dimensional models} \label{sec:Res0D}

\begin{figure}[!ht]
\centering
\hspace{\baselineskip}
\columnname{Aorto-Femoral Healthy}\\
\begin{subfigure}[t]{0.32\textwidth}
	\centering
	\includegraphics[width=\textwidth]{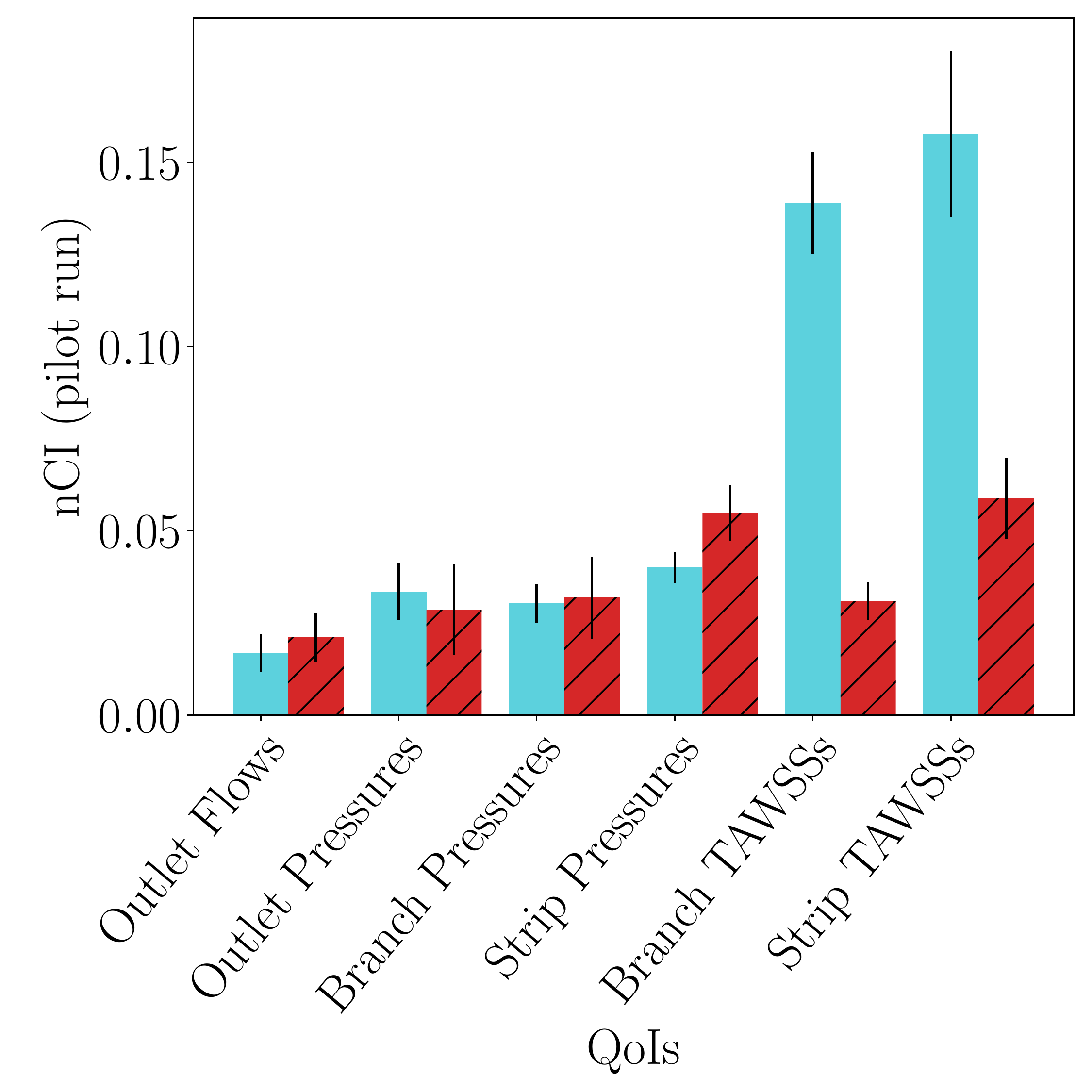}
	\caption{}
	\label{fig:3d1dacc-A}
\end{subfigure}
\begin{subfigure}[t]{0.32\textwidth}
	\centering
	\includegraphics[width=\textwidth]{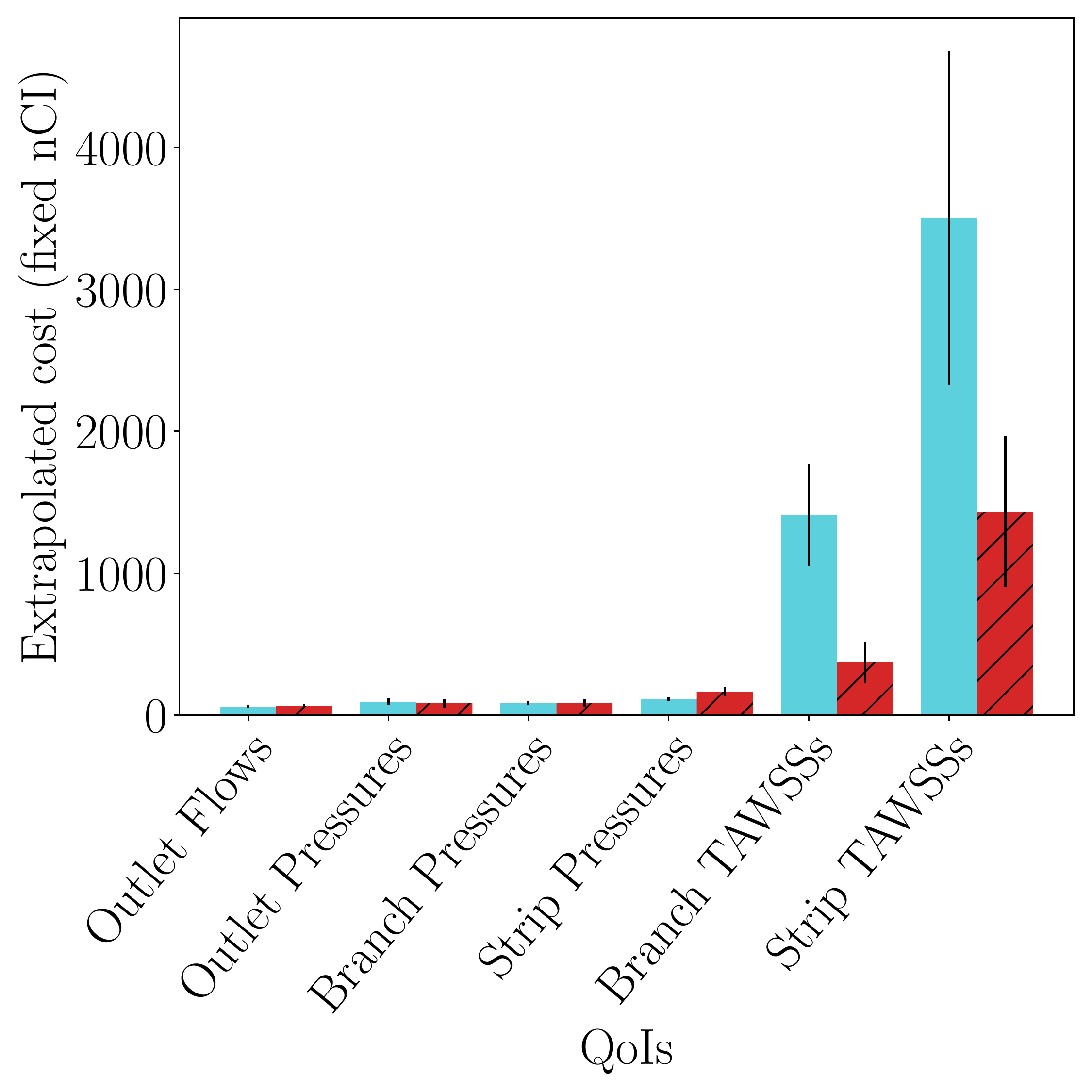}
	\caption{}
	\label{fig:3d1dcost-A} 
\end{subfigure} 
\begin{subfigure}[t]{0.32\textwidth}
	\centering
	\includegraphics[width=\textwidth]{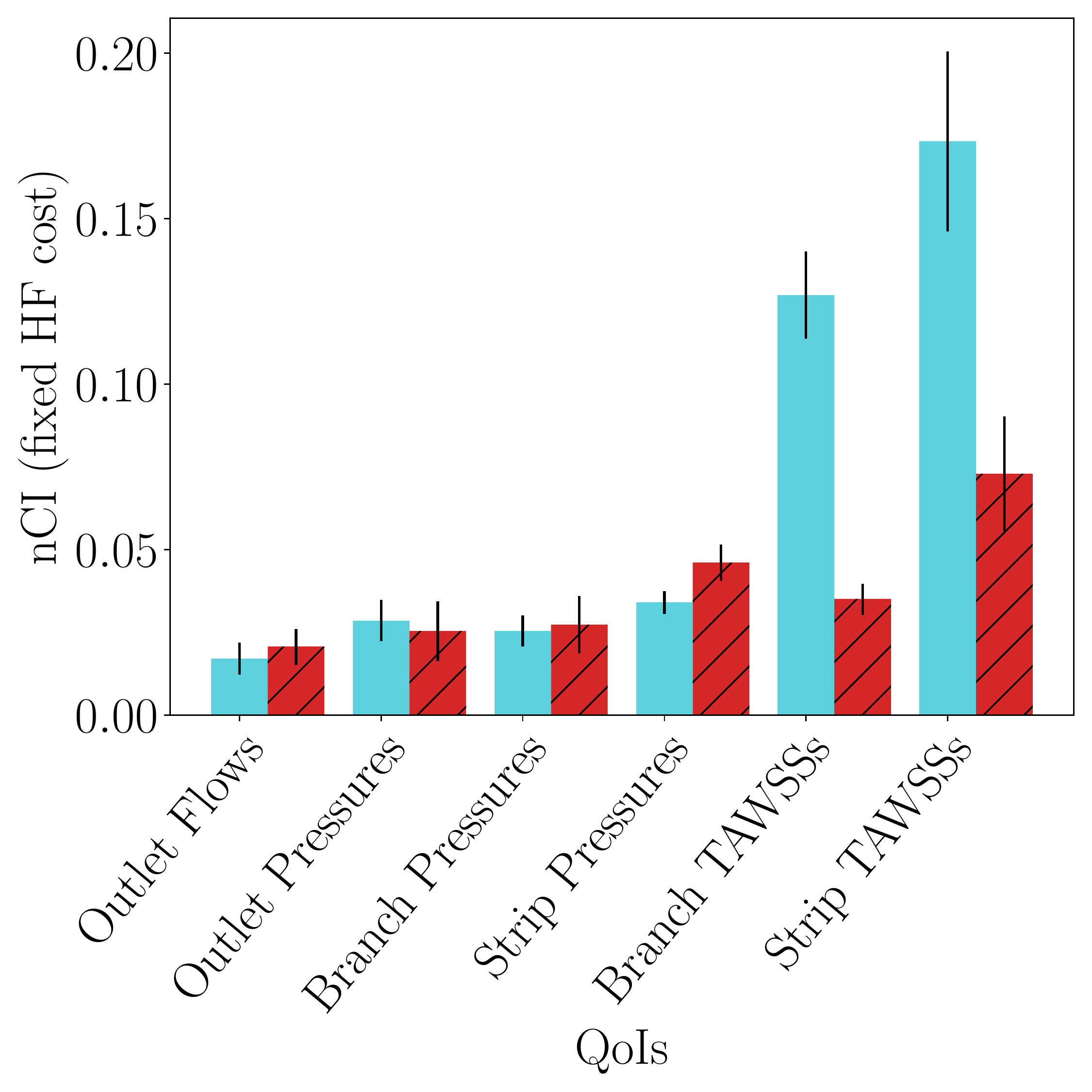}
	\caption{}
	\label{fig:3d1deps-A} 
\end{subfigure} 
\\
\vspace{1em}
\hspace{\baselineskip}
\columnname{Coronary Healthy}\\
\begin{subfigure}[t]{0.32\textwidth} 
	\centering
	\includegraphics[width=\textwidth]{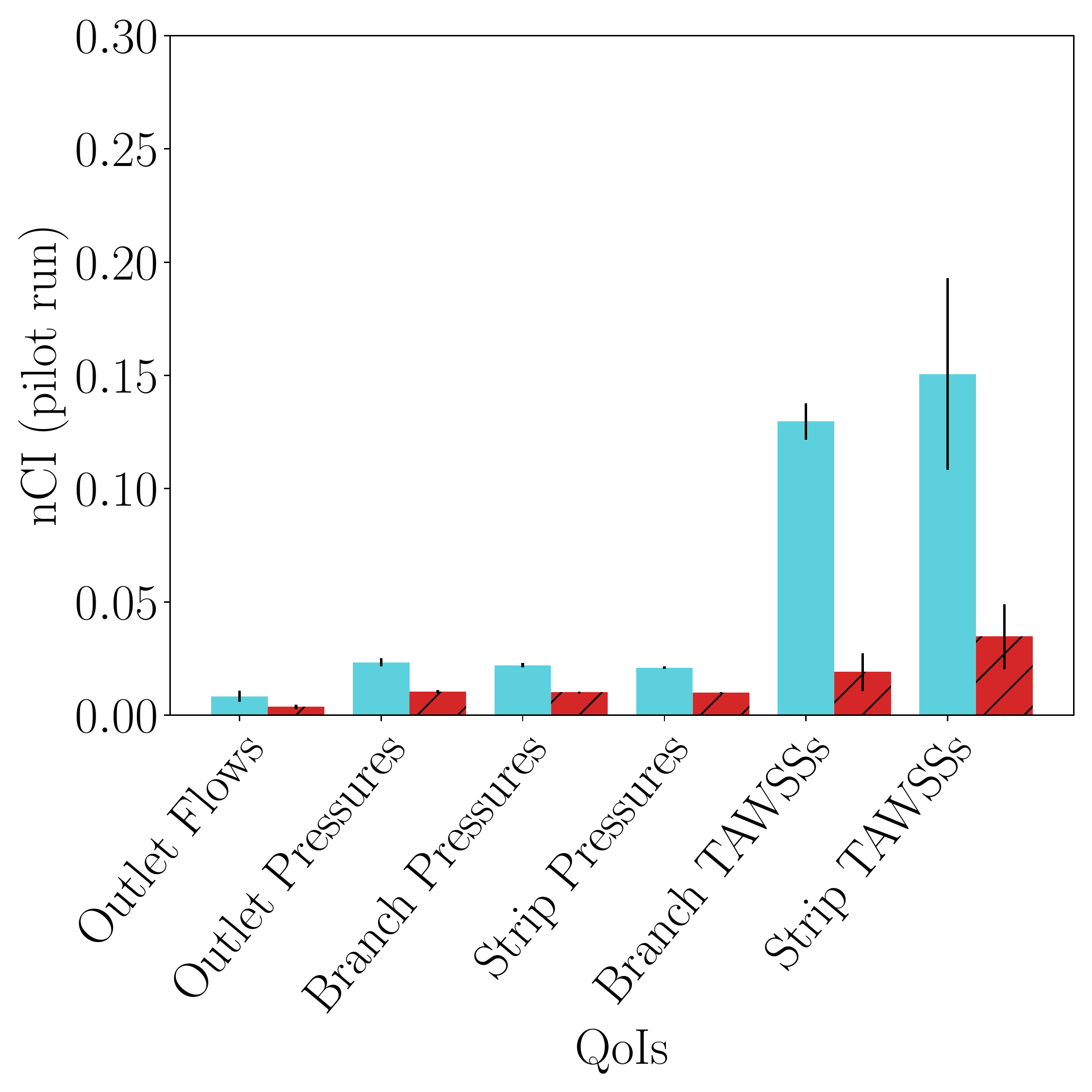}
	\caption{}
	\label{fig:3d1dacc-C}  
\end{subfigure}	 
\begin{subfigure}[t]{0.32\textwidth} 
	\centering
	\includegraphics[width=\textwidth]{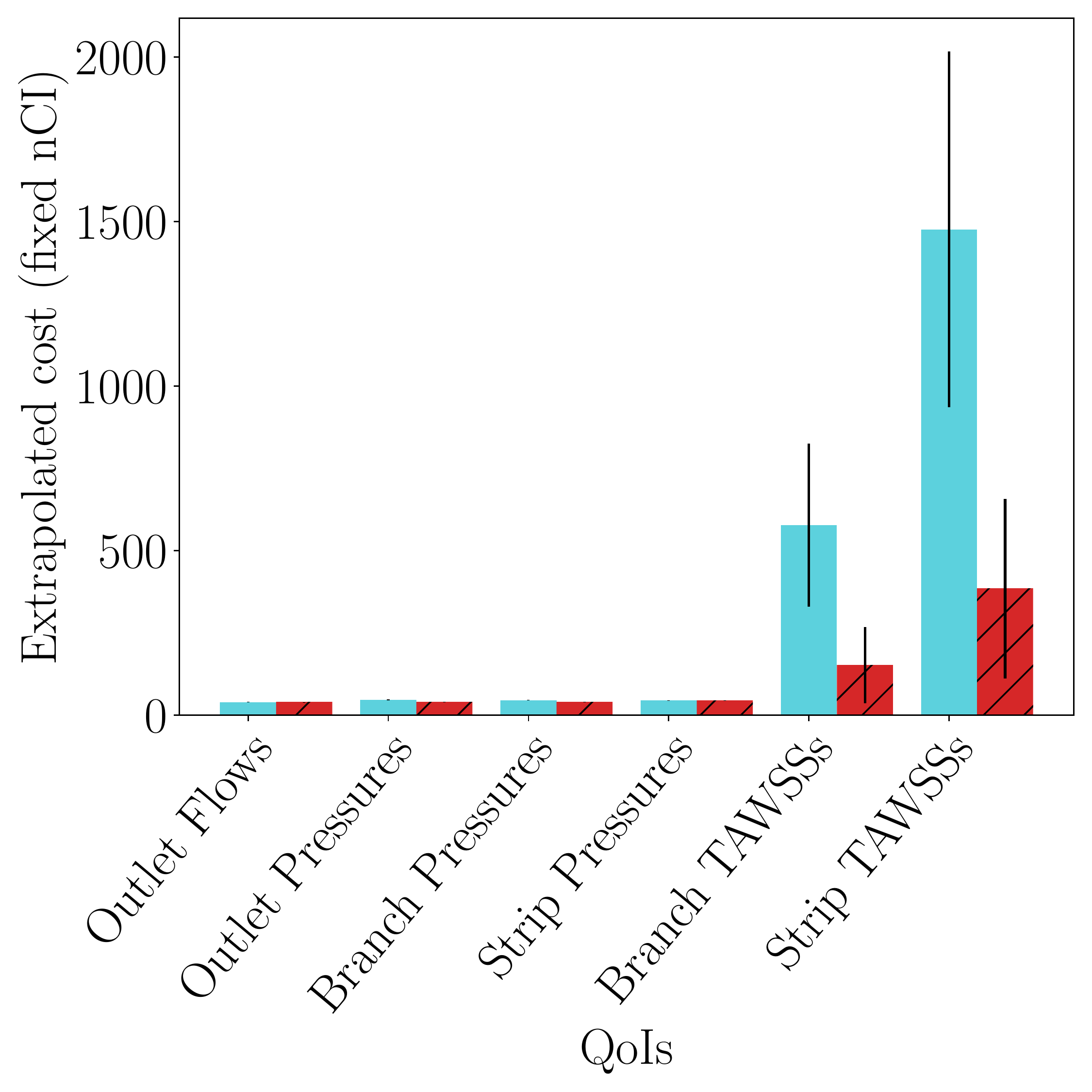}
	\caption{}
	\label{fig:3d1dcost-C}  
\end{subfigure}	 
\begin{subfigure}[t]{0.32\textwidth} 
	\centering
	\includegraphics[width=\textwidth]{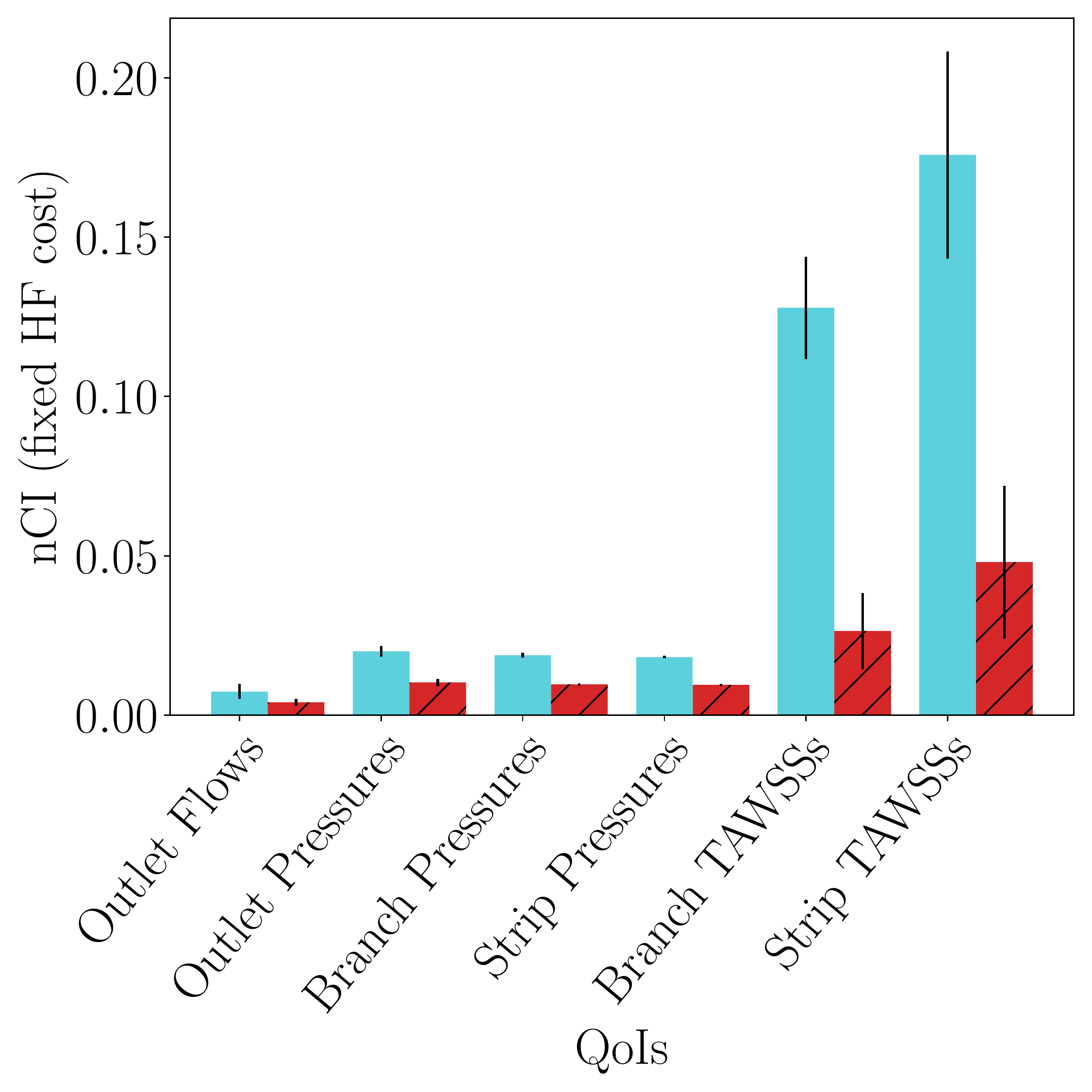}
	\caption{}
	\label{fig:3d1deps-C}  
\end{subfigure}	
\\
\begin{subfigure}[t]{0.7\textwidth} 
	\centering
	\includegraphics[width=\textwidth]{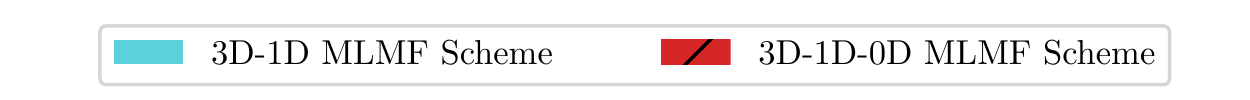}
	\label{fig:3d1dlegend2}  
\end{subfigure}	
\vspace{-1.5em}
\caption{Comparing the performance of MLMF estimators from 3D and 1D models to the MLMF estimators from 3D, 1D, and 0D models for the (top) aorto-femoral and (bottom) coronary healthy models. Averaged nCIs for various groups of QoIs calculated after a fixed pilot run of 250 simulations are shown in plots (a) and (d). Averaged projected extrapolated cost (in units of the equivalent number of 3D fine simulations) to obtain a nCI $\leq 0.01$ are shown in plots (b) and (e). Averaged projected nCI values with the computational budget for all HF simulations equivalent to the cost of 50 3D fine simulations are shown for various QoI categories in plots (c) and (f). Representative results for the healthy anatomies are shown here.}
\label{fig:3D1D0DCompare}
\end{figure}

We now examine the differences between estimators from the MLMF 3D-1D and 3D-1D-0D schemes for the healthy aorto-femoral and coronary models (\autoref{fig:3D1D0DCompare}). These results are presented for the healthy models only for the sake of clarity, and similar trends hold for the diseased cases.

\subsubsection{Normalized confidence intervals after the pilot run}

We assessed the initial nCIs of our estimators from the pilot run samples and estimated how many additional simulations are needed to converge the solution to a desired estimator variance or nCI value for the QoIs. The Dakota workflow continues to automatically manage the UQ study until convergence is achieved.

Calculating the initial nCI from the pilot run, we found that while the global QoIs have similar nCIs for the two schemes, much better nCIs are achieved for local quantities of interest with the 3D-1D-0D scheme than with the 3D-1D scheme for both aorto-femoral and coronary models (\hyperref[fig:3d1dacc-A]{Figures~\ref*{fig:3d1dacc-A}} and~\ref{fig:3d1dacc-C}, respectively). We demonstrate this by computing the average nCI across several global and local QoI categories for both MLMF schemes, including the variability within each group. Local quantities of interest are more difficult to resolve than global quantities, and we find the largest variability occurs within the TAWSS categories. Between the aorto-femoral and coronary model geometries, we see similar nCI values in addition to similar trends between categories.

\subsubsection{Extrapolation to target normalized confidence intervals}

Extrapolating from the pilot run nCIs to obtain $nCI[Q] = 0.01$ for all QoIs, we see that the extrapolated cost is lower for the 3D-1D-0D scheme than for the 3D-1D scheme across most QoI categories for both the aorto-femoral and coronary models (\hyperref[fig:3d1dcost-A]{Figures~\ref*{fig:3d1dcost-A}} and~\ref{fig:3d1dcost-C}, respectively). For the other categories, the costs are equivalent within the margin of variability in each category. As before, we see the most significant difference between the two MLMF methods for the local QoIs. This trend towards lower cost of the 3D-1D-0D scheme follows directly from the fact that this scheme has better nCIs after the pilot run (\hyperref[fig:3d1dacc-A]{Figures~\ref*{fig:3d1dacc-A}} and~\ref{fig:3d1dacc-C}). The extrapolation calculation for the estimated number of additional samples required relies on the nCIs obtained from the pilot run. Therefore, since the 3D-1D-0D scheme had smaller nCI values to begin with, it follows that the 3D-1D-0D scheme also requires a smaller additional cost to decrease that nCI to below 0.01. The extrapolation is also reliant on correlations and the cost ratio between high and low fidelity models. Both the one- and zero-dimensional models have similarly high correlations with the three-dimensional model as well as favorable cost ratios, though the 3D-1D-0D scheme has better cost ratios for the coarse and medium low fidelity models. 

Looking at the global and local QoI categories, we see that it is more expensive, on the order of 10 to 100 times higher, to obtain the target nCI for the local QoIs than the global QoIs. This is also due to the fact that the local quantities generally had the largest (worst) nCIs after the pilot run. We also see the largest intra-category variation in the cost for local TAWSS quantities. As local quantities likely rely more heavily on high fidelity simulations to resolve their estimators, this cost variation is expected, as the high fidelity simulations have larger cost range amongst the levels than the low fidelity simulations.

\subsubsection{Extrapolated estimator normalized confidence intervals with fixed budget}

The results presented in this section thus far may seem to suggest that the 3D-1D-0D scheme is almost always preferable to the 3D-1D scheme. However, it is likely that many UQ studies will be constrained by some computational budget. We needed to investigate if the 3D-1D-0D scheme continued to outperform the 3D-1D scheme in this realistic setting. Extrapolation to a specific nCI may prove computationally prohibitive for some model applications. Instead, with a fixed budget, it is possible to compute the projected nCI for estimators subject to a computational budget constraint. Here, we report the projected nCI values for a high fidelity computational budget equivalent to the cost of 50 3D fine mesh simulations, which we considered to be a realistic UQ scenario. This means that the combined cost of all simulations of all high fidelity modeling levels must have an equivalent to 50. Note that only the HF models are constrained here, as the simulations at all LF levels are obtained after optimization by the algorithm and there is no control over their number nor their total cost. With that said, many thousands of LF simulations can be run for the cost of one HF simulation. With these constrained budgets, for the aorto-femoral model we see the 3D-1D-0D strongly outperforming the 3D-1D scheme for local QoIs, and performing similarly for global QoIs (\autoref{fig:3d1deps-A}). For the coronary model, the 3D-1D-0D scheme again outperforms the 3D-1D scheme, with smaller nCI values achieved across all QoI categories (\autoref{fig:3d1deps-C}). Recall for nCIs, smaller values are preferable as this amounts to higher estimator certainty. When including the overall computational cost, adding in the LF cost, we expect that indeed the 3D-1D-0D should perform better across all QoIs and models. All things equal (i.e. for the same correlation) the optimal $r$ factor controlling the number of LF simulations for each level is equal between 1D and 0D, but the total cost will be far less for 0D simulations.

\subsection{Comparing healthy and diseased estimators} \label{sec:ResHD}

We now examine the difference in estimator performance for healthy and diseased models. We consider local QoIs, such as the spatially-averaged TAWSS value in specific regions on the model, and global QoIs (time-averaged flows and pressures at outlets, spatial- and time-averaged pressure in various regions of the model) separately. Here we use the estimator values and variances from the 3D-1D-0D MLMF scheme, as these are more accurate after the pilot run (see~\autoref{sec:Res0D}). The conclusions drawn from these specific quantities are reflective of the behavior of other QoIs from the same category.

\begin{figure}[!ht]	 
\includegraphics[width=\textwidth]{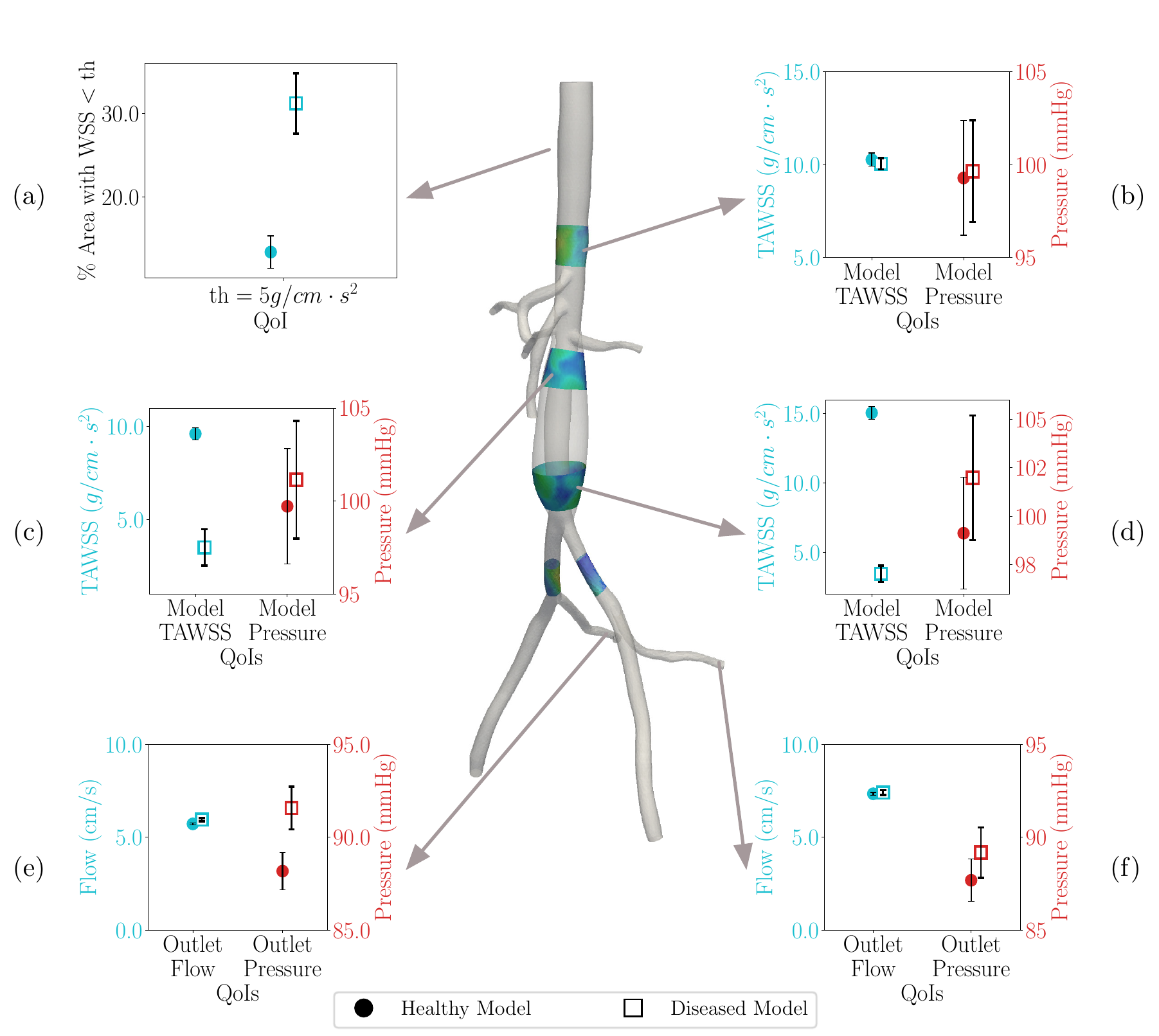}
\caption{Comparison of estimators for healthy and diseased model QoIs. Model is shown with strips of interest colored by TAWSS, with the diseased model including the AAA superimposed over the healthy model. Plot (a) shows the percentage of the model surface area with WSS below a threshold value of $5g/cm\cdot s^2$. Plots (b)-(d) show the spatially-averaged TAWSS and time-averaged pressure in (b) the strip in the descending thoracic aorta, (c) the strip superior to (above) the AAA in the abdominal aorta, and (d) the strip inferior to (below) the AAA in the abdominal aorta. Plots (e) and (f) show the spatial- and time-averaged flow and pressure QoIs at the (e) right internal iliac and (f) left internal iliac outlets. (For interpretation  of  the  references  to  color  in  this  figure  legend,  the  reader  is  referred  to  the  web  version  of  this  article.)}
\label{fig:AortaHD}
\end{figure}

\begin{figure}[!ht]	 
\includegraphics[width=\textwidth]{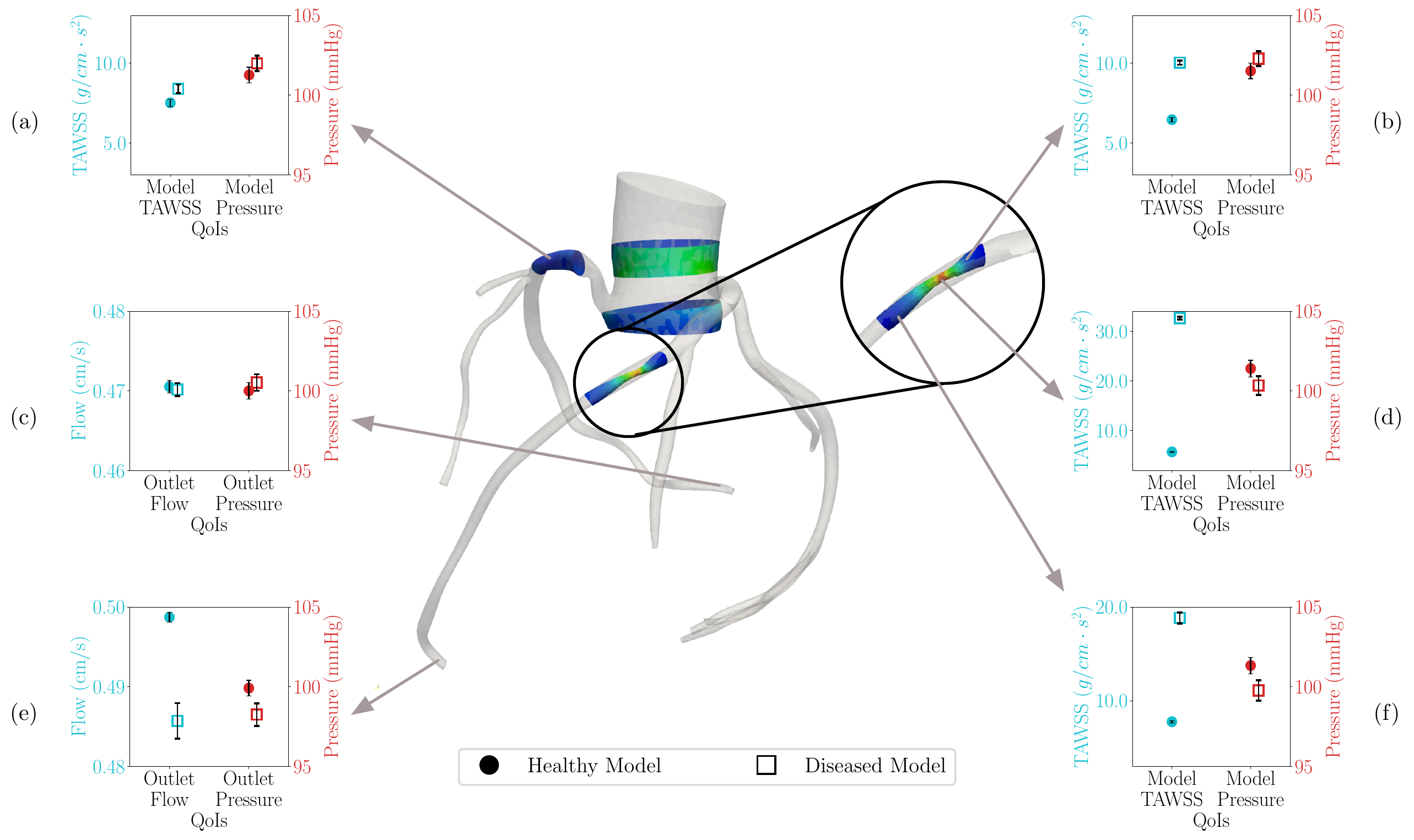}
\caption{Comparison of estimators for healthy and diseased model QoIs. Model is shown with strips of interest colored by TAWSS, with the healthy model superimposed over the diseased (stenosed) model. The stenosed region is shown enlarged in the inset image. Plots (a), (b), (d), and (f) show the spatially-averaged TAWSS and time-averaged pressure in (a) the strip superior to (above) the left and right coronary artery branches, (b) the strip preceding the stenosis of the left anterior descending (LAD) coronary artery (d) the strip within the peak stenosed region, and (f) the strip following the stenosis. Plots (c) and (e) show the spatial- and time-averaged flow and pressure QoIs at the (c) right coronary artery (RCA) and (e) stenosed LAD outlets. (For interpretation  of  the  references  to  color  in  this  figure  legend,  the  reader  is  referred  to  the  web  version  of  this  article.)}
\label{fig:CoronaryHD}
\end{figure}

In the aorto-femoral model, which has a larger diseased region, we find different mean values, with non-overlapping 99.7\% confidence regions, for the local QoIs near the diseased region in the healthy and diseased models (\hyperref[fig:AortaHD]{Figure~\ref*{fig:AortaHD}c and~\ref*{fig:AortaHD}d}). We also observe a large difference in the percentage of vessel wall surface area with low WSS, below a $5 g/cm\cdot s^2$ threshold (\hyperref[fig:AortaHD]{Figure~\ref*{fig:AortaHD}a}). Though it is a function of WSS, this QoI is more of a global quantity as it incorporates the WSS at all vessels of the model. However, in the coronary model, which has a more localized diseased region, the differences in QoI values are only noticeable near the diseased region (\autoref{fig:CoronaryHD}), and most distinct for the local QoIs. 
As global quantities should not be significantly affected by a locally diseased region, this follows our expectations. 
Within the model geometries, we do see results consistent with clinical expectations for each diseased case. For example, in the diseased AAA model we see lower wall shear stresses due to the aneurysm and in stenosed coronary model, near the stenosis, we see increased wall shear stress, both consistent with clinical expectations.

Using the data from the pilot run, each QoI value is shown along with its estimated 99.7\% (unnormalized) confidence interval (i.e. $\mu \pm 3\sigma$, where $\sigma$ is the standard deviation of the estimator value). Recall we use nCIs to facilitate comparisons between model geometries. In general, the nCIs of the diseased model QoIs are slightly larger (less accurate) than the healthy models. 
As diseased models require better local resolution, and therefore more three-dimensional simulations, to resolve quantities in the diseased regions, we expected larger nCI values to arise from the pilot run for the diseased model than for the same QoIs and pilot run size of a healthy model.

The specific QoIs shown were chosen to be representative of each of our four main QoI categories. The diseased AAA model is shown overlaid on the healthy aorto-femoral model (\autoref{fig:AortaHD}), while the healthy coronary model is overlaid on the diseased (stenosed) coronary model, with the stenosed region enlarged in the inset (\autoref{fig:CoronaryHD}). The highlighted strips of interest are colored by the local TAWSS of the diseased models. The QoI examining the regions of model surface area with low WSS (\hyperref[fig:AortaHD]{Figure~\ref*{fig:AortaHD}a}), is generally of interest in cardiovascular hemodynamics as such regions can be predictive of thrombosis~\cite{Zambrano2015}.

\begin{figure}[!ht]
\centering
\hspace{\baselineskip}
\columnname{Aorto-Femoral Models}\\
\begin{subfigure}[t]{0.32\textwidth}
	\centering
	\includegraphics[width=\textwidth]{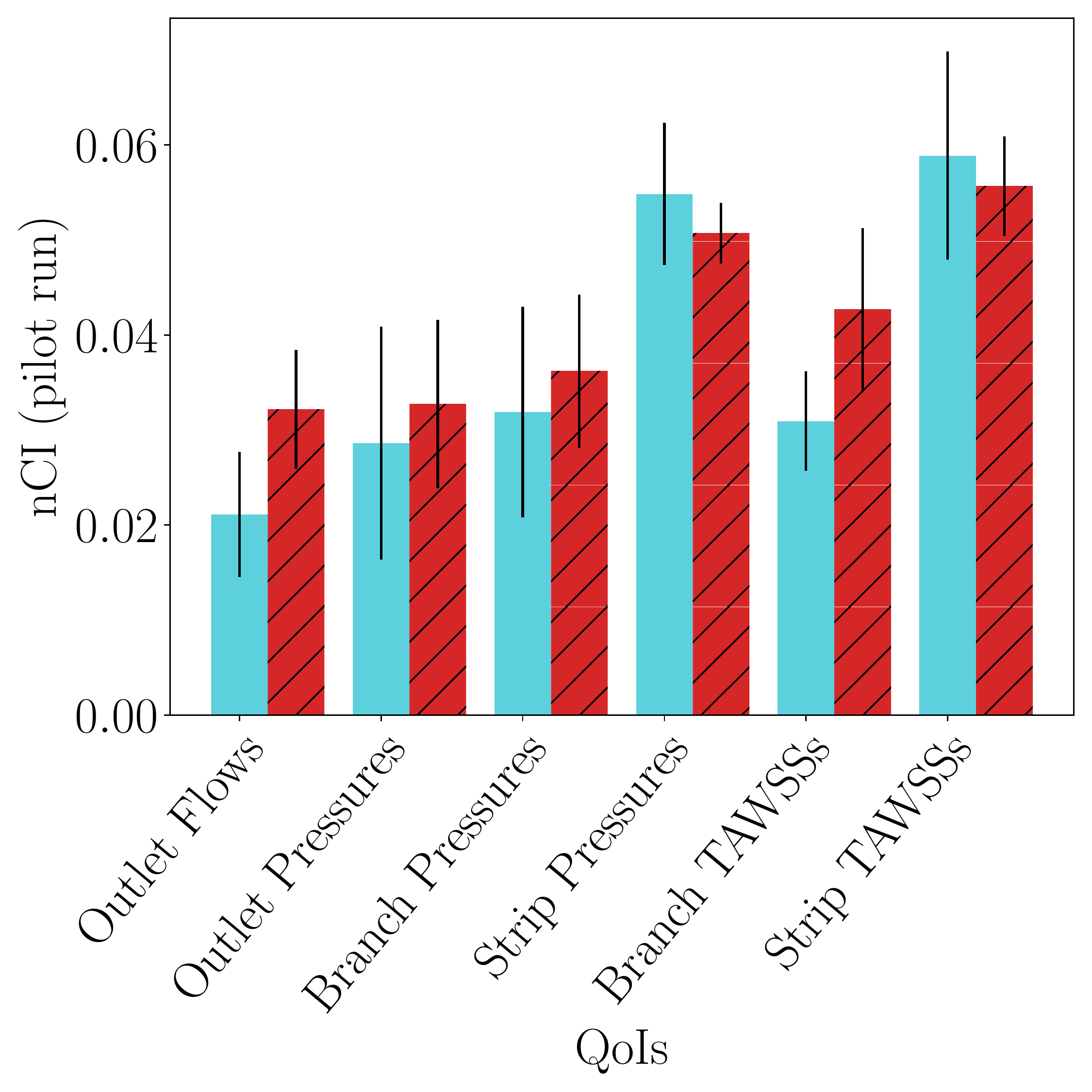}
	\caption{}
	\label{fig:3d1dacc-A2}
\end{subfigure}
\begin{subfigure}[t]{0.32\textwidth}
	\centering
	\includegraphics[width=\textwidth]{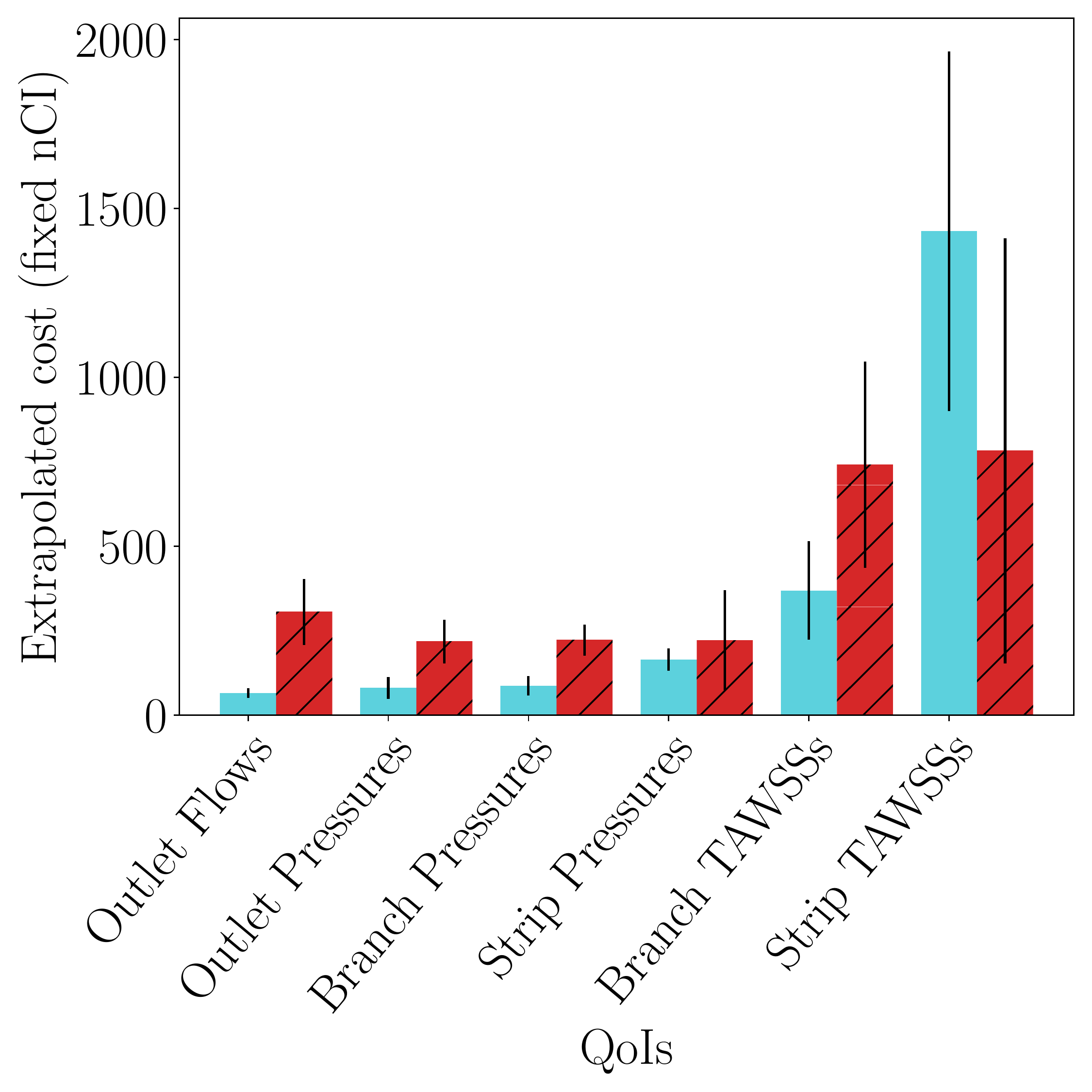}
	\caption{}
	\label{fig:3d1dcost-A2} 
\end{subfigure} 
\begin{subfigure}[t]{0.32\textwidth}
	\centering
	\includegraphics[width=\textwidth]{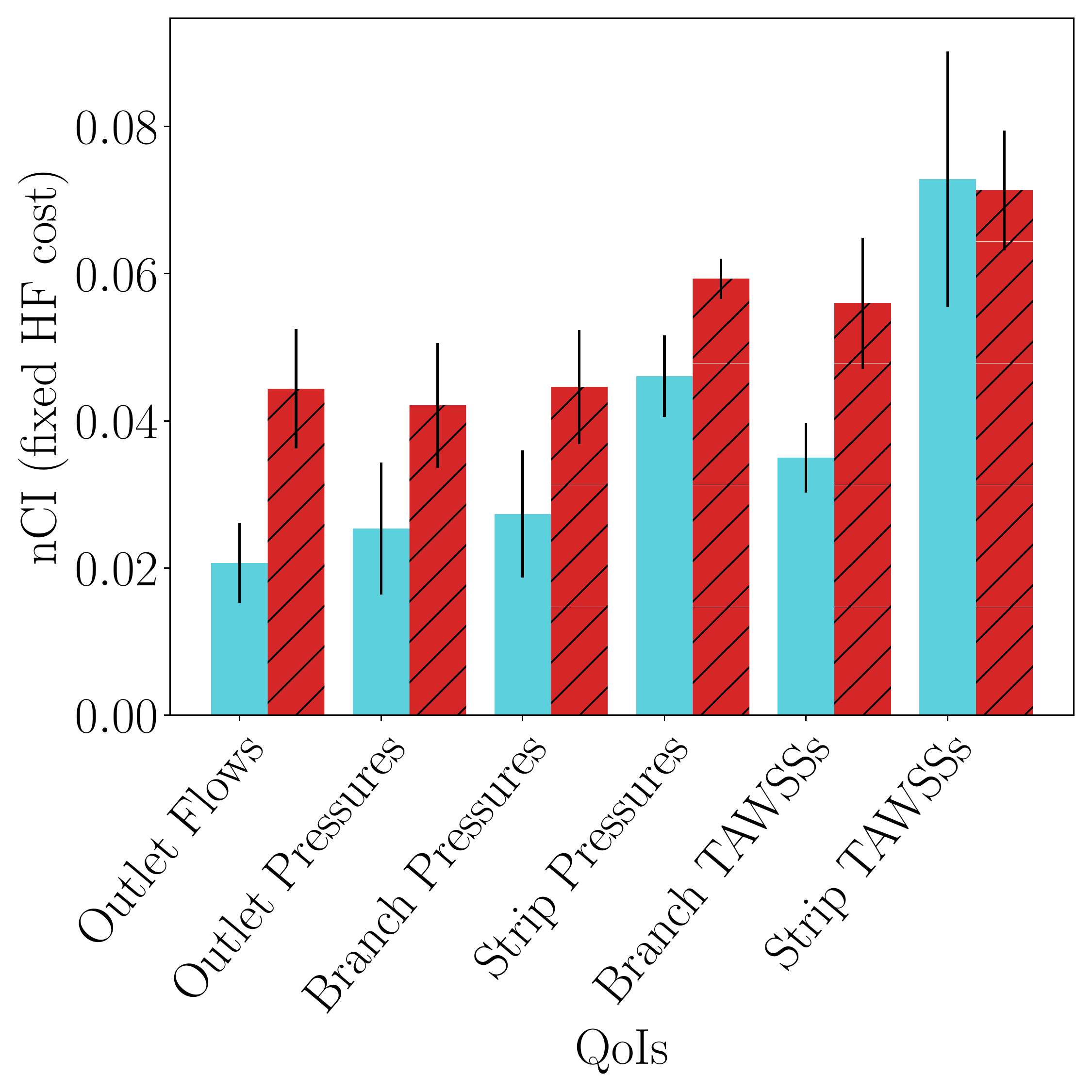}
	\caption{}
	\label{fig:3d1deps-A2} 
\end{subfigure} 
\\
\vspace{1em}
\hspace{\baselineskip}
\columnname{Coronary Models}\\
\begin{subfigure}[t]{0.32\textwidth} 
	\centering
	\includegraphics[width=\textwidth]{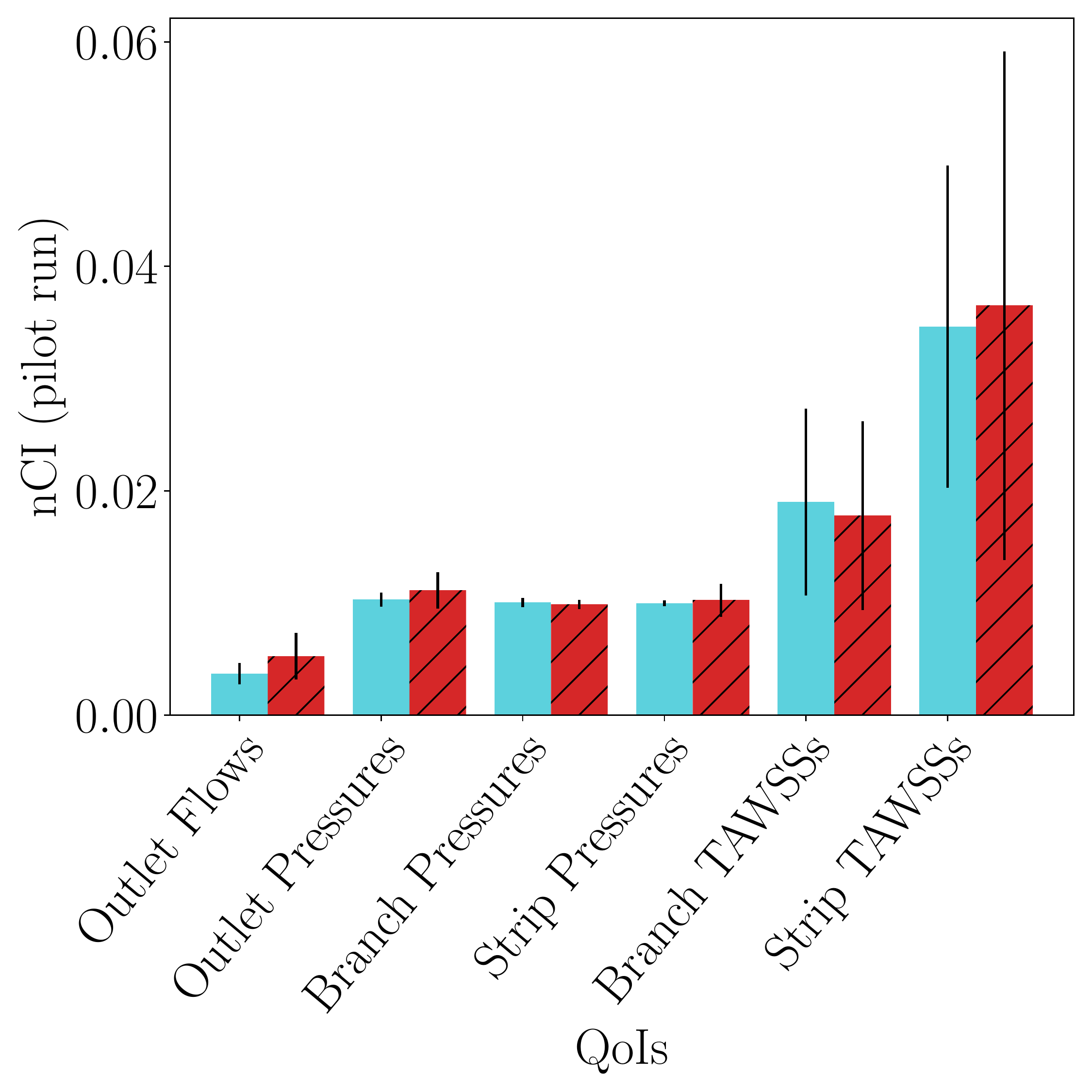}
	\caption{}
	\label{fig:3d1dacc-C2}  
\end{subfigure}	 
\begin{subfigure}[t]{0.32\textwidth} 
	\centering
	\includegraphics[width=\textwidth]{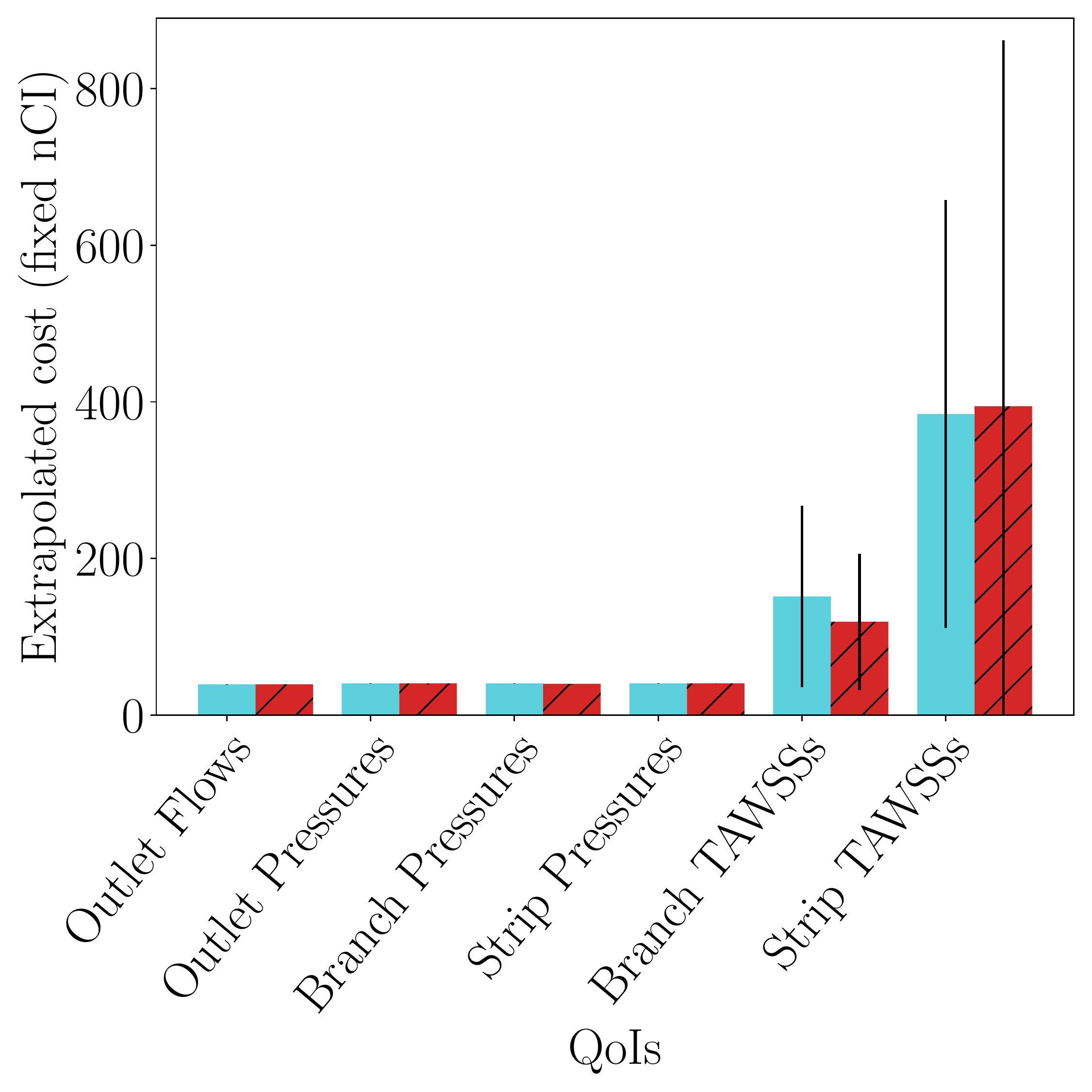}
	\caption{}
	\label{fig:3d1dcost-C2}  
\end{subfigure}	 
\begin{subfigure}[t]{0.32\textwidth} 
	\centering
	\includegraphics[width=\textwidth]{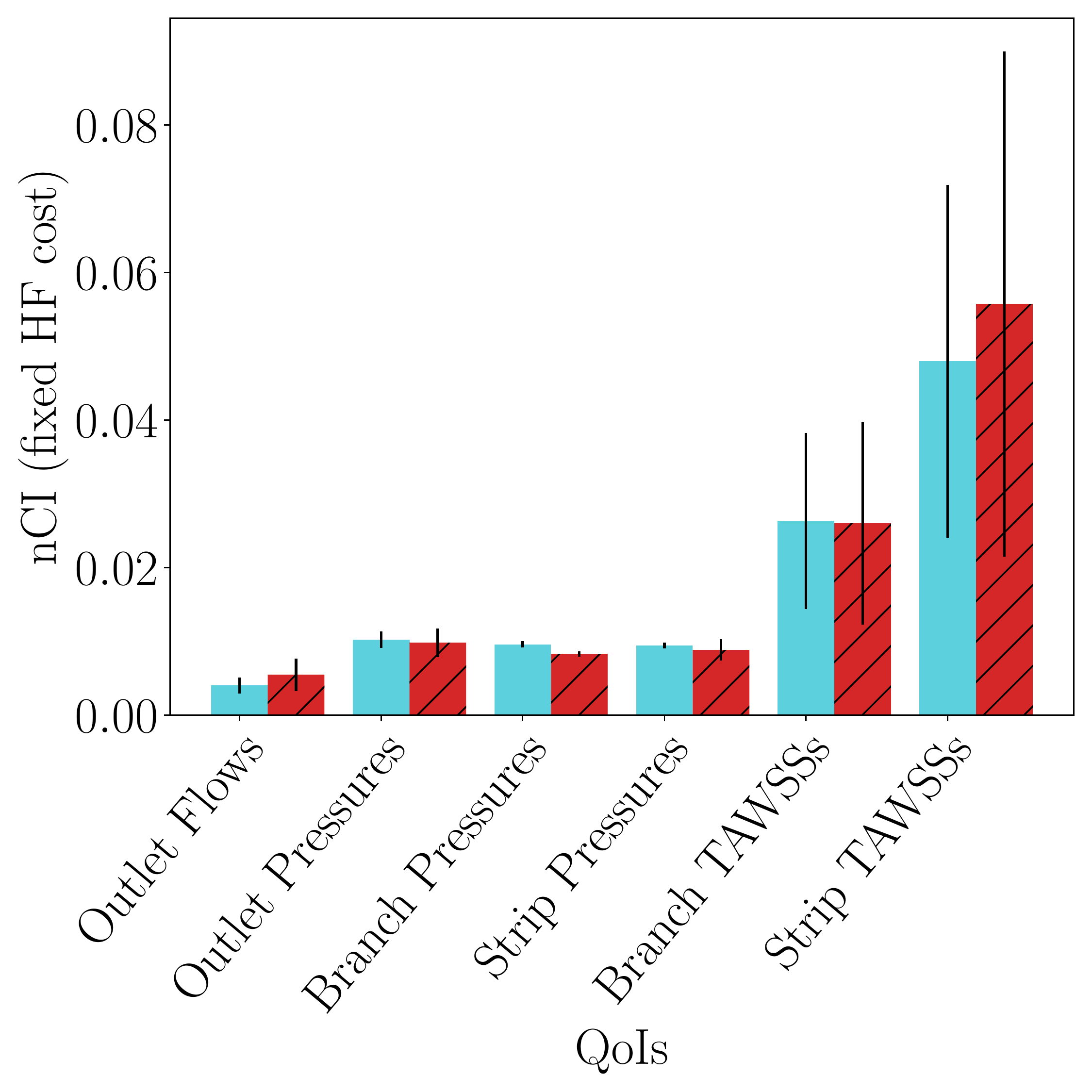}
	\caption{}
	\label{fig:3d1deps-C2}  
\end{subfigure}	
\\
\begin{subfigure}[t]{0.55\textwidth} 
	\centering
	\includegraphics[width=\textwidth]{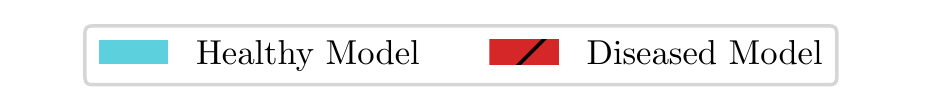}
	\label{fig:3d1dlegend3}  
\end{subfigure}	
\vspace{-1.5em}
\caption{Comparing the performance of 3D-1D-0D MLMF estimators from healthy and diseased geometries for the (top) aorto-femoral and (bottom) coronary models. Averaged nCIs for various categories of QoIs calculated after a fixed pilot run of 250 simulations are shown in plots (a) and (d). Averaged projected extrapolated cost (in units of the equivalent number of 3D fine simulations) to obtain a nCI $\leq 0.01$ are shown in plots (b) and (e). Averaged projected nCI values with the computational budget for all HF simulations equivalent to the cost of 50 3D fine simulations are shown for various QoI categories in plots (c) and (f).}
\label{fig:HDQualitative}
\end{figure}

We can also compare the healthy and diseased geometries in the same way we compared the MLMF estimators with and without zero-dimensional models (\autoref{sec:Res0D}). We again averaged the nCIs from each QoI category for the healthy and diseased models, computed the extrapolated cost of converging to a nCI of 0.01, and predicted the nCI obtained subject to a computational budget constraint (\autoref{fig:HDQualitative}). For both the aorto-femoral and coronary models, we see that the diseased QoIs have similar nCIs to the healthy QoIs across all categories (\hyperref[fig:HDQualitative]{Figures~\ref*{fig:HDQualitative}a} and \hyperref[fig:HDQualitative]{\ref*{fig:HDQualitative}d}). In the aorto-femoral model, which we discussed as a diseased case with slightly more global effects, we do see that most categories have slightly worse averaged nCI for the diseased model. When comparing the expected cost of extrapolation to a nCI of 0.01, we see that the diseased model QoIs are almost always more expensive to converge than the healthy model QoIs for the aorto-femoral model (\hyperref[fig:HDQualitative]{Figure~\ref*{fig:HDQualitative}b}). For the coronary model, many of the QoI nCIs were already below 0.01 after the pilot run, which is why the extrapolated cost is identical here; it is merely the cost of the pilot run (\hyperref[fig:HDQualitative]{Figure~\ref*{fig:HDQualitative}e}). Finally, the projected nCIs for a fixed computational budget are better (lower) or very similar for the healthy model than for the diseased models across all QoI categories of the aorto-femoral model (\hyperref[fig:HDQualitative]{Figure~\ref*{fig:HDQualitative}c}). Again, the coronary model shows similar results between the two cases for the global QoIs, but with a smaller nCI for the healthy model when considering local QoIs (\hyperref[fig:HDQualitative]{Figure~\ref*{fig:HDQualitative}f}). For the local QoIs, this is likely explained by the complex flow features seen in the diseased models which rely more heavily on HF simulations to resolve.

\subsection{Comparing local and global estimators} \label{sec:ResLG}

\begin{figure}[!ht]
\centering
\hspace{\baselineskip}
\columnname{Aorto-Femoral Healthy}\\
\begin{subfigure}[t]{0.32\textwidth}
	\centering
	\includegraphics[trim=10 0 110 75, clip,width=\textwidth]{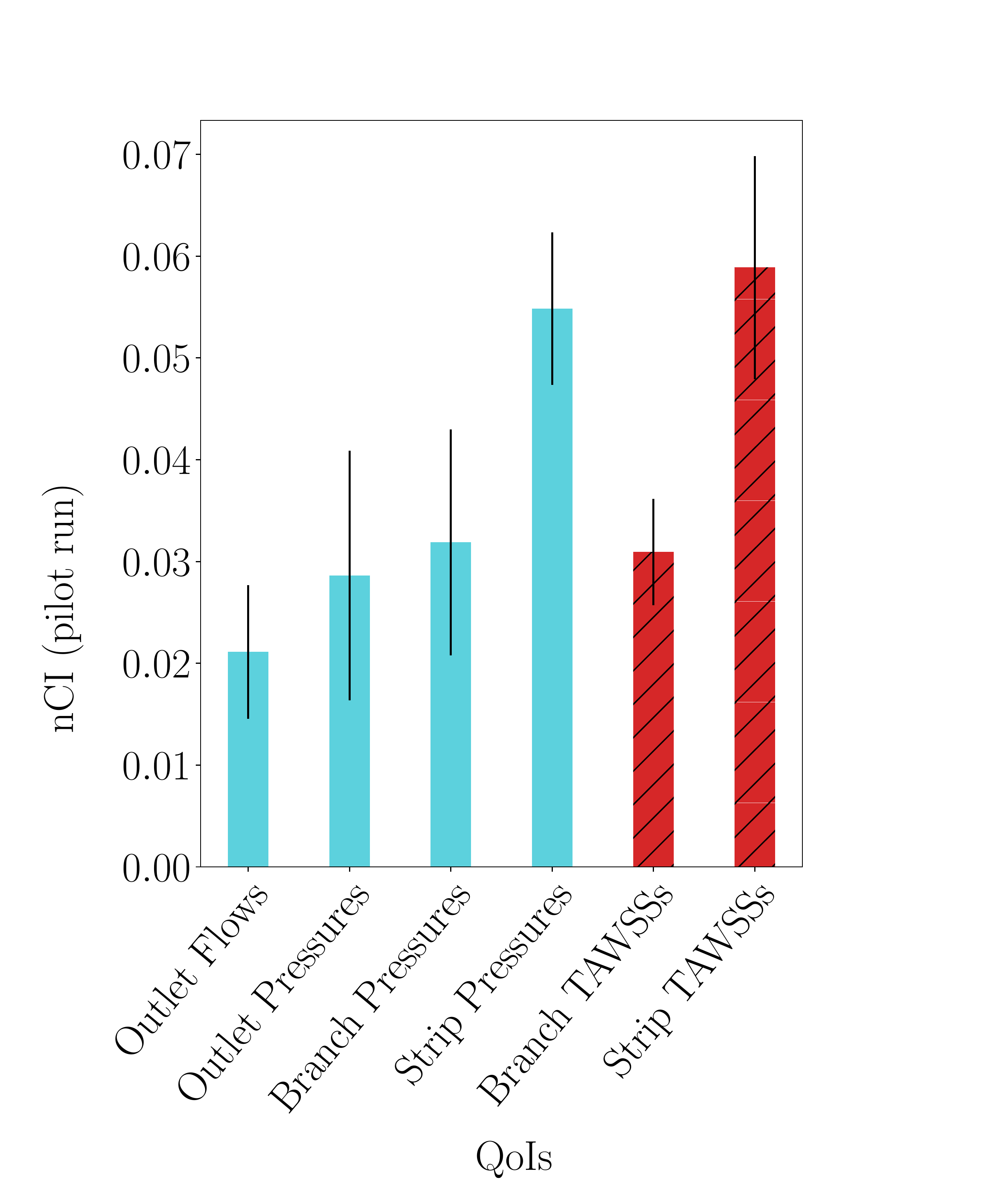}
	\caption{}
	\label{fig:LGacc-A} 
\end{subfigure} 
\begin{subfigure}[t]{0.34\textwidth}
	\centering
	\includegraphics[trim=35 15 70 75, clip,width=\textwidth]{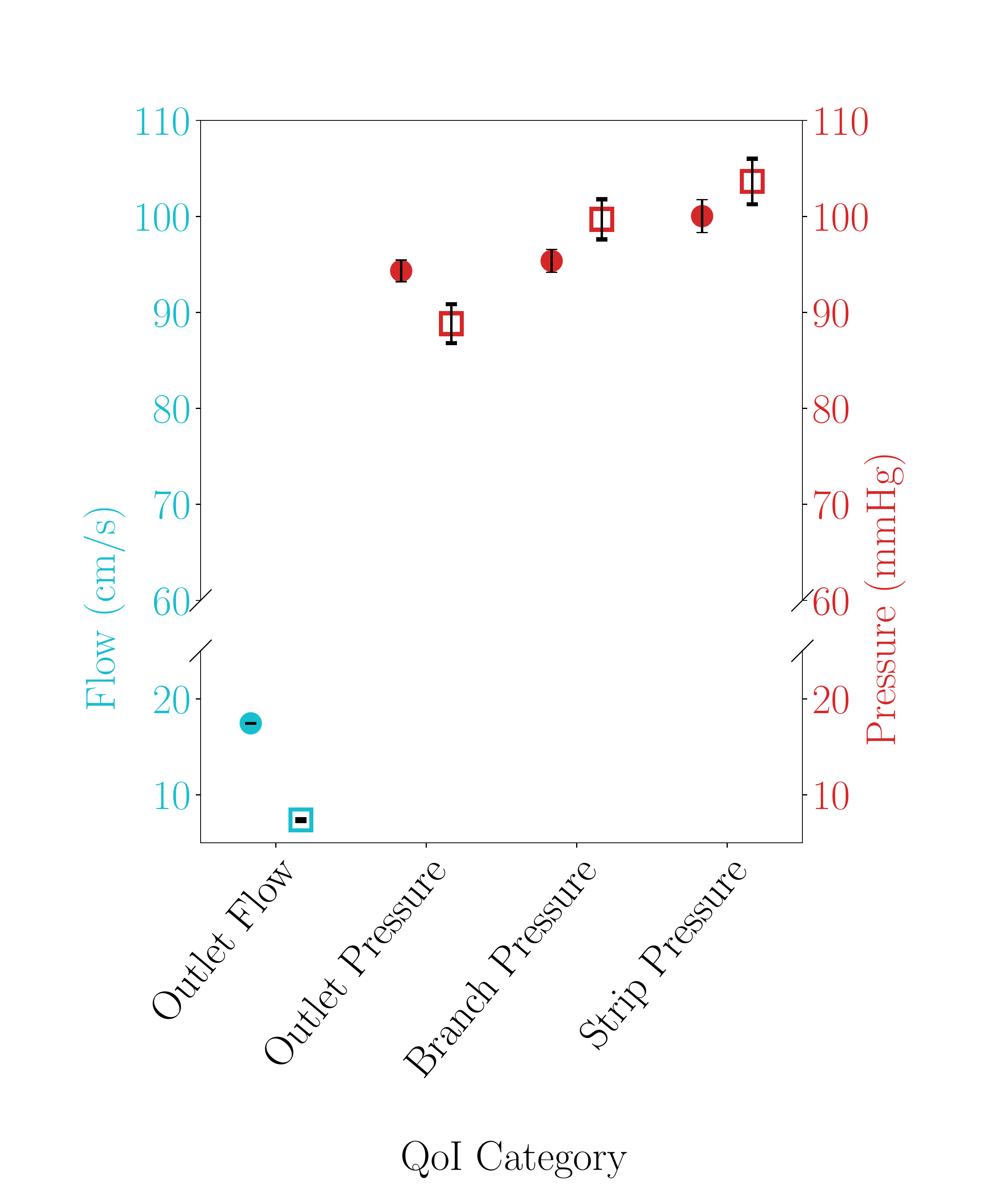}
	\caption{}
	\label{fig:LGglobal-A}
\end{subfigure}
\begin{subfigure}[t]{0.32\textwidth}
	\centering
	\includegraphics[trim=35 15 110 75, clip,width=\textwidth]{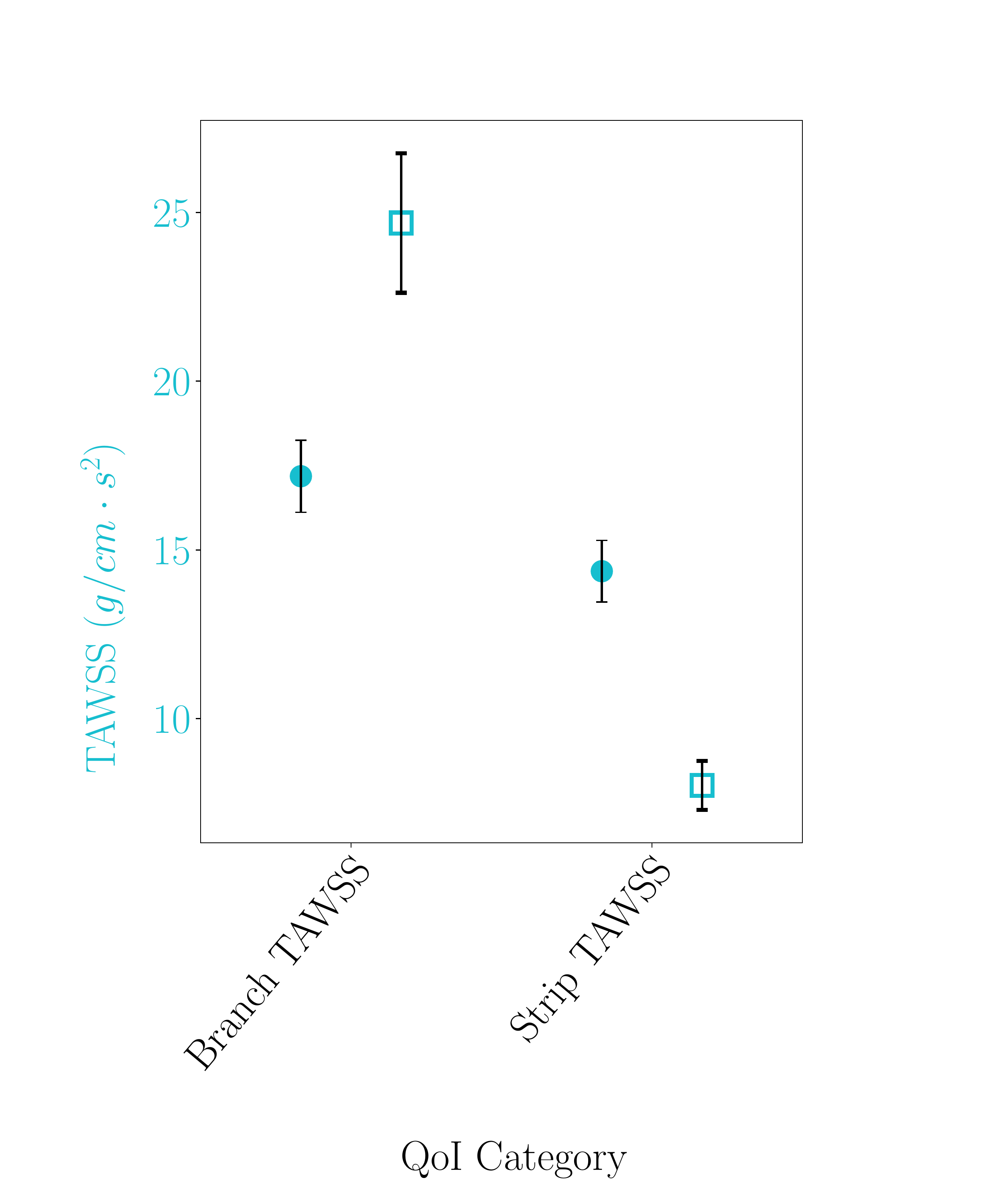}
	\caption{}
	\label{fig:LGlocal-A} 
\end{subfigure} 
\\
\vspace{1em}
\hspace{\baselineskip}
\columnname{Coronary Healthy}\\
\begin{subfigure}[t]{0.32\textwidth}
	\centering
	\includegraphics[trim=10 0 110 75, clip,width=\textwidth]{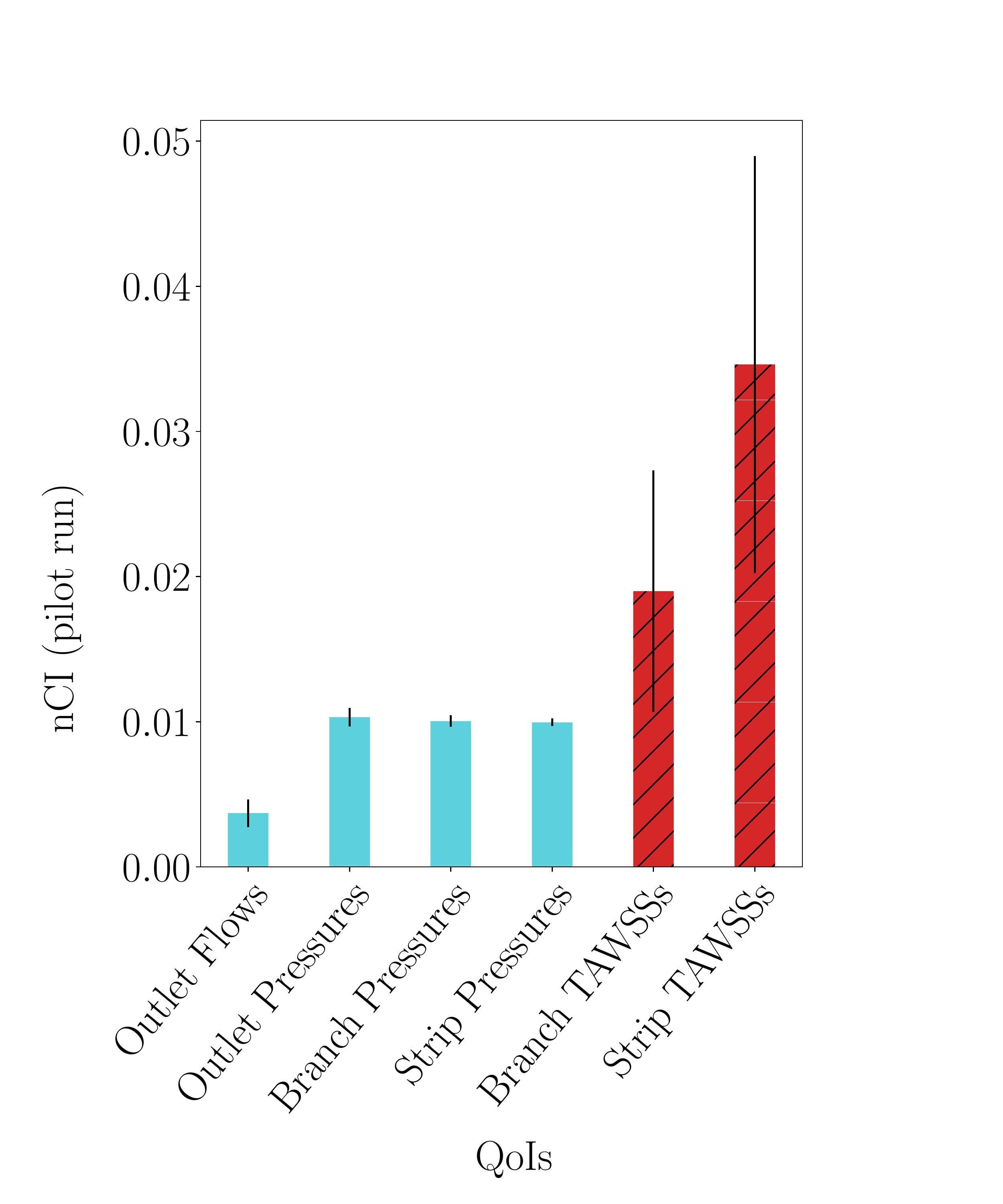}
	\caption{}
	\label{fig:LGacc-C} 
\end{subfigure} 
\begin{subfigure}[t]{0.34\textwidth}
	\centering
	\includegraphics[trim=35 15 50 75, clip,width=\textwidth]{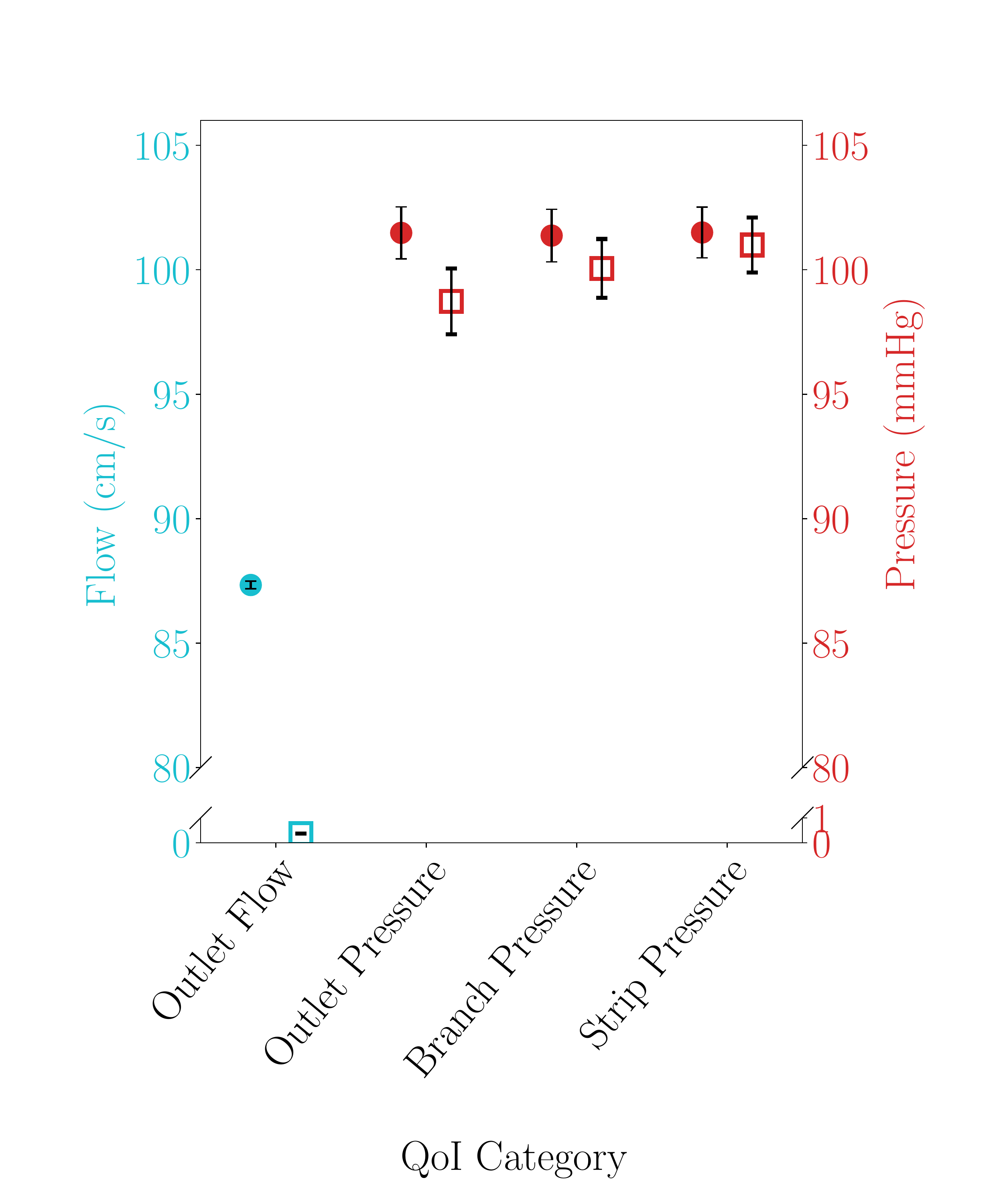}
	\caption{}
	\label{fig:LGglobal-C}
\end{subfigure}
\begin{subfigure}[t]{0.32\textwidth}
	\centering
	\includegraphics[trim=35 15 110 75, clip,width=\textwidth]{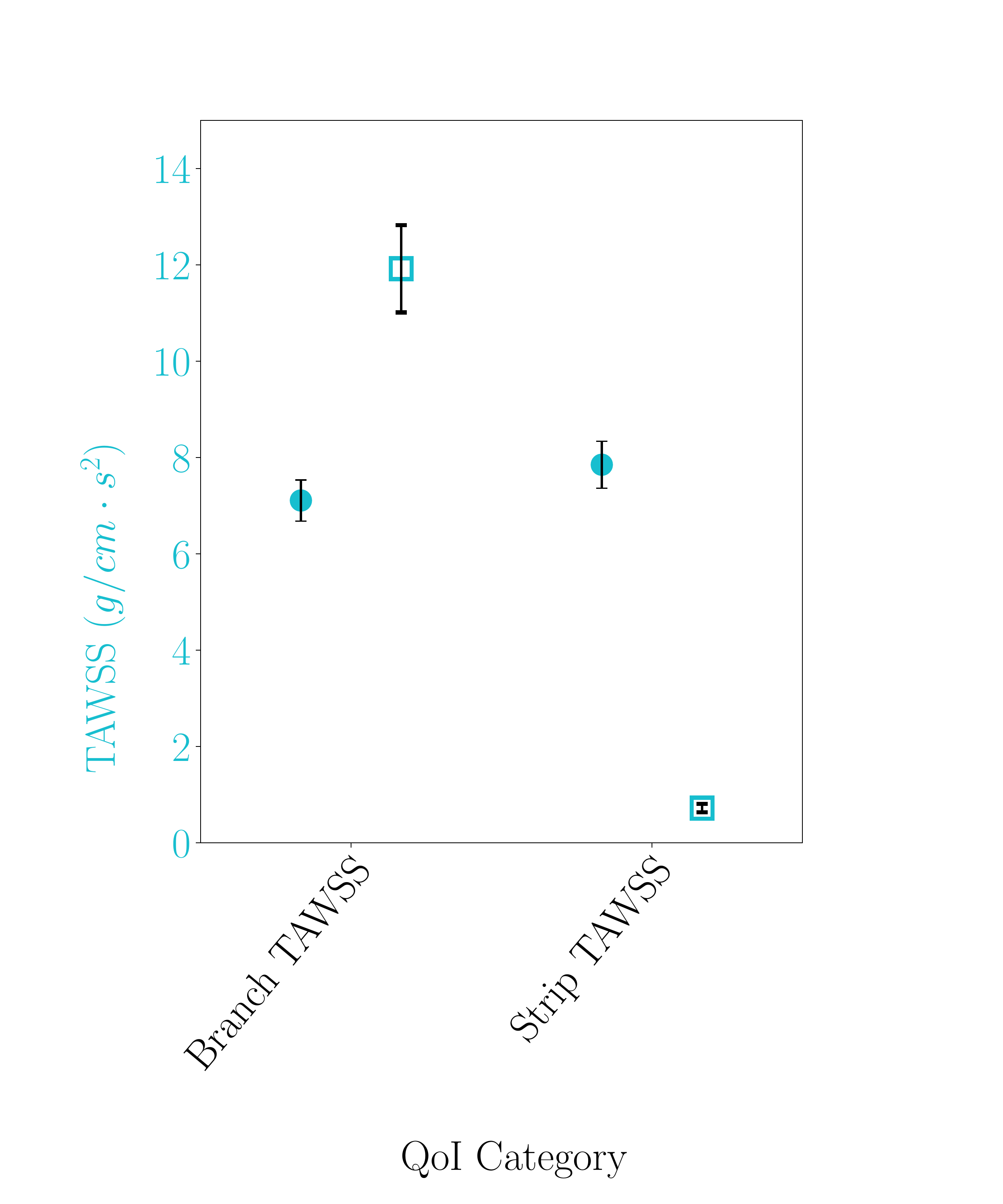}
	\caption{}
	\label{fig:LGlocal-C} 
\end{subfigure} 
\\
\begin{subfigure}[t]{0.33\textwidth} 
	\centering
	\includegraphics[width=\textwidth]{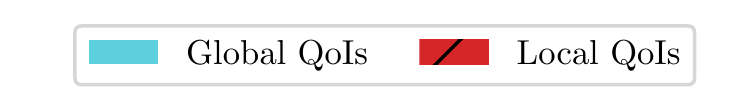}
	\label{fig:LGlegend2}  
\end{subfigure}	
\begin{subfigure}[t]{0.64\textwidth} 
	\centering
	\includegraphics[width=\textwidth]{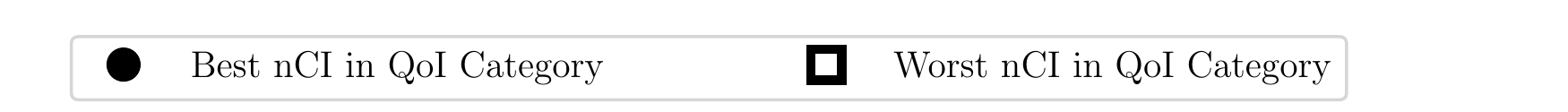}
	\label{fig:LGlegend1}  
\end{subfigure}
\vspace{-1.5em}
\caption{Comparison of estimators for global and local model QoIs for the (top) aorto-femoral and (bottom) coronary models. All values are shown for the healthy models and come from the 3D-1D MLMF scheme. Plots (a) and (d) show the averaged nCIs in both local and global QoI categories. Plots (b) and (e) show the QoI estimators with the best and worst nCI in six global QoI categories. Plots (c) and (f) show the QoI estimators with the best and worst nCI in three local QoI categories.}
\label{fig:LGCompare}
\end{figure}

In~\autoref{sec:ResHD}, we hinted at the difference between estimators for local and global QoIs. We examine these differences more fully here, using the healthy aorto-femoral and coronary models. In general, we see much better (smaller) nCI values for global QoIs than for local QoI estimators (\autoref{fig:LGCompare}).

We first observe a general nCI trend between local and global QoIs by examining average nCI for each QoI category (\hyperref[fig:LGacc-A]{Figures~\ref*{fig:LGacc-A}} and~\ref{fig:LGacc-C} for the aorto-femoral and coronary models, respectively). As before, these values are obtained by computing the sample average and variance of the estimator nCIs from the pilot run for all QoIs of a certain type. The local QoIs have nCIs more than four times larger than the global QoI nCIs calculated from the same pilot run of simulations. This relates to the ability of lower fidelity models to capture the change in global QoIs resulting from changes in the uncertain model parameters. Lower correlations and less accurate estimators are expected for local QoIs such as local flow indicators (as found in our coronary benchmark) or for pathological anatomies. We also see that flows are, in general, the most accurate, followed by pressures.

Since constraining our estimators to be calculated from the fixed pilot run is similar to a budget constraint, we also expect to see better global than local estimator nCIs when calculating estimators with a fixed computational budget. This supports our findings from previous sections. We expect less computational expense to achieve a target nCI level for global QoIs than for local QoIs (\hyperref[fig:3d1dcost-A2]{Figures~\ref{fig:3d1dcost-A2}} and~\ref{fig:3d1dcost-C2}).

Examining specific QoIs reveals the range of nCI values within each QoI category. Identifying the QoIs with the best (smallest) and worst (largest) nCI in each category, we affirm the finding that global QoIs have better best-case and worst-case nCIs than the local QoIs (\hyperref[fig:LGglobal-A]{Figures~\ref*{fig:LGglobal-A}} and~\ref{fig:LGglobal-C} versus~\ref{fig:LGlocal-A}, and~\ref{fig:LGlocal-C}, respectively). More interestingly, we examine the spread in nCI values. The WSS values are not only least accurate, but also differ the most between the best and worst case nCI (\hyperref[fig:LGlocal-A]{Figures~\ref*{fig:LGlocal-A}} and~\ref{fig:LGlocal-C}). The flows are most accurate, as seen in the aggregated trends, and there is also little difference between the best and worst case nCI (\hyperref[fig:LGglobal-A]{Figures~\ref*{fig:LGglobal-A}} and~\ref{fig:LGglobal-C}). 

 We use the estimator values and variances from the 3D-1D MLMF scheme here as we want to observe differences not in converged values of the estimators as in~\autoref{sec:ResHD}, but in the nCIs obtained for local and global QoIs. Since the nCIs were larger (worse) after the pilot run of the MLMF 3D-1D scheme, we consider these here. While these results are presented for the healthy models only, similar trends hold for the diseased cases.

\section{Discussion} \label{sec:Discussion}

This study demonstrated the significant computational benefits which arise from utilizing multifidelity estimators for uncertainty quantification of cardiovascular hemodynamics. By incorporating less accurate inexpensive low fidelity models into the uncertainty quantification workflow, significantly improved estimator confidence can be obtained for the same cost as traditional UQ methods relying only on high fidelity models. We further demonstrated that for this application, the MLMF estimators performed best, achieving the best confidence intervals at the lowest computational cost. Healthy and diseased models of both aorto-femoral and coronary anatomies all demonstrated the computational advantages of the MLMF methods, which supports continued use of these methods in our cardiovascular pipeline. MLMF estimators were shown to be effective for distinct model geometries, suggesting that the method would also be effective for models from other anatomic regions (e.g. pulmonary or cerebral). As the MLMF estimators have proven effective for large problem dimensionality, future work may include many additional uncertain parameters to handle models with more complex anatomy, closed loop LPN boundary conditions, and other FSI models. 

By quantifying the effect of uncertain parameters from both boundary conditions and model material and fluid properties on global and local QoIs, the proposed framework demonstrated robustness for different applications of interest. As this study was conducted for the purpose of demonstrating the power of the MLMF method with an eye toward future cardiovascular modeling applications and clinical questions, there are limitations of this study which can be addressed in future work. Uncertain parameters were assumed to be distributed uniformly according to literature targets. In future work, these uncertain mean values could be tuned to patient-specific target values. In this case, a more sophisticated uncertain distribution, such as a Gaussian, may more accurately reflect prior knowledge. It would also be possible in the future to assimilate uncertain distributions directly from patient data if given repeated readings of patient-specific targets such as heart rate and blood pressure.

To possibly further reduce the number of high-fidelity simulations, future work can investigate a more ideal distribution of simulations in the pilot study. When using 25 simulations at each level as we did in this study, we obtain nCIs on the order of 0.01 for flow and pressure QoIs, suggesting we are likely utilizing too many 3D simulations. We could reshape the pilot size using tools already included in Dakota. This will result in higher initial nCIs after the pilot study, but we will eventually obtain nCIs on the order of 0.01 with possibly fewer overall high-dimensional simulations (though more low-fidelity simulations) following or extrapolation and convergence simulations.

Additional uncertain parameters could also be explored in future studies. The three-dimensional Navier-Stokes equations~\eqref{equ:3DNS} include at least one parameter we did not explore in this study: the vessel wall parameter for thickness $\zeta$. From previous work~\cite{TRAN_2019402}, we know that vessel wall thickness uncertainties do propagate to have a significant impact on QoIs such as wall strain, though their impact on other quantities is less significant. These uncertainties should be considered in the future. With that said, model geometry is likely the source of most of our unaccounted for uncertainties. The models used in this study were constructed using the SimVascular workflow. This workflow requires semi-automated lumen segmentation of vessels, which is an operator-dependent method. This leads to inherent uncertainties in the model geometry which were not addressed in the current study. Future work should explore the effect of these modeling uncertainties. One possible approach could rely on machine learning techniques for model building to assimilate uncertain distributions for the vessel lumina. 

In this study, many quantities of interest were tested, but a more targeted approach could be developed with a specific clinical question in mind. For instance, QoIs could be concentrated in aneurysms or other diseased regions, rather than distributed throughout the model as chosen for the purpose of exploration in this study. Additionally, quantities of interest were chosen to be single values obtained from temporally averaging over an entire cardiac cycle. In clinical applications, we are often interested in time-dependent QoIs. The current methods can be easily extended to provide estimators at multiple time points. This would give full waveforms and provide enhanced clinical information. Similarly, time-averaged model pressure and TAWSS values could be computed along vessel centerlines, with uncertainties computed at each point along this centerline, instead of one estimator per model branch or strip of interest.

We assumed a consistent parameterization among models for this work. However, in general this may not be the case. To handle that scenario of inconsistent parameterization, we will explore the possibility of adopting dimension reduction strategies, for instance an active subspace mapping as described in~\cite{Geraci_2018}. This would allow for additional low-fidelity modeling approaches. Recent developments in reduced order modeling (ROM) are one way to expand our low-fidelity modeling approach, as new model generation approaches could easily be fit into the MLMF framework as additional fidelity levels, or used to replace one or both of our low fidelity models. Integration of ROMs within the multifidelity control variate framework has recently been demonstrated~\cite{Blonigan2019,Blonigan2020}, and could be extended to the MLMF approach. One such ROM method is the so-called ``hierarchical model reduction'' approach~\cite{Perotto2010,Perotto2013,Aletti2018}, which allows for more flexibility in fidelity and discretization levels through the use of Fourier-like expansions and has been effectively demonstrated for general cardiovascular problems~\cite{MansillaAlvarez2017,BLANCO2015,Alvarez2018} as well as specifically for cardiovascular UQ applications~\cite{GUZZETTI2020}. Another method approximated the Navier-Stokes partial differential equations with a reduced basis and neural networks~\cite{Pegolotti2019}. Such ROMs may help address some of the limitations we see with our current one-dimensional modeling, such as extension to multi-inlet and highly tapered anatomies, and could, for example, enable reduced modeling of more complex models that include ventricular flow. Certain quantities of interest fall outside the bounds of any of our low-fidelity modeling capabilities. New methods such as~\cite{Pegolotti2019} could extend our UQ methods to include thrombus modeling or QoIs such as residence time and flow recirculation.

In addition to development aimed at addressing such concerns, there is active development on automated generation of the one- and zero-dimensional low fidelity models used in this study, including zero-dimensional models constructed directly from three-dimensional lofted models as the one-dimensional models are. An automated process to generate low fidelity models would allow the proposed uncertainty quantification framework to be implemented more easily across a range of patient data, anatomies, and diseases.

\section{Conclusion} \label{sec:Conclusion}

As clinicians continue to incorporate computational simulations into patient treatment, it is imperative that robust methods of uncertainty quantification are readily available. To assist in widespread adoption, these UQ methods need to be adaptable to practical computational budgets and be relatively easy to use. By using MLMF estimators, we have demonstrated enormous cost savings compared to traditional UQ approaches such as Monte Carlo, which rely only on the highest fidelity three-dimensional models. By incorporating the existing Dakota framework, our workflow is streamlined and partially automated, avoiding the need for continual monitoring by the user. As such, estimators with high levels of confidence can be obtained with a reasonable computational budget in a semi-automated workflow. This study demonstrated the effectiveness of the method on multiple patient-specific model geometries for both healthy and diseased anatomies. By substituting LF one-dimensional models with zero-dimensional LPNs which only loosely approximate the HF model geometry, we can obtain even higher estimator confidence improvements for the same computational cost. If we are interested in global quantities of interest, we require fewer simulations than if we are interested in local quantities of interest. This is due to the degraded spatial resolution of the LF models, resulting in local QoIs with reduced correlation between model fidelities. This study examined the effect of uncertain boundary conditions and model parameters on a wide range of quantities of interest, demonstrating the robustness of MLMF estimators. Uncertainty quantification is essential for providing clinicians with not only simulated predictions, but the confidence of those predictions. In this study we investigated the possibility of employing multifidelity approaches in the future to improve the confidence of targeted clinical QoIs while using uncertainties assimilated from patient data. Several additional multifidelity approaches more general than MLMF estimators are now available that will be the subject of future studies. These include generalized approximate control variate approaches~\cite{Gorodetsky_2020,Geraci_2019AIAA} and latent variables~\cite{Gorodetsky_2018}.

\section*{Acknowledgments}
This work was supported by NIH grants R01-EB018302 and R01-HL123689 (P.I. Alison Marsden), the National Science Foundation under grant  NSF CDSE CBET 1508794, and a Center for Computing Research Summer Fellowship at Sandia National Laboratories (Casey Fleeter). This work used computational resources from the Extreme Science and Engineering Discovery Environment (XSEDE)~\cite{Towns_2014}, which is supported by National Science Foundation grant number ACI-1548562, and the Stanford Research Computing Center (SRCC). We thank Mahidhar Tatineni for his assistance building Dakota on the Comet cluster and advice on efficiency in the workflow, which was made possible through the XSEDE Extended Collaborative Support Service (ECSS) program~\cite{Wilkins_2016}. We thank Michael S. Eldred of Sandia National Laboratories for his ongoing support with Dakota. We also acknowledge the open source SimVascular project at \url{http://www.simvascular.org} and the open source Dakota toolkit at \url{https://dakota.sandia.gov}.

Sandia National Laboratories is a multimission laboratory managed and operated by National Technology \& Engineering Solutions of Sandia, LLC, a wholly owned subsidiary of Honeywell International Inc., for the U.S. Department of Energy's National Nuclear Security Administration under contract DE-NA0003525. This paper describes objective technical results and analysis. Any subjective views or opinions that might be expressed in the paper do not necessarily represent the views of the U.S. Department of Energy or the United States Government.

\section*{Disclosures}

The authors have no conflicts of interest to disclose.

\bibliography{CaseyFleeter-Short}

\end{document}